\documentclass[twocolumn]{aastex7}

\expandafter\def\csname editcolor1\endcsname{magenta}
\expandafter\def\csname editcolor2\endcsname{blue}  
\expandafter\def\csname editcolor3\endcsname{violet} 

\usepackage{newtxtext,newtxmath}
\usepackage[T1]{fontenc}
\usepackage{graphicx}	
\usepackage{amsmath}	
\usepackage{float}
\usepackage{bookmark} 
\bookmarksetup{numbered, open}
\usepackage{enumitem}
\setlist[enumerate]{itemsep=0mm}
\usepackage{hyperref}
\usepackage{subfigure}
\usepackage{xspace}
\usepackage{xcolor} 
\usepackage{multirow}
\usepackage{tabularx}

\newcommand{\Msol}{\ensuremath{M_{\odot}}\xspace}

\newcommand{\kms}{km~s\ensuremath{^{-1}}\xspace}

\newcommand{\Halpha}{H\ensuremath{\alpha}\xspace}
\newcommand{\Hbeta}{H\ensuremath{\beta}\xspace}
\newcommand{\Hgamma}{H\ensuremath{\gamma}\xspace}
\newcommand{\Hdelta}{H\ensuremath{\delta}\xspace}

\newcommand{\Nifs}{\ensuremath{^{56}}Ni\xspace}

\newcommand{\mic}{\ensuremath{\mu}m\xspace}

\newcommand{\snii}{SN~II\xspace}
\newcommand{\sneii}{SNe~II\xspace}
\newcommand{\sniip}{SN~IIP\xspace}

\newcommand{\sneiin}{SNe~IIn\xspace}
\newcommand{\ixf}{SN~2023ixf\xspace}
\newcommand{\acko}{SN~2022acko\xspace}
\newcommand{\ggi}{SN~2024ggi\xspace}

\newcommand{\jwst}{{\it JWST}\xspace}

\newcommand{\spitzer}{{\it Spitzer}\xspace}
\newcommand{\hlwl}[3]{{#1}~\ensuremath{{#2}}~({#3}~\mic)}
\newcommand{\hl}[2]{{#1}~\ensuremath{{#2}}\xspace}
\newcommand{\uhl}[2]{\ion{H}{1}~(\ensuremath{{#1}})~({#2}~\mic)}

\submitjournal{\apj}

\graphicspath{{./}}

\begin{document}

\title{{\it JWST} Observations of SN~2023ixf I: Completing the Early Multi-Wavelength Picture with Plateau-phase Spectroscopy}

\correspondingauthor{James M DerKacy}

\author[0000-0002-7566-6080]{J.~M.~DerKacy}
\affiliation{Space Telescope Science Institute, 3700 San Martin Drive, Baltimore, MD 21218-2410, USA}
\affiliation{Department of Physics, Virginia Tech, Blacksburg, VA 24061, USA}
\email{jderkacy@stsci.edu}

\author[0000-0002-5221-7557]{C.~Ashall}
\affiliation{Institute for Astronomy, University of Hawai’i at Manoa, 2680 Woodlawn Dr., Hawai’i, HI 96822, USA}
\affiliation{Department of Physics, Virginia Tech, Blacksburg, VA 24061, USA}
\email{cashall@hawaii.edu}

\author[0000-0001-5393-1608]{E.~Baron}
\affiliation{Planetary Science Institute, 1700 East Fort Lowell Road, Suite 106, Tucson, AZ 85719-2395, USA}
\affiliation{Hamburger Sternwarte, Gojenbergsweg 112, D-21029 Hamburg, Germany}
\email{ebaron@psi.edu}

\author[0000-0001-7186-105X]{K.~Medler}
\affiliation{Institute for Astronomy, University of Hawai’i at Manoa, 2680 Woodlawn Dr., Hawai’i, HI 96822, USA}
\affiliation{Department of Physics, Virginia Tech, Blacksburg, VA 24061, USA}
\email{kmedler@hawaii.edu}

\author[0000-0001-5888-2542]{T.~Mera}
\affiliation{Department of Physics, Florida State University, 77 Chieftan Way, Tallahassee, FL 32306, USA}
\email{tbm20x@fsu.edu}

\author[0000-0002-4338-6586]{P.~Hoeflich}
\affiliation{Department of Physics, Florida State University, 77 Chieftan Way, Tallahassee, FL 32306, USA}
\email{}

\author[0000-0002-9301-5302]{M.~Shahbandeh}
\altaffiliation{STScI Fellow}
\affiliation{Space Telescope Science Institute, 3700 San Martin Drive, Baltimore, MD 21218-2410, USA}
\email{mshahbandeh@stsci.edu}

\author[0000-0003-4625-6629]{C.~R.~Burns}
\affiliation{Observatories of the Carnegie Institution for Science, 813 Santa Barbara Street, Pasadena, CA 91101, USA}
\email{cburns@carnegiescience.edu}

\author[0000-0002-5571-1833]{M.~D.~Stritzinger}
\affiliation{Department of Physics and Astronomy, Aarhus University, Ny Munkegade 120, DK-8000 Aarhus C, Denmark}
\email{max@phys.au.dk}

\author[0000-0002-2471-8442]{M.~A.~Tucker}
\altaffiliation{CCAPP Fellow}
\affiliation{Center for Cosmology and AstroParticle Physics, The Ohio State University, 191 W. Woodruff Ave., Columbus, OH 43210, USA}
\email{tuckerma95@gmail.com}

\author[0000-0003-4631-1149]{B.~J.~Shappee}
\affiliation{Institute for Astronomy, University of Hawai’i at Manoa, 2680 Woodlawn Dr., Hawai’i, HI 96822, USA}
\email{shappee@hawaii.edu}

\author[0000-0002-4449-9152]{K.~Auchettl}
\affiliation{School of Physics, The University of Melbourne, Parkville, VIC 3010, Australia}
\affiliation{Department of Astronomy and Astrophysics, University of California, Santa Cruz, CA 95064, USA}
\email{katie.auchettl@unimelb.edu.au}

\author[0000-0002-4269-7999]{C.~R.~Angus}
\affiliation{Astrophysics Research Centre, School of Mathematics and Physics, Queen’s University Belfast, Belfast BT7 1NN, UK}
\affiliation{DARK, Niels Bohr Institute, University of Copenhagen, Jagtvej 128, DK-2200 Copenhagen {\O} Denmark}
\email{c.angus@qub.ac.uk}

\author[0000-0002-2164-859X]{D.~D.~Desai}
\affiliation{Institute for Astronomy, University of Hawai’i at Manoa, 2680 Woodlawn Dr., Hawai’i, HI 96822, USA}
\email{dddesai@hawaii.edu}

\author[0000-0003-3429-7845]{A.~Do}
\affiliation{Institute of Astronomy and Kavli Institute for Cosmology, Madingley Road, Cambridge CB3 0HA, UK}
\email{ajmd6@cam.ac.uk}

\author[0000-0001-9668-2920]{J.~T.~Hinkle}
\altaffiliation{FINNEST FI}
\affiliation{Institute for Astronomy, University of Hawai’i at Manoa, 2680 Woodlawn Dr., Hawai’i, HI 96822, USA}
\email{jhinkle6@hawaii.edu}

\author[0000-0003-3953-9532]{W.~B.~Hoogendam}
\altaffiliation{NSF Graduate Research Fellow}
\affiliation{Institute for Astronomy, University of Hawai’i at Manoa, 2680 Woodlawn Dr., Hawai’i, HI 96822, USA}
\email{willemh@hawaii.edu}

\author[0000-0003-1059-9603]{M.~E.~Huber}
\affiliation{Institute for Astronomy, University of Hawai’i at Manoa, 2680 Woodlawn Dr., Hawai’i, HI 96822, USA}
\email{mehuber7@hawaii.edu}

\author[0000-0003-3490-3243]{A.~V.~Payne}
\affiliation{Space Telescope Science Institute, 3700 San Martin Drive, Baltimore, MD 21218-2410, USA}
\email{apayne@stsci.edu}

\author[0000-0002-6230-0151]{D.~O.~Jones}
\affiliation{Institute for Astronomy, University of Hawai’i, 640 N. A’ohoku Pl., Hilo, HI 96720, USA}
\email{dojones@hawaii.edu}

\author[0009-0008-3724-1824]{J.~Shi}
\affiliation{School of Physics, The University of Melbourne, Parkville, VIC 3010, Australia}
\email{jennifer.shi@student.unimelb.edu.au}

\author[0009-0005-5121-2884]{M.~Y.~Kong}
\affiliation{Institute for Astronomy, University of Hawai’i at Manoa, 2680 Woodlawn Dr., Hawai’i, HI 96822, USA}
\email{ykong2@hawaii.edu}

\author[0009-0003-8153-9576]{S.~Romagnoli}
\affiliation{School of Physics, The University of Melbourne, Parkville, VIC 3010, Australia}
\email{romagnolis@student.unimelb.edu.au}

\author[0009-0000-6821-9285]{A.~Syncatto}
\affiliation{Institute for Astronomy, University of Hawai’i at Manoa, 2680 Woodlawn Dr., Hawai’i, HI 96822, USA} 
\affiliation{Institute for Astronomy, University of Hawai’i, 200 W Kawili St, Hilo, HI 96720, USA}
\email{jaq45@hawaii.edu}

\author[0000-0001-5221-0243]{S.~Moran}
\affiliation{School of Physics and Astronomy, University of Leicester, University Road, Leicester LE1 7RH, UK}
\email{shane.moran@leicester.ac.uk}

\author[0009-0001-9148-8421]{E.~Fereidouni}
\affiliation{Department of Physics, Florida State University, 77 Chieftan Way, Tallahassee, FL 32306, USA}
\email{ef22g@fsu.edu}

\author[0000-0001-6272-5507]{P.~J.~Brown}
\affiliation{George P. and Cynthia Woods Mitchell Institute for Fundamental Physics and Astronomy, Texas A\&M University, Department of Physics and Astronomy, College Station, TX 77843, USA}
\email{pbrown801@tamu.edu}

\author[0000-0003-0209-674X]{M.~Engesser}
\affiliation{Space Telescope Science Institute, 3700 San Martin Drive, Baltimore, MD 21218-2410, USA}
\email{mengesser@stsci.edu}

\author[0000-0003-2238-1572]{O.~D.~Fox}
\affiliation{Space Telescope Science Institute, 3700 San Martin Drive, Baltimore, MD 21218-2410, USA}
\email{ofox@stsci.edu}

\author[0000-0002-1296-6887]{L.~Galbany}
\affiliation{Institute of Space Sciences (ICE, CSIC), Campus UAB, Carrer de Can Magrans, s/n, E-08193 Barcelona, Spain}
\affiliation{Institut d’Estudis Espacials de Catalunya (IEEC), E-08034 Barcelona, Spain}
\email{lgalbany@ice.csic.es}

\author[0000-0003-1039-2928]{E.~Y.~Hsiao}
\affiliation{Department of Physics, Florida State University, 77 Chieftan Way, Tallahassee, FL 32306, USA}
\email{ehsiao@fsu.edu}

\author[0000-0001-6069-1139]{T.~de~Jaeger}
\affiliation{Institute for Astronomy, University of Hawai'i at Manoa, 2680 Woodlawn Dr., Hawai'i, HI 96822, USA}
\email{thomas.dejaeger@lpnhe.in2p3.fr}

\author[0000-0001-8367-7591]{S.~Kumar}
\affiliation{Department of Astronomy, University of Virginia, 530 McCormick Road, Charlottesville, VA 22904, USA}
\email{bsw2dc@virginia.edu}

\author[0000-0002-3900-1452]{J.~Lu}
\affiliation{Department of Physics \& Astronomy, Michigan State University, East Lansing, MI, USA}
\email{lujing8@msu.edu}

\author[0000-0002-5529-5593]{M.~Matsuura}
\affiliation{Cardiff Hub for Astrophysical Research and Technology (CHART), School of Physics and Astronomy, Cardiff University, The Parade, Cardiff CF24 3AA, UK}
\email{MatsuuraM@cardiff.ac.uk}

\author[0000-0001-6876-8284]{P.~Mazzali}
\affiliation{Astrophysics Research Institute, Liverpool John Moores University, UK}
\affiliation{Max-Planck Institute for Astrophysics, Garching, Germany}
\email{p.mazzali@ljmu.ac.uk}

\author[0000-0003-2535-3091]{N.~Morrell}
\affiliation{Las Campanas Observatory, Carnegie Observatories, Casilla 601, La Serena, Chile}
\email{nmorrell@carnegiescience.edu}

\author[0000-0002-7305-8321]{C. M. Pfeffer}
\altaffiliation{NSF Graduate Research Fellow}
\affiliation{Institute for Astronomy, University of Hawai’i at Manoa, 2680 Woodlawn Dr., Hawai’i, HI 96822, USA}
\email{cpfeffer@hawaii.edu}

\author[0000-0003-2734-0796]{M.~M.~Phillips}
\affiliation{Las Campanas Observatory, Carnegie Observatories, Casilla 601, La Serena, Chile}
\email{mmp@lco.cl}

\author[0000-0002-4410-5387]{A.~Rest}
\affiliation{Space Telescope Science Institute, 3700 San Martin Drive, Baltimore, MD 21218-2410, USA}
\email{arest@stsci.edu}

\author[0000-0001-6107-0887]{S.~Shiber}
\affiliation{Department of Physics, Florida State University, 77 Chieftan Way, Tallahassee, FL 32306, USA}
\email{sshiber@fsu.edu}

\author[0000-0002-7756-4440]{L.~Strolger}
\affiliation{Space Telescope Science Institute, 3700 San Martin Drive, Baltimore, MD 21218-2410, USA}
\email{strolger@stsci.edu}

\author[0000-0002-8102-181X]{N.~B.~Suntzeff}
\affiliation{George P. and Cynthia Woods Mitchell Institute for Fundamental Physics and Astronomy, Texas A\&M University, Department of Physics and Astronomy, College Station, TX 77843, USA}
\email{nsuntzeff@tamu.edu}

\author[0000-0001-7380-3144]{T.~Temim}
\affiliation{Department of Astrophysical Sciences, Princeton University, Princeton, NJ 08544, USA}
\email{temim@astro.princeton.edu}

\author[0000-0002-1481-4676]{S.~Tinyanont}
\affiliation{Department of Astronomy and Astrophysics, University of California, Santa Cruz, CA 95064, USA}
\email{samaporn@NARIT.OR.TH}

\author[0000-0001-5233-6989]{Q.~Wang}
\affiliation{Department of Physics and Kavli Institute for Astrophysics and Space Research, Massachusetts Institute of Technology, 77 Massachusetts Avenue, Cambridge, MA 02139, USA}
\email{qnwang12@gmail.com}

\author[0000-0002-4000-4394]{R.~Wesson}
\affiliation{Cardiff Hub for Astrophysical Research and Technology (CHART), School of Physics and Astronomy, Cardiff University, The Parade, Cardiff CF24 3AA, UK}
\affiliation{Department of Physics and Astronomy, University College London (UCL), Gower Street, London WC1E 6BT, UK}
\email{WessonR1@cardiff.ac.uk}

\author[0000-0001-7488-4337]{S.~H.~Park}
\affiliation{Department of Physics and Astronomy, Seoul National University, Gwanak-ro 1, Gwanak-gu, Seoul, 08826, South Korea}
\email{rogersh0125@snu.ac.kr}

\author[0000-0003-3643-839X ]{J.~Rho}
\affiliation{SETI Institute, 339 Bernardo Ave., Ste. 200, Mountain View, CA 94043, USA}
\email{jrho@seti.org}

\begin{abstract}
We present and analyze panchromatic (0.35--14~\mic) spectroscopy of 
the Type II supernova 2023ixf, including near- and mid-infrared 
spectra obtained 33.6 days after explosion during the plateau-phase,
with the {\it James Webb Space Telescope} (\jwst). This is the first 
in a series of papers examining the evolution of SN~2023ixf with \jwst 
spanning the initial 1000 days after explosion, monitoring the formation 
and growth of molecules and dust in ejecta and surrounding environment.
The \jwst infrared spectra are overwhelmingly dominated by H lines, 
whose profiles reveal ejecta structures, including flat tops,
blue notches, and red shoulders, unseen in the optical spectra. We 
characterize the nature of these structures, concluding that they likely 
result from a combination of ejecta geometry, viewing angle, and opacity 
effects. We find no evidence for the formation of dust precursor 
molecules such as carbon-monoxide (CO), nor do we observe an infrared 
excess attributable to dust. These observations imply that the 
detections of molecules and dust in SN~2023ixf at later epochs arise 
either from freshly synthesized material within the ejecta or 
circumstellar material at radii not yet heated by the supernova at 
this epoch.
\end{abstract}


\section{Introduction} \label{sec:intro}

Nearby supernovae (SNe) provide valuable insight into the late stages 
of stellar evolution and explosion physics, which cannot be replicated 
by observations of more distant objects. Their proximity enables 
earlier detection and long-duration follow-up campaigns, detailed 
studies of the surrounding environment, and (when data exists)
investigation of the pre-explosion nature of the progenitor star.

\ixf was discovered in Messier 101 (M~101, $d = 6.85$~Mpc) on 2023 May 19.73 UT 
(MJD=60083.73) by Kōichi Itagaki \citep{Itagaki2023}. Rapid spectroscopic 
observations revealed \ixf to be a Type II supernova (\snii) with 
multiple flash ionization features \citep{Perley2023}. Due to the 
rarity of SNe at $d<7$~Mpc, a global, ground- and space-based follow-up 
campaign constrained early-time physics of the explosion spanning  
$\gamma$-ray \citep{Sarmah2023,Muller2023}, 
X-ray \citep{Grefenstette2023,Chandra2023,Panjkov2023,Nayana2024}, 
ultraviolet \citep{Hosseinzadeh2023,SinghTeja2023,Zimmerman2023,Bostroem2024},
optical \citep{Yamanaka2023,Stritzinger2023,Jacobson-Galan2023,
Hosseinzadeh2023,Smith2023,Bostroem2023,Hiramatsu2023,Michel2025}, 
near-infrared (\citealp[NIR;][]{Yamanaka2023,VanDyk2024,Park2025}), 
and radio \citep{Berger2023,Iwata2024} wavelengths.
Upper limits on multi-messenger signals from neutrinos 
\citep{Guetta2023,Kheirandish2023} and gravitational waves 
\citep{LIGO2024} were also studied. 

\ixf is a rapidly declining \snii 
($s_2 = 1.85$~mag~(100 days)$^{-1}$)\footnote{$s_{2}$ is defined 
in \citet{Anderson2014} as the decline rate in $V$-band magnitude 
per 100~days during the ``plateau'' phase.} whose peak luminosity 
is enhanced by circumstellar interaction \citep{Zimmerman2023,
Bostroem2024,Singh2024}. The structure of this circumstellar 
material is multi-faceted; with an outer, low-density region 
\citep{Bostroem2024}, and an inner region of enhanced mass loss 
ejected in the final few years before the explosion 
\citep{Bostroem2024,Iwata2024}, which may be disk-like or toroidal 
in shape \citep{Vasylyev2023,Singh2024}. The completeness of this 
early data has enabled detailed modeling of the progenitor; its 
surroundings; and the early light curve, including shock breakout
\citep{Niu2023,Zhang2023,Soker2023,Martinez2023,Bersten2023,
Li2023,Moriya2024,Hu2024}. 

The proximity of \ixf also enabled searches for both pre-explosion 
variability and direct detection of the progenitor in archival images 
of both ground- and space-based telescopes from the ultraviolet through
mid-infrared (\citealp[MIR;][]{Flinner2023,Dong2023,Kilpatrick2023}). 
While there is consensus that the progenitor star was a dusty red 
supergiant~(RSG), estimates of the progenitor mass cover both the low mass 
($M \lesssim 12\Msol$; \citealp{Kilpatrick2023,Pledger2023,Neustadt2023,
VanDyk2023}), and high mass ($M \gtrsim 17\Msol$; \citealp{Jencson2023,
Soraisam2023,Niu2023,Liu2023,Qin2023,Ransome2023}) ends of plausible 
\snii progenitors \citep{Smartt2015}. 

Early observations at IR wavelengths are crucial for 
understanding the formation of molecules and dust in \sneii.
Much of the dust observed in the early universe \citep{Bertoldi2003,
Maiolino2004,Dwek2007,Li2020} is thought to have formed 
in the ejecta of core collapse SNe \citep{Cernuschi1967,Hoyle1970}. 
This is because the AGB stars which produce this dust in the 
local universe are not yet old enough to have produced the 
observed dust masses in these high-$z$ galaxies 
\citep{Dwek1998,Ferrarotti2006,Gall2011,DiCriscienzo2013,
DellAgli2015}. The formation of molecules in the
SN ejecta provides both an important cooling mechanism and 
the necessary nucleation sites for the later formation and 
survival of dust grains.

The most prominent of these molecules in the NIR and MIR 
wavelengths are carbon-monoxide~(CO) 
and silicon-monoxide (SiO). The timing, location, and 
amount of CO and SiO formation are related to the He-core
mass of the progenitor, which determines the relative 
conditions and abundances within the progenitor 
star at the time of explosion \citep{Woosley2002,Sarangi2013,
Muller2016,Brooker2022,Dessart2025}. \sneii progenitors 
from systems with strong binary interaction (e.g., mergers) may 
have different He-core masses than those from single-star systems 
\citep{Zapartas2021,Tsuna2025}.
CO and/or SiO have been detected in SN~1987A 
\citep{Catchpole1988,Spyromilio1988,Meikle1989,Wooden1993}, 
multiple \sneii observed by the {\it Spitzer Space Telescope}
(\citealp[e.g.,][]{Kotak2006,Szalai2013}), and well-studied 
\sneii with ground-based NIR time series 
(\citealp[e.g.][]{Kotak2005,Fox2010,Rho2018,Szalai2019,Davis2019}). 
The first overtone of CO has been detected in 
ground-based NIR spectroscopy of \ixf starting $+199$~days after
explosion \citep{Park2025}. Both the first overtone and the 
CO fundamental have been detected in time-series \jwst 
spectroscopy spanning $\sim250-720$~days \citep{Medler2025_23ixf}.
Recent \jwst observations of the most nearby \sneii demonstrate 
its ability to trace the formation and evolution of dust over 
decades \citep{Larsson2023,Jones2023,Shahbandeh2023,
Shahbandeh2025_05ip}. These long baseline observations are 
critical for determining how the dust mass grows over time in 
\sneii \citep{Gall2014,Dwek2019}, but rely on the 
upper limits of surviving molecules and dust in both the nearby 
circumstellar medium (CSM) and interstellar medium 
(ISM) determined at early times.

Long-baseline observations are especially important for 
understanding dust formation in \sneii with dense CSM 
(e.g., SNe~IIL/P with early flash-features) or \sneii 
which show long-lived interaction (e.g., \sneiin). The shocks which 
form as a result of the interaction between the ejecta and the
CSM can destroy interstellar dust through evaporative 
collisions between grains and thermal sputtering 
\citep{Barlow1978a,Barlow1978b,Barlow1978c,Jones1996,Jones2004,Slavin2015,Slavin2020}.
However, these shocks are also responsible for forming 
the cold dense shell (CDS), the most likely site of new dust
formation within the SN ejecta \citep{Pozzo2004,Meikle2011}. 
Dust may also form in the surrounding dense CSM \citep{Smith2008,
Miller2010}, and pre-existing dust grains may serve as 
condensation sites for additional dust growth \citep{Fox2010,Fox2011}. 
Distinguishing between these sources of dust is important to 
understanding how shocks influence dust formation 
\citep{Gall2014,Matsuura2019}, and whether \sneiin are more likely 
to form dust with different characteristics than other subsets 
of \sneii (\citealp[e.g.,][]{Pozzo2004,Smith2009,Serrano-Hernandez2025}).

Here, we present plateau-phase \jwst spectra of the nearby \ixf
obtained with the Near-Infrared Spectrograph (NIRSpec; 
\citealp{Jakobsen2022,Boker2023}) and the Low Resolution 
Spectrograph (LRS; \citealp{Kendrew2015}) of the Mid-Infrared 
Instrument (MIRI), and contemporaneous ground-based spectral 
observations in the optical and NIR. This is the first paper 
in a series of papers documenting the evolution of \ixf with 
\jwst spectroscopy obtained by the MidInfared SuperNovA 
Collaboration (MIRSNAC) under programs 
JWST-DD-4522 \citep{Ashall2023_cycle1_23ixf}, JWST-DD-4575 
\citep{Ashall2023_cycle2_23ixf} and JWST-GO-5290 
\citep{Ashall2024_cycle3_ixf}. Paper II \citep{Medler2025_23ixf} 
focuses on the panchromatic evolution and NIR+MIR 
spectroscopic properties of \ixf during the nebular phase.
The observations presented here and in Paper II lay the 
groundwork for future efforts to model the full panchromatic 
SED and emission-line properties. Scheduled observations in 
upcoming cycles will (when combined with this dataset) offer 
unprecedented insight into the location and conditions under 
which molecules and dust form in \sneii, and set the stage for 
continued observations of \ixf throughout the lifetime of \jwst.

In \autoref{sec:obs}, we present our observations and reduction
procedures. We identify the strong lines in the spectrum in 
\autoref{sec:line_ids}, and compare them to previous IR observations 
in\autoref{sec:compare}. We discuss the overall SED in \autoref{sec:sed},
while \autoref{sec:lines} analyzes the velocities and profiles 
of the identified features. \autoref{sec:modeling} showcases our 
modeling efforts, including placing limits on the amount of dust 
pre-cursor molecules present in the ejecta. We summarize our findings 
in \autoref{sec:conclusion}.

\section{Observations} \label{sec:obs}

\begin{deluxetable}{lcc}[t!]
  \tablecaption{Properties of SN~2023ixf and Messier 101 \label{tab:props}} 
  \tablehead{\colhead{Parameter} & \colhead{Value} & \colhead{Source}}
  \startdata
    \multicolumn{3}{c}{SN 2023ixf} \\
    \hline
    R.A. & 14$^{h}$03$^{m}$38$^{s}$.562 & (1) \\ 
    Dec. & $+$54\arcdeg18\arcmin41\arcsec.94 & (1) \\ 
    Discovery (MJD) & 60083.73 & (2) \\ 
    $T_{\rm exp}$ (MJD) & $60082.75$ & (3) \\ 
    $V_{\rm max}$ (mag) & $\sim-18.4$ & (4) \\ 
    $E(B-V)_{MW}$ (mag) & $0.0077 \pm 0.0002$ & (5) \\ 
    $E(B-V)_{Host}$ (mag) & $0.031 \pm 0.012$ & (6) \\ 
    \hline
    \multicolumn{3}{c}{Messier 101} \\
    \hline
    R.A. & 14$^{h}$03${^m}$12$^{s}$.544 & (7) \\ 
    Dec. & $+$54\arcdeg20\arcmin56\arcsec.22 & (7) \\ 
    Morphology & SAB(rs)cd & (8) \\ 
    $v_{\rm helio}$ (\kms) & $ 241 \pm 2$ & (7) \\ 
    $v_{\rm rot}$ (\kms) & $ 7 \pm 1 $ & (6) \\ 
    $z$ & 0.000804 & (9) \\
    $\mu$ & $ 29.18 \pm 0.04 $ & (9) \\ 
    $d_L$ (Mpc) & $ 6.85 \pm 0.15 $ & (9) \\ 
  \enddata
  \tablerefs{(1) \href{https://www.wis-tns.org/object/2023ixf}{TNS}, 
  (2) \cite{Itagaki2023} (3) \cite{Hosseinzadeh2023}, 
  (4) \cite{Zimmerman2023}, (5) \cite{Schlafly2011}, (6) \cite{Smith2023}, 
  (7) NED, (8) \cite{deVaucouleurs1991}, (9) \cite{Riess2022}}
\end{deluxetable}

\subsection{JWST Observations}

Observations of \ixf with \jwst were obtained through our program 
DD-JWST-4522 \citep{Ashall2023_cycle1_23ixf}, using both NIRSpec and 
MIRI/LRS beginning at 2023 June 21.33 UT. Consistent with other works, 
we adopt an explosion time of $\text{MJD}=60082.75$ as derived 
from the midpoint of the earliest reported detection and the 
latest deep non-detection \citep{Mao2023,Yaron2023,Hosseinzadeh2023,
Zimmerman2023}. This places our \jwst observations $+33.6$~days 
after explosion. Key properties related to \ixf  and its host galaxy 
M~101 used throughout this paper are summarized in \autoref{tab:props}. 
NIRSpec observations were performed with the F170LP/G235M 
($\sim1.66-3.07$~\mic) and F290LP/G395M ($\sim2.87-5.10$~\mic) 
filter/grating combinations, providing continuous coverage from 
1.7--5.1~\mic at $R \sim 1000$. MIRI/LRS observations span the 
$\sim5-14$~\mic range at a wavelength-dependent resolution 
$R \approx 50-200$. The data were reduced using the \jwst Science 
Calibration Pipeline (\citealp[version 1.18.0;][]{Bushouse2025}) and 
CRDS version \texttt{jwst\_1364.pmap}. These data can be accessed 
via \dataset[doi: 10.17909/ekjp-5b33]{\doi{10.17909/ekjp-5b33}}.
Full details of the observational set-up are provided in \autoref{tab:obs}.

\begin{deluxetable}{lccc}
  \tablecaption{Observation Details \label{tab:obs}} 
  \tablehead{\colhead{Parameter} & \colhead{Value} & \colhead{Value} }
  \startdata
    \multicolumn{3}{c}{NIRSpec Acquisition Image} \\
    \hline
    Filter & \multicolumn{2}{c}{F140X} \\
    Exp Time (s) & \multicolumn{2}{c}{0.08} \\
    Readout Pattern & \multicolumn{2}{c}{NRSRAPID} \\
    \hline
    \multicolumn{3}{c}{NIRSpec Spectral Observations} \\
    \hline
    Slit & \multicolumn{2}{c}{S400A1} \\
    Subarray & \multicolumn{2}{c}{SUBS400A1} \\
    Grating/Filter & G235M/F170LP & G395M/F290LP \\
    $T_{\rm obs}$ (MJD) & 60116.34 & 60116.33 \\
    Phase from Exp. (days) & $+33.58$ & $+33.57$ \\
    Exp Time (s) & 60.8 & 98.2 \\
    Groups per Integration & 3 & 5 \\
    Integrations per Exp. & 1 & 1 \\
    Exposures per Dither & 1 & 1 \\
    Total Dithers & 3 & 3 \\
    Readout Pattern & NRS & NRS \\
    \hline
    \multicolumn{3}{c}{MIRI Acquisition Image} \\
    \hline
    Filter & \multicolumn{2}{c}{F560W} \\
    Exp Time (s) & \multicolumn{2}{c}{11.1} \\
    Readout Pattern & \multicolumn{2}{c}{FAST} \\
    \hline
    \multicolumn{3}{c}{MIRI Spectral Observations}  \\
    \hline
    Mode & \multicolumn{2}{c}{LRS} \\
    Exp Time (s) & \multicolumn{2}{c}{260.9} \\
    $T_{\rm obs}$ (MJD) & \multicolumn{2}{c}{60116.35} \\
    Phase from Exp. (days) & \multicolumn{2}{c}{$+33.59$}\\
    Groups per Integration & \multicolumn{2}{c}{15} \\
    Integrations per Exp. & \multicolumn{2}{c}{3} \\
    Exposures per Dither & \multicolumn{2}{c}{1} \\
    Total Dithers & \multicolumn{2}{c}{2} \\
    \hline
    \multicolumn{3}{c}{Ground-based Optical Spectra} \\
    \hline
      Telescope & \multicolumn{2}{c}{NOT} \\
      Instrument & \multicolumn{2}{c}{ALFOSC} \\
      $T_{\rm obs}$ (MJD) & \multicolumn{2}{c}{60117.08} \\
      Phase from Exp. (days) & \multicolumn{2}{c}{$+34.32$} \\
      Exp Time (s) & \multicolumn{2}{c}{180} \\ 
    \hline
    \multicolumn{3}{c}{Ground-based NIR Spectra} \\
    \hline
    Telescope & \multicolumn{2}{c}{IRTF} \\
    Instrument & \multicolumn{2}{c}{Spex} \\
    $T_{\rm obs}$ (MJD) & \multicolumn{2}{c}{60115.48} \\
    Phase from Exp. (days) & \multicolumn{2}{c}{$+32.72$} \\
    Exp Time (s) & \multicolumn{2}{c}{169.6} \\
  \enddata
\end{deluxetable}

\subsection{Ground-based Observations}

Ground-based follow-up of \ixf at optical and NIR wavelengths
were obtained to complete the spectral energy distribution.
Optical spectroscopic follow up observations spanning 
2.60--64.57~days after explosion were made with both the 
the SuperNova Integral Field Spectrograph (\citealp[SNIFS;][]{Lantz2004}) 
on the University of Hawai`i 88-in telescope (UH88) and the
the Alhambra Faint Object Spectrograph and Camera (ALFOSC) 
on the 2.5-m Nordic Optical Telescope (NOT).
UH88 spectra were obtained by the Spectroscopic Classification 
of Astronomical Transients collaboration (\citealp[SCAT;][]{Tucker2022}). 
These data were reduced following the methods outlined in 
\citet{Tucker2022}.
NOT spectroscopic observations were obtained as part of a 
followup campaign of \ixf led by the NOT Un-biased Transient 
Survey (NUTS2)\footnote{\url{https://nuts.sn.ie/}}. 
The data were reduced following standard methods; including
bias subtraction, flat-fielding of the two-dimensional images, 
wavelength calibration of the extracted spectrum from arc lamp 
exposures, and the removal and correction of telluric features 
and cosmic rays. Three individual, high signal-to-noise 180s 
exposures were obtained and median combined. The midpoint of 
the three exposures (MJD=60117.08; $+34.32$~days) is adopted 
as the time of the observation.

NIR time-series spectra of \ixf were obtained by HISS \citep{Medler2025_hiss}
between $+8.58$ and $+32.72$~days from the explosion with Keck-II/NIRES 
and IRTF/SpeX. Details on the instrument configurations and associated
reduction procedures can be found in \citet{Medler2025_hiss}.

For the purposes of constructing a contemporaneous panchromatic 
SED spanning the optical to MIR, we combine the NOT optical 
spectra obtained on MJD=60117.08 and the IRTF spectra from 
MJD=60115.48 with our \jwst data.
A full log of the ground-based observations can be found in 
\autoref{tab:gbspec}.

\begin{deluxetable}{cccc}
  \tablecaption{Log of spectroscopic observations \label{tab:gbspec}}
  \tablehead{\colhead{Date (UT)} & \colhead{MJD} &
    \colhead{Epoch\tablenotemark{a}} & \colhead{Exp. Time (s)} }
  \startdata
    \multicolumn{4}{c}{UH88/SNIFS Optical Spectra}\\
    \hline
    2023 May 21.34 & 60085.34 & 2.59 & 2000 \\
    2023 May 23.25 & 60087.25 & 4.52 & 2000 \\
    2023 May 25.34 & 60089.34 & 6.61 & 1800 \\
    2023 Jun 04.41 & 60099.41 & 16.68 & 1800 \\
    2023 Jun 05.25 & 60100.25 & 17.54 & 1800 \\
    2023 Jun 10.44 & 60105.44 & 22.69 & 1800 \\
    2023 Jun 12.44 & 60107.44 & 24.69 & 1800 \\
    2023 Jun 16.33 & 60111.33 & 28.59 & 1800 \\
    2023 Jun 20.29 & 60115.29 & 32.55 & 1800 \\
    2023 Jun 24.30 & 60119.30 & 36.56 & 1800 \\
    2023 Jun 26.28 & 60121.28 & 38.55 & 1800 \\
    2023 Jun 28.27 & 60123.27 & 40.53 & 1800 \\
    2023 Jun 30.29 & 60125.29 & 42.56 & 1800 \\
    2023 Jul 04.41 & 60129.41 & 46.66 & 1800 \\
    2023 Jul 06.43 & 60131.43 & 48.68 & 1800 \\
    2023 Jul 08.29 & 60133.29 & 50.55 & 1800 \\
    2023 Jul 10.29 & 60135.29 & 52.55 & 2400 \\
    2023 Jul 12.29 & 60137.29 & 54.56 & 1800 \\
    2023 Jul 14.26 & 60139.26 & 56.53 & 1800 \\
    2023 Jul 16.28 & 60141.28 & 58.54 & 1800 \\
    2023 Jul 22.31 & 60147.31 & 64.57 & 1800 \\
    \hline
    \multicolumn{4}{c}{Near-infrared Spectra} \\
    \hline
    2023 May 27.33 & 60091.33 & 8.58 & 40 \\
    2023 Jun 04.24 & 60099.24 & 16.49 & 3415.82 \\
    2023 Jun 06.24 & 60101.24 & 18.49 & 5213.62 \\
  \enddata
  \tablecomments{\tablenotemark{a}Rest frame days relative to explosion 
  on MJD$=60082.75$ \citep{Hosseinzadeh2023}.}
\end{deluxetable}

\section{Line Identifications} \label{sec:line_ids}

\begin{figure*}[th!]
    \centering
    \includegraphics[width=\textwidth]{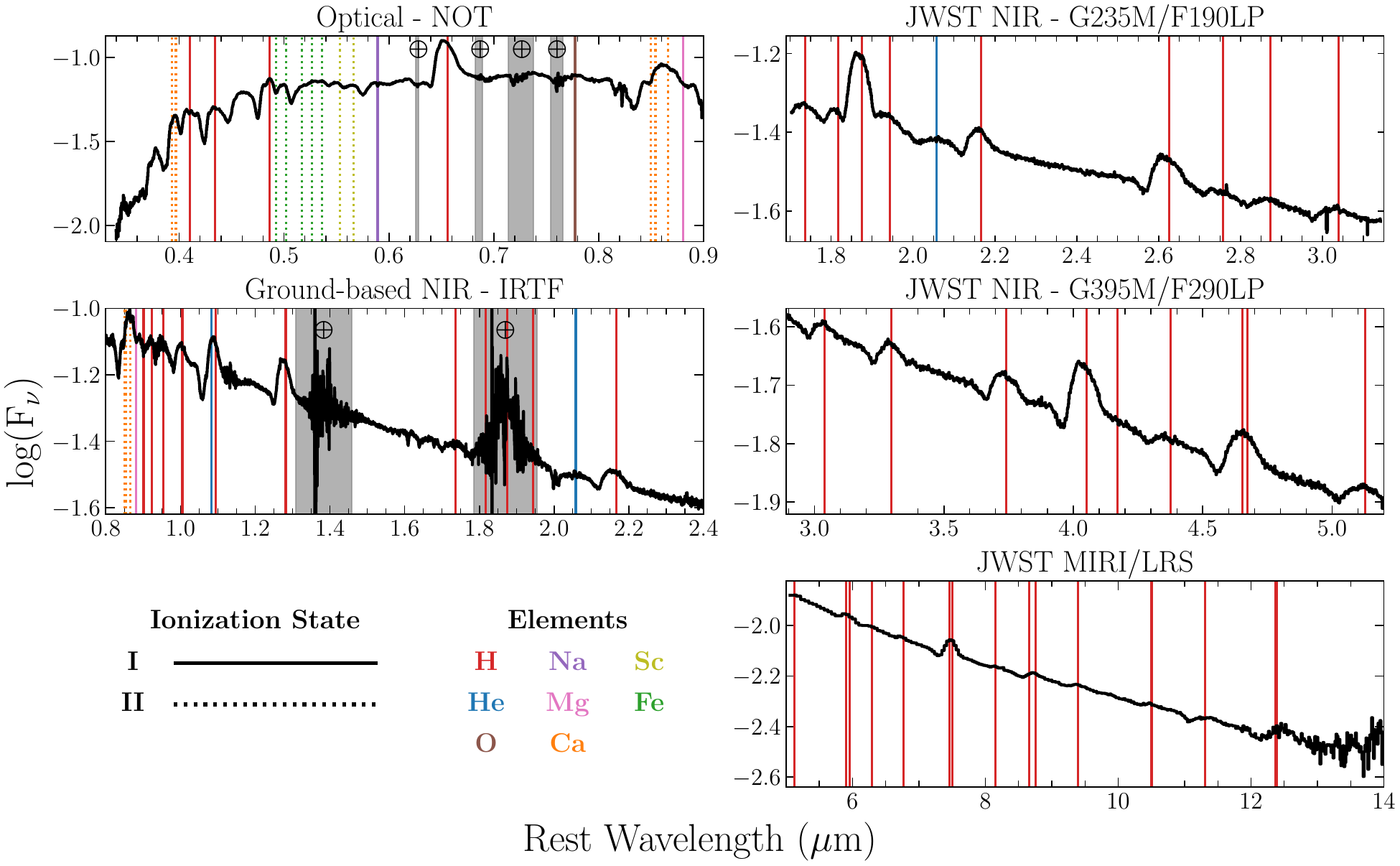}
    \caption{Line identifications based on known transitions 
    common to \sneii (see \autoref{tab:line_ids}). The spectra shown are arranged by telescope 
    and grating combination. Based on Monte Carlo fits (see \autoref{sec:vel_fits}), 
    the absorption troughs are shifted by up to $-7500$~\kms for 
    hydrogen lines, and $\sim -6100$~\kms for all other lines. Strong 
    telluric regions in the ground-based optical and NIR data are marked
    in grey.}
    \label{fig:line_ids}
\end{figure*}

\autoref{fig:line_ids} shows the lines identified in the 
combined spectra of \ixf, with the individual transitions
listed in \autoref{tab:line_ids}. These identifications
were compiled from a list of plausible lines seen in 
previous analyses of \sneii, including: 
\citet{Mazzali1992,Baron2003,Gutierrez2017,Davis2019,
Shahbandeh2022,Shahbandeh2024_22acko}, and sources therein.

Similar to the \jwst observations of \acko \citep{Shahbandeh2024_22acko}, 
the spectra during the plateau phase are dominated by strong 
hydrogen lines from the Balmer, Paschen, Brackett, Pfund, 
Humphreys, and other unnamed higher order series. 
Strong lines (e.g., the $\alpha$, $\beta$, and $\gamma$ 
transitions) within the named H series appear with well-defined P~Cygni 
shapes. Weaker lines (e.g., the $\epsilon$, $\zeta$, and $\eta$ 
transitions) show a larger diversity in their line profiles, often 
appearing with weaker relative emission components or only in absorption. 
\citet{Baron2025} find similar behavior in both their observations
and NLTE model of \ggi, and discuss what ejecta conditions lead 
to the formation of the different line profiles. In the line 
identifications presented below, we consider weaker hydrogen 
lines as identified components of blends if: 
(1) another strong line originating from the same upper energy 
state is seen elsewhere in the spectrum, or 
(2) if an additional line from the same series originating in a 
higher energy level is clearly seen (e.g., as in the case of 
\hl{Hu}{\gamma}). 

\begin{deluxetable}{lc|lc}
    \tablecaption{Line Identifications \label{tab:line_ids}}
    \tablehead{\colhead{Line} & \colhead{Wavelength (\mic)} & \colhead{Line} & \colhead{Wavelength (\mic)}}
    \startdata
    \multicolumn{4}{c}{Optical Lines (0.35--0.9~\mic)} \\
    \hline
    \ion{Ca}{2} & 0.3934 & \ion{Fe}{2} & 0.5531 \\
    \ion{Ca}{2} & 0.3968 & \ion{Sc}{2} & 0.5531 \\
    \ion{H}{1} (\Hdelta) & 0.4102 & \ion{Sc}{2} & 0.5663 \\
    \ion{Fe}{2} & 0.4303 & \ion{Na}{1} & 0.5983 \\
    \ion{H}{1} (\Hgamma) & 0.4340 & \ion{H}{1} (\Halpha) & 0.6563 \\
    \ion{H}{1} (\Hbeta) & 0.4861 & \ion{O}{1} & 0.7774 \\
    \ion{Fe}{2} & 0.4924 & \ion{Ca}{2} & 0.8498 \\
    \ion{Fe}{2} & 0.5018 & \ion{Ca}{2} & 0.8542 \\
    \ion{Fe}{2} & 0.5169 & \ion{Ca}{2} & 0.8662 \\
    \ion{Fe}{2} & 0.5267 & \ion{Mg}{1}\tablenotemark{a} & 0.8807 \\
    \ion{Fe}{2} & 0.5363 & \ion{H}{1} (\hl{Pa}{\eta)} & 0.9014 \\  
    \hline
    \multicolumn{4}{c}{Ground-based NIR Lines (0.9--1.7~\mic)} \\
    \hline
    \ion{H}{1} (\hl{Pa}{\zeta}) & 0.923 & \ion{He}{1} & 1.083 \\
    \ion{H}{1} (\hl{Pa}{\epsilon}) & 0.955 & \ion{H}{1} (\hl{Pa}{\gamma}) & 1.094 \\
    \ion{H}{1} (\hl{Pa}{\delta}) & 1.005 & \ion{H}{1} (\hl{Pa}{\beta}) & 1.282\\
    \hline
    \multicolumn{4}{c}{NIRSpec Lines (1.7--5~\mic)} \\
    \hline
    \ion{H}{1} (\hl{Br}{\zeta}) & 1.737 & \ion{H}{1} (\hl{Pf}{\epsilon}) & 3.039 \\
    \ion{H}{1} (\hl{Br}{\epsilon}) & 1.817 & \ion{H}{1} (\hl{Pf}{\delta}) & 3.297 \\
    \ion{H}{1} (\hl{Pa}{\alpha}) & 1.875 & \ion{H}{1} (\hl{Pf}{\gamma}) & 3.741 \\
    \ion{H}{1} (\hl{Br}{\delta}) & 1.944 & \ion{H}{1} (\hl{Br}{\alpha}) & 4.052 \\
    \ion{He}{1}\tablenotemark{a} & 2.058 & \ion{H}{1} (\hl{Hu}{\eta}) & 4.171 \\
    \ion{H}{1} (\hl{Br}{\gamma}) & 2.166 & \ion{H}{1} (\hl{Hu}{\zeta}) & 4.376\\
    \ion{H}{1} (\hl{Br}{\beta}) & 2.626 & \ion{H}{1} (\hl{Pf}{\beta}) & 4.654 \\
    \ion{H}{1} (\hl{Pf}{\theta}) & 2.675 & \ion{H}{1} (\hl{Hu}{\epsilon}) & 4.673 \\ 
    \ion{H}{1} (\hl{Pf}{\eta}) & 2.758 & \ion{H}{1} (\hl{Hu}{\delta}) & 5.129 \\
    \ion{H}{1} (\hl{Pf}{\zeta}) & 2.873  & \nodata & \nodata \\
    \hline
    \multicolumn{4}{c}{MIRI/LRS Lines (5--14~\mic)} \\
    \hline
    \ion{H}{1} (\hl{Hu}{\gamma}) & 5.908 & \ion{H}{1} (8--14)\tablenotemark{a} & 8.665 \\
    \ion{H}{1} (7--14)\tablenotemark{a} & 5.957 & \ion{H}{1} (7--10)\tablenotemark{a} & 8.760 \\
    \ion{H}{1} (7--13)\tablenotemark{a} & 6.292 & \ion{H}{1} (8--13)\tablenotemark{a} & 9.392 \\
    \ion{H}{1} (7--12)\tablenotemark{a} & 6.772 & \ion{H}{1} (8--12)\tablenotemark{a} & 10.503 \\
    \ion{H}{1} (\hl{Pf}{\alpha}) & 7.459 & \ion{H}{1} (7--9)\tablenotemark{a} & 11.309 \\
    \ion{H}{1} (\hl{Hu}{\beta}) & 7.502 & \ion{H}{1} (\hl{Hu}{\alpha}) & 12.372 \\
    \ion{H}{1} (7--11)\tablenotemark{a} & 7.508 & \ion{H}{1} (8--11)\tablenotemark{a} & 12.387 \\
    \ion{H}{1} (8--15)\tablenotemark{a} & 8.155 & \nodata & \nodata \\
    \enddata
    \tablecomments{\tablenotemark{a}Tentative}
\end{deluxetable}

\subsection{NIRSpec (1.7-5~\mic)} \label{sec:nirspec_lines}

The combined NIR spectrum is primarily dominated by H lines of the Brackett,
Pfund, and Humphries series, all of which have remarkably similar P~Cygni 
profiles. The identified lines include: \hlwl{Br}{\zeta}{1.737},
\hlwl{Br}{\epsilon}{1.817}, \hlwl{Pa}{\alpha}{1.875}, \hlwl{Br}{\delta}{1.944}, 
\hlwl{Br}{\gamma}{2.166}, \hlwl{Br}{\beta}{2.626}, \hlwl{Pf}{\theta}{2.675}, 
\hlwl{Pf}{\eta}{2.758}, \hlwl{Pf}{\zeta}{2.873}, \hlwl{Pf}{\epsilon}{3.039}, 
\hlwl{Pf}{\delta}{3.297}, \hlwl{Pf}{\gamma}{3.741}, \hlwl{Br}{\alpha}{4.052}, 
\hlwl{Hu}{\eta}{4.171}, \hlwl{Hu}{\zeta}{4.376}, \hlwl{Pf}{\beta}{4.654},
\hlwl{Hu}{\epsilon}{4.673}, and \hlwl{Hu}{\delta}{5.129}. 

There is an additional weak, broad feature near 2.05~\mic which 
we tentatively identify as the He~I 2.0581~\mic line. This 
identification is supported by the presence of the 
\ion{He}{1}~1.083~\mic line in the ground-based NIR data (see 
\autoref{sec:nir_lines}) despite the lack of strong He lines 
in the optical wavelengths. This is consistent with the optical 
He lines being more difficult to excite than those in the 
NIR \citep{Harkness1987,Lucy1991}.

At this phase, there is no evidence for CO emission from either the
fundamental ($\sim$4.2--6~\mic) or first overtone ($\sim$2.1--2.6~\mic) 
rovibrational bands. Upper limits on the amount of CO are further 
explored in \autoref{sec:modeling}. See \citet{Park2025} 
and \citet{Medler2025_23ixf} for discussions on the detection of CO
at later phases in \ixf.

\subsection{MIRI/LRS (5-14~\mic)} \label{sec:miri_lines}

In the MIR spectrum, two features are particularly strong. The first 
is the blend of \hlwl{Pf}{\alpha}{7.459} and \hlwl{Hu}{\beta}{7.502};
the second feature being the \hlwl{Hu}{\alpha}{12.372} line, which 
itself is weakly blended with \hlwl{\ion{H}{1}}{(8-11)}{12.387}. 
Based on the detection of \hl{Hu}{\delta} in both the LRS and NIRSpec 
data, we identify the \hlwl{Hu}{\gamma}{5.908} line, albeit with a 
significantly undersampled and possibly blended profile.

Identification of additional features becomes more difficult as the 
strength of the features above the continuum decreases. Furthermore,
the low resolution of the spectrograph ($R \approx 50-200$) 
undersamples
the line profiles by spreading them over only a small handful ($2-5$) 
of pixels. When combined with the different line profile
shapes found in MIR models of \sneii \citep{Baron2025}, the observed
line profiles may appear different for lines which should otherwise 
appear nearly identical. This can be clearly seen in the 
\hl{Hu}{\gamma} line, which appears weak in both absorption 
and emission, compared to both the \hl{Hu}{\alpha} line at 
higher-resolution in the LRS data (a strong P  Cygni profile)
and the \hl{Hu}{\delta} line (a "detached" profile with
weak emission; \citealp{Baron2025}) in the NIRSpec data 
(see \autoref{sec:lines}).

Relying on the spectroscopic models of a companion paper on the 
plateau-phase \jwst spectroscopy of \ggi \citep{Baron2025}, 
we are able to tentatively identify several additional hydrogen 
lines. These identifications are supported by the models showing 
hydrogen features with similar profiles and strengths above the 
continuum. These lines include: 
\uhl{7-14}{5.957}, \uhl{7-13}{6.292}, \uhl{7-12}{6.772}, \uhl{7-11}{7.508}, 
\uhl{8-15}{8.155}, \uhl{8-14}{8.665}, \uhl{7-10}{8.760}, \uhl{8-13}{9.392}, 
\uhl{8-12}{10.503}, \uhl{7-9}{11.309}, and \uhl{8-11}{12.387}.

Beyond $\sim$13~\mic, the decreased throughput of the LRS mode results
in noise levels exceeding the feature strengths above the continuum, 
making further line identifications difficult.

\subsection{Optical (0.35-0.9~\mic)} \label{sec:opt_lines}

The NOT optical spectrum of \ixf shows features typical of normal \sneii
roughly 30 days after explosion. The spectrum displays a prominent
\Halpha line with a defined P~Cygni profile, while other Balmer lines
including \Hbeta, \Hgamma, and \Hdelta show only absorption components,
as expected at these phases \citep{Gutierrez2017}. Other strong features
identified include the \ion{Ca}{2} H~\&~K lines; \ion{Fe}{2} lines at
0.4861~\mic, 0.4924~\mic, 0.5018~\mic, 0.5169~\mic, 0.5267~\mic,
0.5363~\mic, 0.5531~\mic; the \ion{Na}{1} D doublet; the \ion{Ca}{2} 
NIR triplet; and \hlwl{Pa}{\eta}{0.9014}.  

Some weaker features also appear in the optical spectrum. Tentative 
evidence exists for absorption from \ion{O}{1}~$\lambda7774$, however 
it is strongly contaminated by telluric absorption. Weak \ion{Sc}{2} 
features are seen at 0.5531~\mic (blended with the \ion{Fe}{2} transition 
at the same wavelength) and 0.5663~\mic, with absorption minima matching 
that of the unblended \ion{Fe}{2}~0.5169~\mic line ($-6120 \pm 420$~\kms), 
which is taken to represent the photosphere. 
No evidence is seen for an absorption minimum corresponding to 
\ion{Sc}{2} 0.6247~\mic at the photospheric velocity, suggesting that 
the trough to the blue of the \Halpha P~Cygni is a ``Cachito'' feature 
\citep{Gutierrez2017}. Consistent with other works \citep{Singh2024}
we find the most likely origin of the Cachito in \ixf to be 
high-velocity~(HV) \Halpha, based on the velocity measures described 
in \autoref{sec:vel_fits}.

\subsection{Ground-based NIR (0.9-1.7~\mic)} \label{sec:nir_lines}

Ground-based NIR spectra from IRTF show good agreement with the \jwst NIRSpec 
data; e.g., the spectrum is dominated by Paschen series lines on top 
of a blackbody-like continuum. In the 0.9-1.7~\mic region not covered 
by optical data nor the \jwst data, we see \hlwl{Pa}{\zeta}{0.923}, 
\hlwl{Pa}{\epsilon}{0.955}, \hlwl{Pa}{\delta}{1.005}, \hlwl{Pa}{\gamma}{1.094},
and \hlwl{Pa}{\beta}{1.282}. The increased strength of the absorption 
trough of the Pa~$\gamma$ feature relative to those of other 
Paschen lines in the ground-based and \jwst NIR spectra is due to the
blending of the \ion{He}{1} $\lambda1.083$ line within the feature. This 
identification is confirmed by the weak P~Cygni \ion{He}{1} line seen in 
the $+8.58$~day NIR spectrum prior to the emergence of strong H P~Cygni 
lines in both the optical and NIR, and the measured velocity of the 
\ion{He}{1} $\lambda1.083$ line matching that of the \hl{Pa}{\gamma} line 
(see \autoref{sec:vel_fits} for details). These line identifications are 
consistent with those found in other ground-based NIR spectral time series
observations \citep{Park2025}.

The strong presence of the \ion{He}{1} $\lambda1.083$ line
further supports the identification of the weak, broad emission of 
the feature near 2.05~\mic as \ion{He}{1}; as the broad emission is also 
seen in the HISS data presented here and the ground-based NIR time-series 
spectra of \citet{Park2025}. We note that several other potential confounding 
lines could contribute to such a blend, including C, Mg, Si, and Sr lines 
\citep{Davis2019,Shahbandeh2022}. However, we consider these lines 
as unlikely contributors because, if present, their velocities would be 
inconsistent with the measured photospheric velocity. 
The spectrum shows high qualitative agreement with ``strong \sneii'' 
within the scheme of \citet{Davis2019}, and a ``strong'' classification 
is consistent with the measured value of $s_2$, and the absence of 
observed \ion{Sr}{2} features at $\lambda1.033$~\mic. Such a classification 
is important because ``strong'' and ``weak'' \sneii show differences in both 
the presence of \ion{Sr}{2} and the formation timescales of CO \citep{Davis2019}. 
Importantly, the first overtone of CO is observed as early as $\sim100$~days 
in other ``strong'' \sneii, and signals the arrival of ejecta conditions
favorable for the formation of dust 
\citep{Gerardy2000,Woosley2002,Kotak2005,Kotak2006,Davis2019}.

\section{Comparisons to Previous MIR Observations} \label{sec:compare}

\subsection{JWST/NIR Comparisons} \label{sec:nir_compare}

\begin{figure*}[th!]
    \centering
    \includegraphics[width=0.9\textwidth]{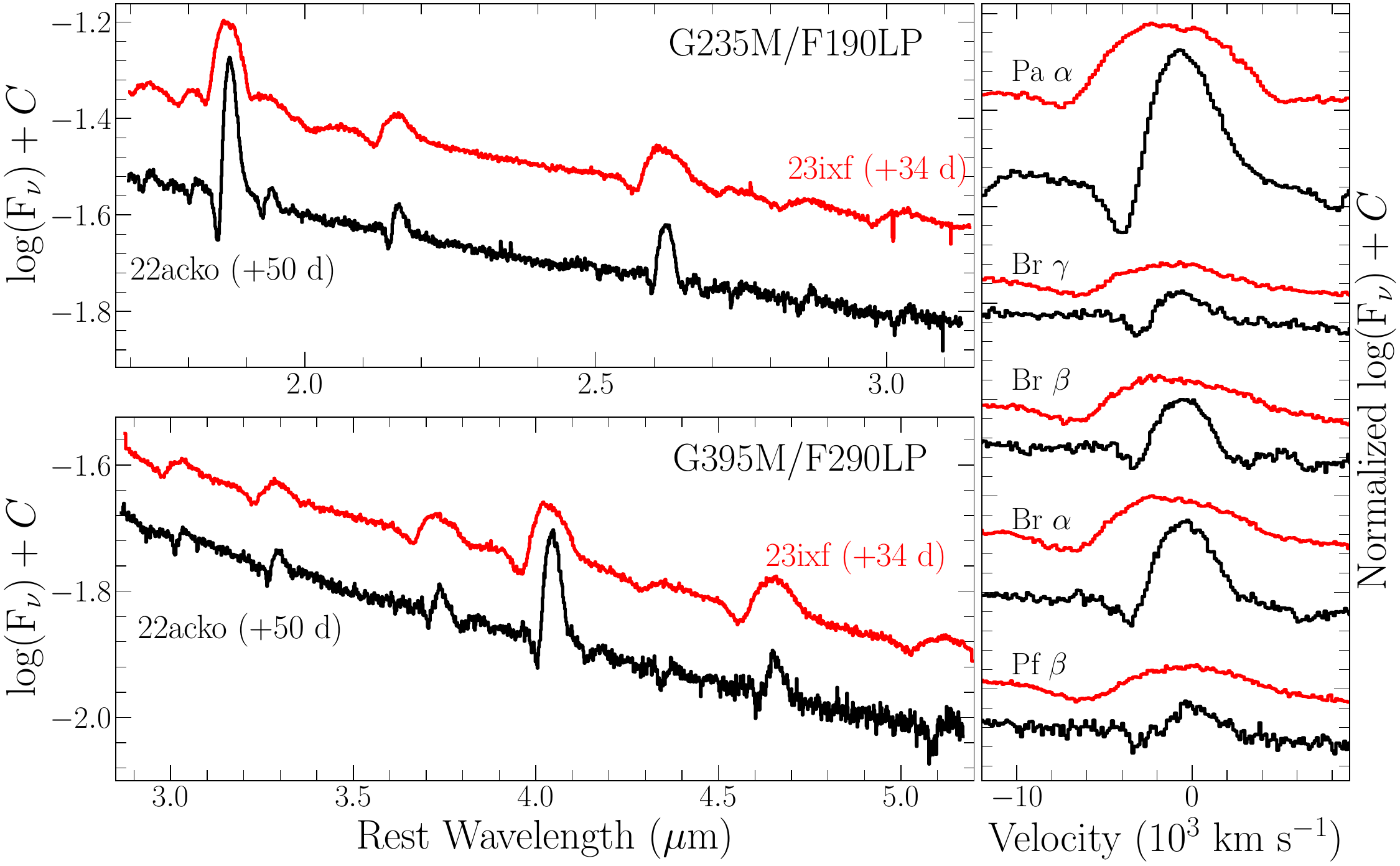}
    \caption{Comparison of \ixf to \acko NIRSpec data (left panels). The line 
    profiles of the strongest H lines in the NIRSpec coverage (\hl{Pa}{\alpha},
    \hl{Br}{\gamma}, \hl{Br}{\beta}, \hl{Br}{\alpha}, and \hl{Pf}{\beta}), are 
    shown in velocity space on the right; highlighting the broader emission and
    weaker absorption found in \ixf relative to \acko.}
    \label{fig:nir_compare}
\end{figure*}

\autoref{fig:nir_compare} shows the comparison of \ixf to 
the $+50$~day spectrum of \acko also taken with JWST/NIRSpec 
\citep{Shahbandeh2024_22acko}. Both spectra were obtained roughly 
halfway through their respective plateau phases. The spectra 
show the same hydrogen lines are present, 
with the lines appearing faster, broader, and more rounded in 
\ixf relative to \acko. This behavior is commonly observed in 
\sneii \citep{Hamuy2002,deJaeger2020}, where brighter objects 
like \ixf ($V_{\rm max} \approx -18.4$~mag) show higher \Hbeta 
velocities than dimmer \sneii such as \acko ($V_{\rm max} = -15.4$~mag).

\subsection{MIR Comparisons} \label{sec:mir_compare}

\begin{figure}[th!]
    \centering
    \includegraphics[width=\columnwidth]{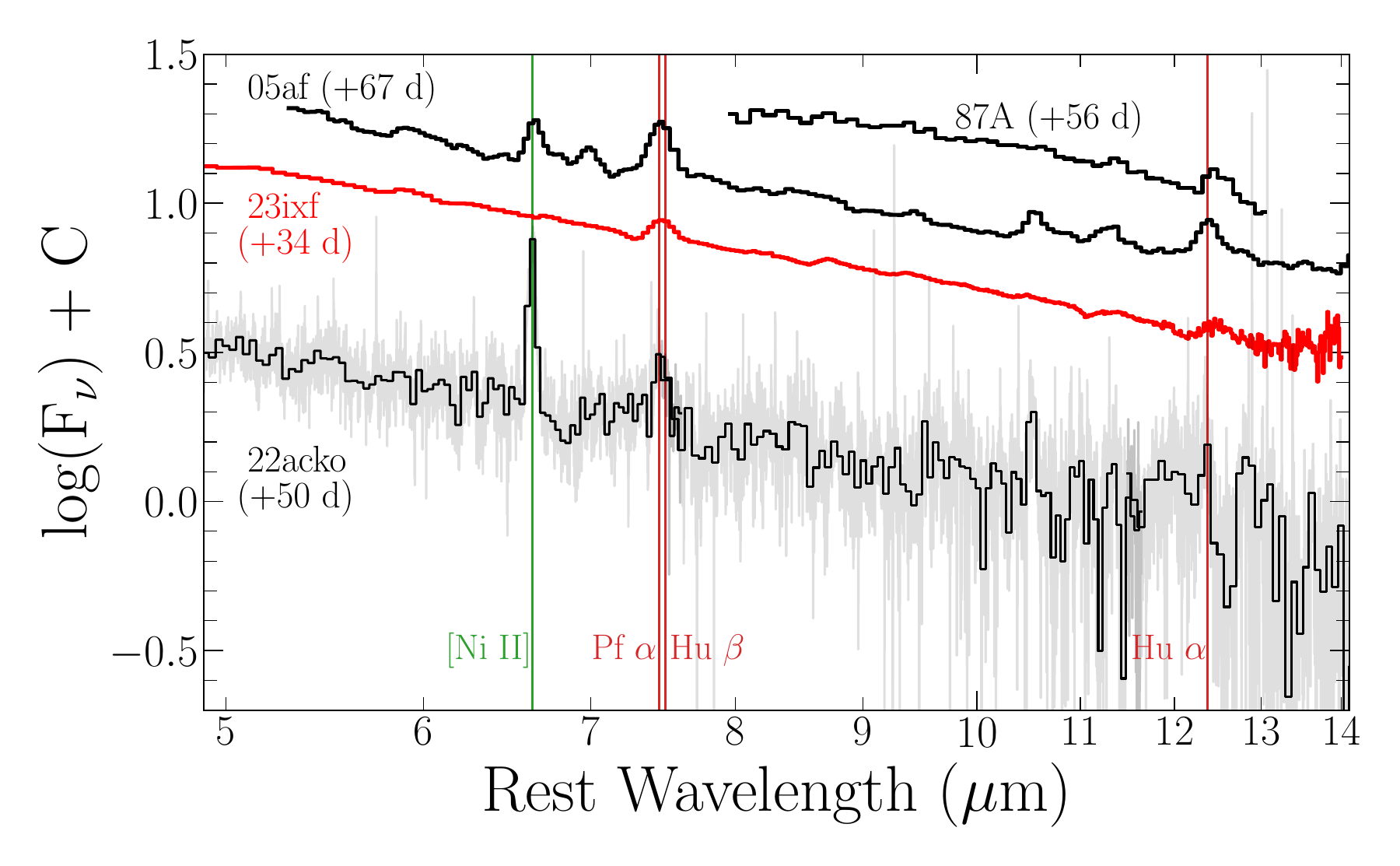}
    \caption{Comparison of \ixf MIRI/LRS data to MIR spectra of SNe~1987A 
    \citep{Aitken1988b}, 2005af \citep{Kotak2006}, and 2022acko 
    \citep{Shahbandeh2024_22acko} at similar epochs, with strong lines common
    to multiple SNe highlighted. The \acko observations have 
    been smoothed to $R = 200$ to better match the low-resolution 
    data of the other observations.} 
    \label{fig:mir_compare}
\end{figure}

\autoref{fig:mir_compare} shows the MIR spectra of \ixf and \acko, 
along with MIR spectra of SNe~1987A \citep{Aitken1988b} and 2005af 
\citep{Kotak2006} at similar phases from explosion. We note that 
the {\it Spitzer} observations of SN~2005af likely occurred 
after it left the plateau phase \citep{Kotak2006}, while 
SN~1987A observations were taken during its rise to secondary 
maximum. The \acko spectrum presented here has been re-reduced using the 
\texttt{AstroBkgInterp}\footnote{https://github.com/brynickson/AstroBkgInterp}
package (\citealp[Nickson et al., in preparation;][]{Shahbandeh2025_05ip}).
Relative to the reduction presented in \citet{Shahbandeh2024_22acko}, the
use of \texttt{AstroBkgInterp} for the background subtraction provides 
a higher S/N for this observation by modeling the underlying background 
in each image slice independently by a two-dimensional polynomial 
extrapolation (in this instance with third-degree polynomials) to 
the background enclosed in an annulus neighboring the extraction 
aperture surrounding the SN. This modeled background is then subtracted 
from the \texttt{s3d} files produced by Stage 2 of the \jwst Pipeline
before feeding the background subtracted data cube back into Stage 3, 
where the final extraction is performed channel-by-channel as normal 
with the \texttt{Extract1D} function. 

The two strongest features are the \hl{Pf}{\alpha} plus \hl{Hu}{\beta} 
blend and the \hl{Hu}{\alpha} line, which are also seen in the other 
three SNe\footnote{The \hl{Pf}{\alpha} blend in SN~1987A falls outside 
of the observed range of the \citet{Aitken1988b} observations shown in 
\autoref{fig:mir_compare}, but are clearly visible in the $+$60~day 
spectrum in Fig.~1 of both \citet{Rank1988} and \citet{Wooden1993}.}. 
The subset of weaker (e.g., $n_l = 7,8$ series) hydrogen lines identified 
in the MIR spectra varies between objects, due to instrumental 
sensitivities and blending.

In our new reduction of \acko narrow emission from 
6.636~\mic [\ion{Ni}{2}] is now seen as it was in SN~2005af, 
but is absent in SNe~1987A and 2023ixf. 
The 6.985~\mic [\ion{Ar}{2}] line is not detected in 
either \acko and \ixf, nor is it in seen in the $+60$~day spectrum 
of SN~1987A in \citet{Rank1988}. 
Observations of SN~2004dj after the plateau phase (roughly $+$106 
and $+$129 days after explosion) also show 6.636~\mic 
[\ion{Ni}{2}] and 7.50~\mic [\ion{Ni}{1}] \citep{Kotak2005}.
The lack of forbidden emission lines in \ixf supports 
the conclusion that the photosphere in the IR still resides 
within the H-rich envelope at this phase. Observations 
of SNe~2022acko (T.~Mera et al., in preparation), 2023ixf 
\citep{Medler2025_23ixf} and additional \sneii after their 
plateau phases with \jwst will allow the community to investigate 
differences in the forbidden lines in the MIR.

\section{Spectral Energy Distribution} \label{sec:sed}

\begin{figure*}[th!]
    \centering
    \includegraphics[width=0.9\textwidth]{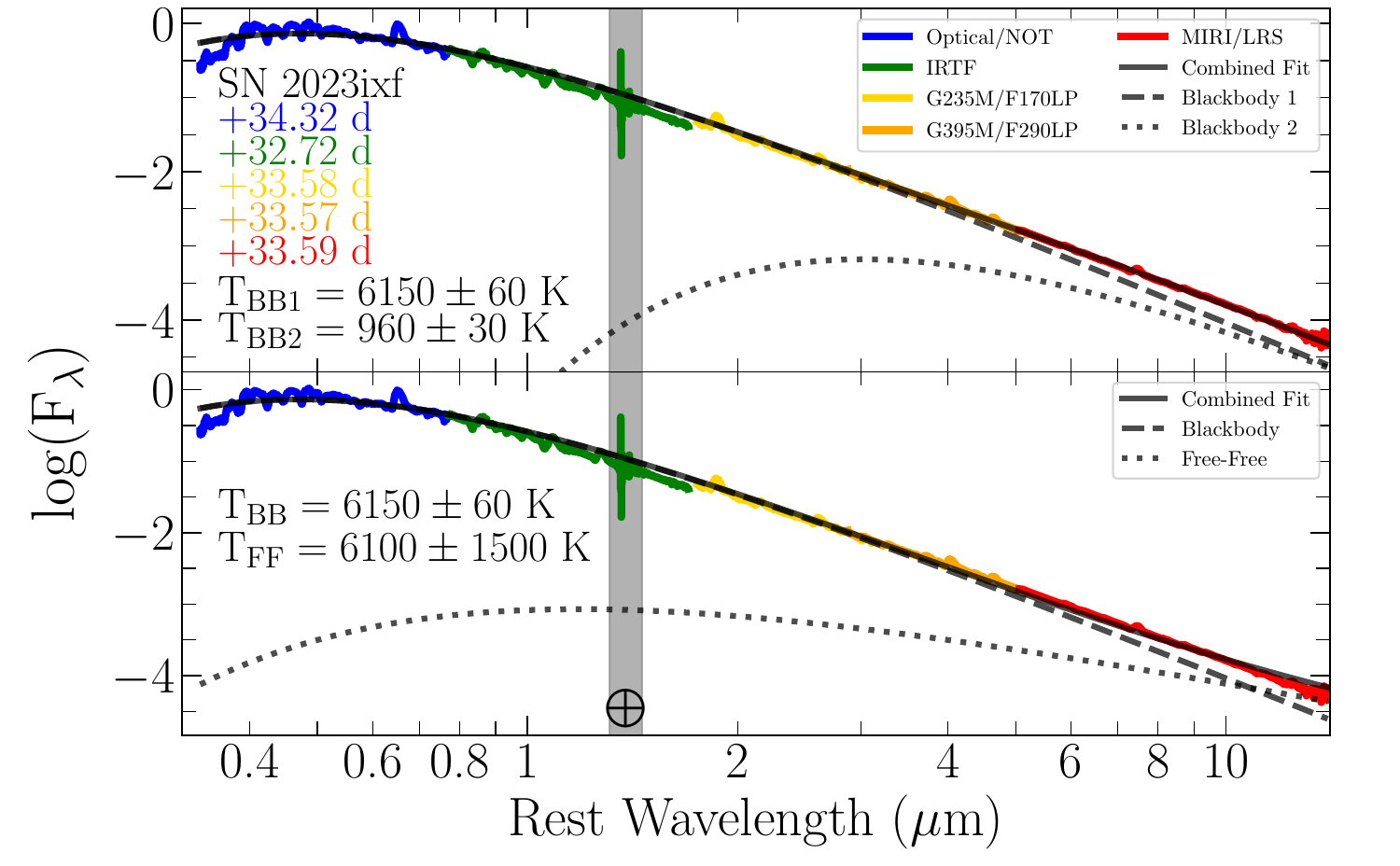}
    \caption{The optical through MIR SED of \ixf compared to the 
    simultaneous multi-component Monte Carlo fits. The top panel shows 
    a fit comprised of two blackbodies, while the bottom panel replaces
    the second, cooler blackbody with free-free emission. Both fits are 
    able to reproduce the emission at $\lambda > 4$~\mic, but are 
    overfit according to $\chi^2$. Based on the physical process which
    are occurring in the SN ejecta, we rule out warm dust as a source 
    of the IR excess.}
    \label{fig:sed}
\end{figure*}

\autoref{fig:sed} shows the optical through MIR SED of 
\ixf. The data have been corrected for extinction and 
redshift, and the ground-based optical and NIR spectra 
have been scaled to match the \jwst spectra in the 
overlap regions. At optical and NIR 
wavelengths ($0.4$~\mic $\lesssim \lambda \lesssim 4$~\mic), 
the underlying supernova continuum can be reasonably 
well approximated by a blackbody. The SED at 
bluer wavelengths is known to deviate from a blackbody 
approximation due to line blanketing from iron-group 
elements \citep{Hauschildt1995,Baron2003,Dessart2005,Bostroem2024}.
However, at longer wavelengths, emission 
from additional processes becomes important, rendering 
a single blackbody unable to capture the continuum behavior 
of the spectrum. These processes include: 
(1) the increasing fraction of free-free emission at 
longer wavelengths \citep{Aitken1988b}, 
(2) bluer flux redistributed to the IR by line scattering,
(3) heated circumstellar dust located within the unshocked 
CSM \citep{Aitken1988a,Roche1993,Sarangi2018_sn2010jl},
or some combination of the above \citep{Wooden1993}.

We performed a series of Monte Carlo (MC) fits utilizing 
a pair of two component models. The first model is the 
sum of a blackbody modeling the peak of the emission and  
a free-free emission component that provided excess 
emission at longer wavelengths. The second model replaces 
the free-free emission with a second blackbody component. 
The errors on the fit parameters reported below include 
both the fit error and the errors derived from MC 
distributions added in quadrature.
At the time of our observations ($+33.6$~days), the 
optical+NIR continuum of the combined spectrum is well fit 
by a blackbody with temperature $T_{BB} = 6150 \pm 60$~K, 
broadly consistent with the $T_{BB} = 5900 \pm 100$~K found 
by \citet{Zimmerman2023} derived from {\it UBVRI} photometry 
at $t = 34.06$~days. The addition of a free-free component 
($T = 6100 \pm 1500$~K) accounts for the under-prediction 
of the flux at $\lambda > 4$~\mic, however an additional blackbody 
at $T_{BB} = 960 \pm 30$, does just as well, if not slightly better, 
statistically at reproducing the observed continuum flux. 
A Markov Chain MC analysis shows that the parameters of the 
free-free fit are, in fact, insensitive to the data. Interestingly, 
the SED of \ggi at $\sim 55$~d is well fit by a single blackbody 
with $T_{BB} = 5000 \pm 100$~K. \citet{Baron2025} were 
able to model the continuum and line emission from \ggi, redward of 
$1$~\mic with a full NLTE model using a simple power-law density 
structure. This model includes the physics associated with scattering,
line transitions, and free-free emission in a self-consistent manner, 
where the multi-component fits presented here are only sensitive to 
the flux of the pseudo-continuum.

The comparable fit quality of the free-free fit and a cold blackbody 
to reproduce the residual emission renders us unable to address 
whether contributions from a dust continuum are necessary to match 
the observed flux at IR wavelengths. Such a dust component has 
been inferred as early as 60~days in SN~1987A \citep{Wooden1993} 
and at 87~days in the case of the Type IIn SN~2010jl 
\citep{Sarangi2018_sn2010jl}, but is not seen in SN~2004et at
64~days \citep{Kotak2009_04et}. \citet{VanDyk2024} fit the early 
($\sim 4$~d) CSM emission of \ixf with the combination of a 26,600~K 
blackbody and a 1600~K blackbody, attributing the IR excess to 
either dust or CO emission. 

However, most observations of \sneii 
at MIR wavelengths have focused on the nebular phases, leaving the 
early phases relatively unexplored \citep{Szalai2019} both in terms
of temporal coverage and the source of the IR emission (e.g., SN 
ejecta, free-free emission, or heated CSM dust).
The lack of a large excess at longer IR wavelengths
suggests that any dust component is either significantly dimmer than
the supernova itself (the pre-explosion flux values of the 
progenitor and its surroundings were $30.5 \pm 1.2$~$\mu$Jy and 
$22.1 \pm 1.0$~$\mu$Jy, respectively, in the 3.6~\mic and 4.5~\mic\
\spitzer bands, \citealp{VanDyk2023}), or most of the nearby dust has 
been destroyed by the interaction and subsequent shocks. 
\citet{Medler2025_23ixf} find 
evidence for dust emission decreasing with time in \ixf beginning 
at their earliest epoch ($t\sim 253$~d). The dust emission at 10~\mic 
in their data is $\sim 5.6$~mJy, whereas our fit gives a flux at this 
early time of only $\sim 2$~mJy, making it unlikely that the excess 
flux is really due to emission from warm dust. 

Additional plateau-phase data and full NLTE modeling, including 
free-free emission and other physical processes, are necessary to 
draw conclusions about the nature of any pre-existing molecules or 
dust in \sneii, in order to accurately estimate the amount of 
newly formed dust at later epochs. Physically, there is a contribution 
to the MIR flux from free-free emission and the underlying continuum 
is not actually that of a blackbody, even if the multi-component
fits are not statistically sensitive to it. Therefore, we reject the 
interpretation that the MIR excess is due to warm dust.

\section{Line Velocities and Profiles} \label{sec:lines}

\begin{figure*}[ht!]
    \centering
    \includegraphics[trim=5.3cm 0cm .35cm 2cm,clip=true,width=\textwidth]{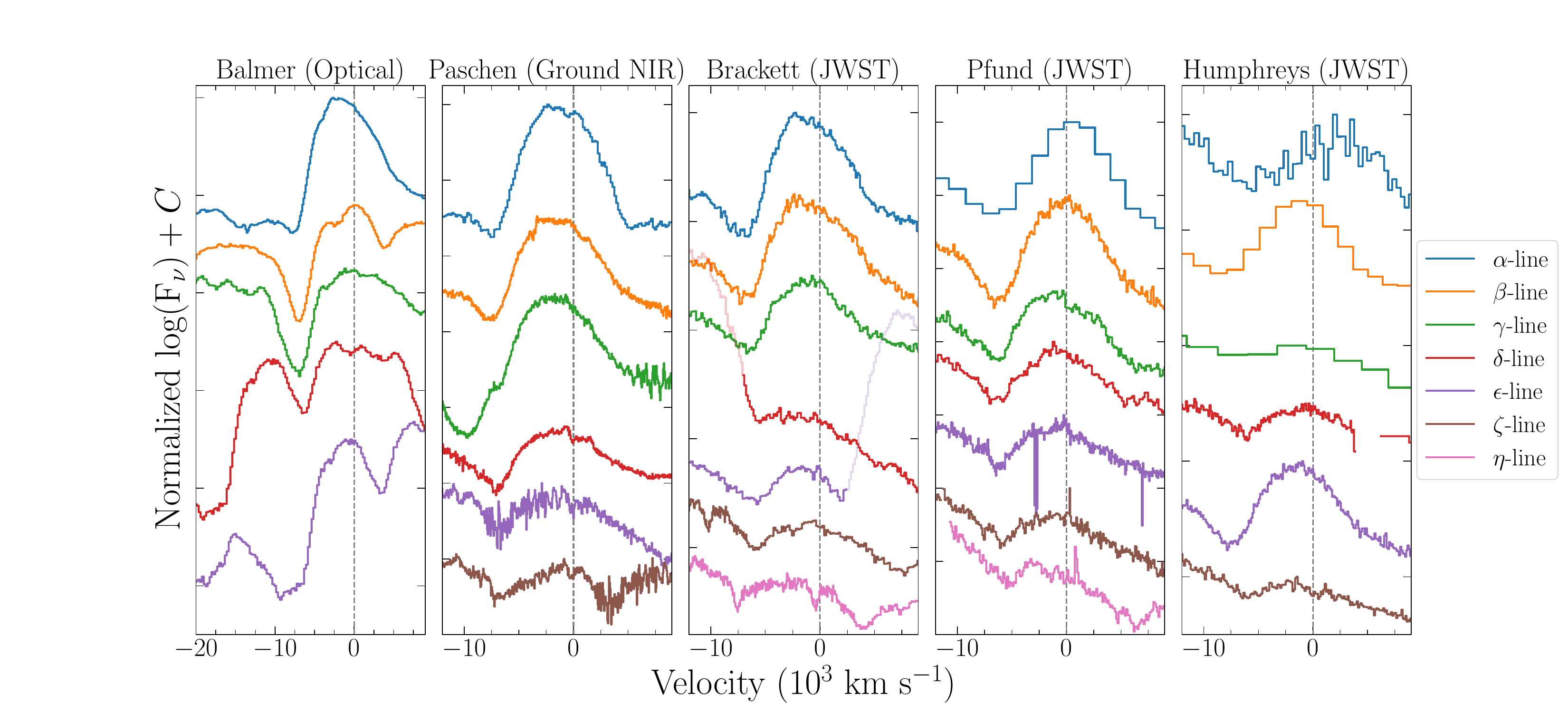}
    \caption{Hydrogen lines separated by series. Wavelengths of the individual 
    transitions can be found in \autoref{sec:line_ids} and their measured
    velocities in \autoref{tab:hvels}. Strongly blended lines include:
    \hl{Pa}{\alpha} with \hl{Br}{\delta} and \hl{Br}{\epsilon},
    \hl{Pf}{\alpha} with \hl{Hu}{\beta}, \hl{Pf}{\beta} with \hl{Hu}{\epsilon},
    and possibly \hl{Hu}{\gamma} with unknown lines.}
    \label{fig:hlines_byseries}
\end{figure*}

\autoref{fig:hlines_byseries} shows the strong, clearly 
identified hydrogen lines separated by series. In general, 
the widths of the line profiles are remarkably consistent 
across each individual series, except where the lines of 
the Brackett, Pfund, and Humphreys series blend and overlap. 
These widths are also consistent when compared according to 
the corresponding transition (e.g., alpha line, beta line, etc.) 
within each series. Individual lines of the Humphreys series 
are also subject to undersampling due to the lower resolution of the 
MIRI/LRS data relative to the NIRSpec data, particularly in 
the \hl{Hu}{\gamma} line. Weak and tentatively identified H 
lines in the MIRI data are excluded in the following 
analysis due to these issues. 

\subsection{Hydrogen Velocities} \label{sec:vel_fits}

To fit the absorption velocities, a hand-selected linear continuum is 
fit using nearby regions free of other lines. Where the lines show a 
defined P~Cygni shape, this continuum is subtracted, and the continuum
level used to separate the line into absorption and emission components. 
In lines without a clear P~Cygni profile or which are contaminated by 
strong blending (such as the \Hbeta, \Hgamma, and \Hdelta lines), this 
continuum is fit ``peak-to-peak'', and the fitting procedure is identical 
to that of \citet{Davis2019}, \citet{Shahbandeh2022}, and 
\cite{Shahbandeh2024_22acko}. Each measurement is repeated 500 times, 
where the flux is resampled from a normal distribution with its standard 
deviation equal to the flux error of each individual spectral point.
The values are shown in \autoref{tab:hvels}, where the reported errors 
include both the fit error and the resolution error of the spectrum 
added in quadrature. In most cases, the resolution error dominates over
the fitting error.

Superior fits were achieved for several lines by including a secondary 
absorption component. In select cases, (e.g., \Hbeta and \hl{Pa}{\gamma})
these secondary components could be attributed to blending by additional
lines (\ion{Fe}{2} $\lambda4303$ and \ion{He}{1} $\lambda1.083$, respectively).
However, in most cases, this second component was only added to better 
approximate the absorption component of the P~Cygni profile, which is known
to be non-Gaussian \citep{Teffs2020,Shahbandeh2022}. 
In these instances, the absorption minimum was obtained from the 
multi-component fit, with the reported error determined in the same 
manner as in the single Gaussian case, as the resolution error 
consistently dominates over the fit errors.
We also fit the \ion{Fe}{2} $\lambda5169$ line, as its velocity is 
commonly taken to represent of the photospheric velocity in \sneii. 
We find the absorption minima to be $-6120\pm420$~\kms.

From the results of these fits, we see that the absorption minima follow 
a near-monotonic trend, where the lower energy transitions (e.g., alpha-lines)
have faster absorption minima than high energy lines within their own series. 
For hydrogen lines with strong, unblended emission components (such as 
\Halpha and \hl{Br}{\alpha}), the emission peaks are blue-shifted by 
$\sim 3000$~\kms. Models of the emission lines in \sneii require
steep density profiles (power laws with indexes $n \lesssim -8$) in order 
to achieve this blue-shift \citep{Duschinger1995,Dessart2005}.
The observed properties of \ixf are consistent with the trends 
found in \citet{Anderson2014} that larger blue-shifts are correlated 
brighter $V$-band maximums and a larger decline rate during the 
plateau ($s_{2}$) in \sneii. From these observations, we therefore 
conclude that the ejecta of \ixf must also have a steep density profile.

\begin{deluxetable}{cccc}
    \tablecaption{Hydrogen Absorption Velocities \label{tab:hvels}}
    \tablehead{\colhead{Line} & \colhead{Rest Wavelength} & 
               \colhead{Velocity}& \colhead{Error} \\
                & \colhead{(\mic)} & \colhead{(\kms)} & \colhead{(\kms)}}
    \startdata
      \hline 
      \multicolumn{4}{c}{Balmer Series} \\
      \hline
      \Hdelta & 0.4102 & $-6310$ & 420 \\
      \Hgamma & 0.4340 & $-6480$ & 420 \\
      \Hbeta & 0.4861 & $-6790$ & 420 \\
      \Halpha & 0.6563 & $-7630$ & 420 \\
      \hline 
      \multicolumn{4}{c}{Paschen Series} \\
      \hline
      \hl{Pa}{\eta} & 0.902 & $-6850$ & 400 \\
      \hl{Pa}{\zeta} & 0.923 & $-6870$ & 400 \\
      \hl{Pa}{\epsilon} & 0.955 & $-6940$ & 400 \\
      \hl{Pa}{\delta} & 1.005 & $-6940$ & 400 \\
      \hl{Pa}{\gamma} & 1.094 & $-6860$ & 400 \\
      \hl{Pa}{\beta} & 1.282 & $-7890$ & 400 \\
      \hl{Pa}{\alpha} & 1.875 & $-7720$ & 380 \\
      \hline 
      \multicolumn{4}{c}{Brackett Series} \\
      \hline
      \hl{Br}{\gamma} & 2.166 & $-6760$ & 330 \\
      \hl{Br}{\beta} & 2.625 & $-7170$ & 270 \\
      \hl{Br}{\alpha} & 4.051 & $-7470$ & 290 \\
      \hline 
      \multicolumn{4}{c}{Pfund Series} \\
      \hline
      \hl{Pf}{\zeta} & 2.873 & $-6540$ & 250 \\
      \hl{Pf}{\epsilon} & 3.039 & $-6520$ & 230 \\
      \hl{Pf}{\delta} & 3.297 & $-6700$ & 360 \\
      \hl{Pf}{\gamma} & 3.741 & $-6830$ & 320 \\
      \hl{Pf}{\beta} & 4.654 & $-6750$  & 260 \\
      \hl{Pf}{\alpha} & 7.46 & $-7970$ & 2980 \\
      \hline 
      \multicolumn{4}{c}{Humphreys Series} \\
      \hline
      \hl{Hu}{\zeta} & 4.376 & $-6270$ & 300 \\
      \hl{Hu}{\delta} & 5.129 & $-6520$ & 240 \\
      \hl{Hu}{\alpha} & 12.372 & $-6760$ & 1400 \\
    \enddata
    \tablecomments{The following \ion{H}{1} lines are too blended or
    at too low a resolution to accurately measure: \hl{Br}{\delta}, 
    \hl{Br}{\epsilon}, \hl{Hu}{\beta}, \hl{Hu}{\gamma}, \hl{Hu}{\epsilon}, 
    \hl{Hu}{\eta}.}
\end{deluxetable}

\subsection{IR Hydrogen Line Profiles} \label{sec:Halphas}

The line profiles of key transitions are known to encode 
important information about the ejecta, environment, and 
evolution of \sneii. Well-known examples of this structure 
include (but are not limited to): high velocity absorptions
arising from both the CDS and the forward/reverse shocks 
\citep{Chugai2007,Dessart2022}, the ``Cachito'' feature 
\citep{Gutierrez2017}, the Bochum event in SN~1987A (i.e., 
a \Nifs bullet, \citealp{Larson1987,Phillips1989,Utrobin1995}), 
dust absorption \citep{Lucy1989,Smith2008,Gall2014}, and 
clumpy material \citep{Singh2024}. Here, we systematically 
examine several structures seen in the five named series 
producing strong hydrogen lines across the spectrum.

\subsubsection{High Velocity Features}

\begin{figure}
    \centering
    \includegraphics[width=\columnwidth]{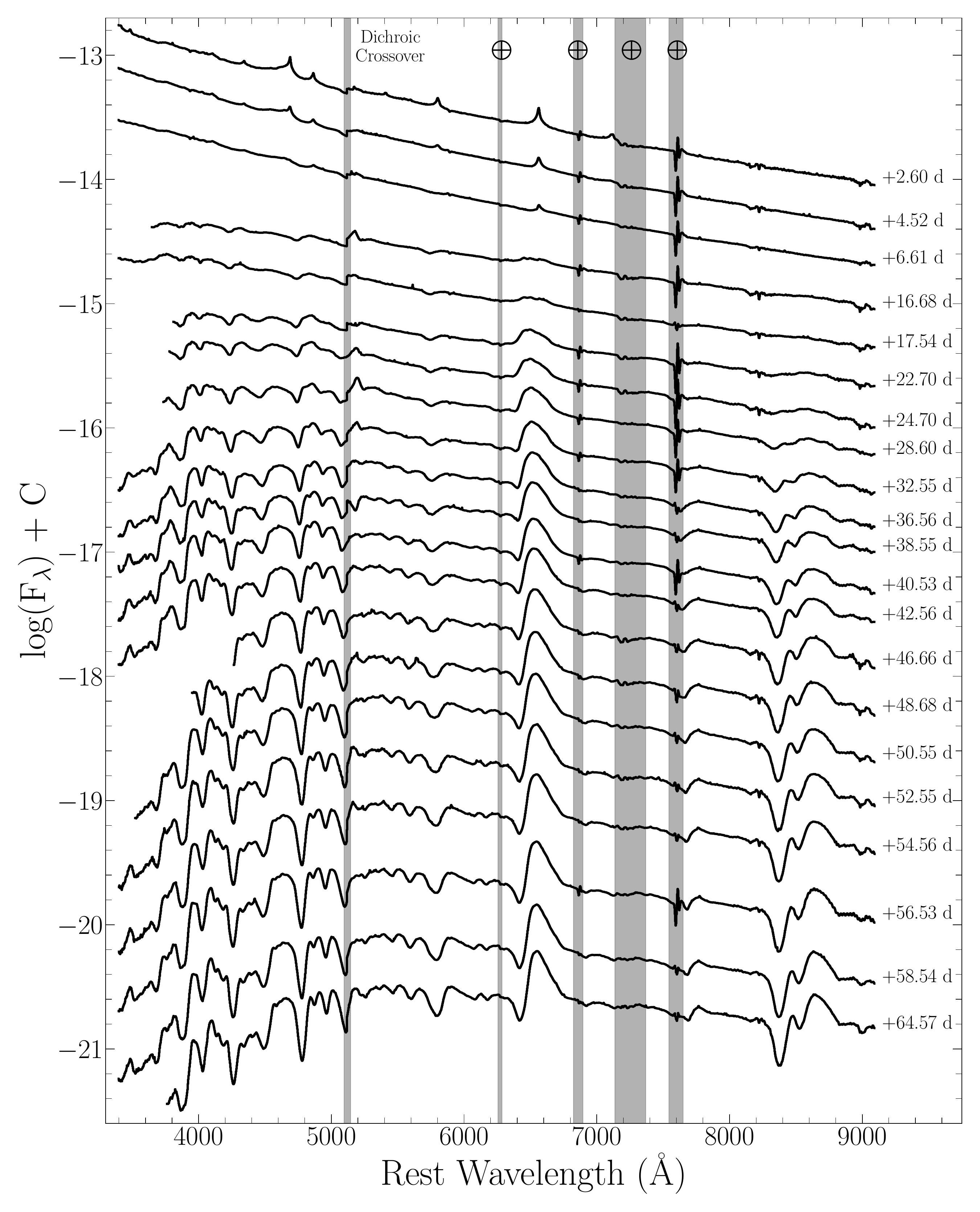}
    \caption{Ground-based time-series optical spectra obtained 
    with UH88/SNIFS by the SCAT collaboration. Regions of strong 
    telluric absorption and the dichroic crossover region at 
    $\sim 5100$~\AA\ are marked in grey.}
    \label{fig:scat_opt}
\end{figure}

\begin{figure}
    \centering
    \includegraphics[width=\columnwidth]{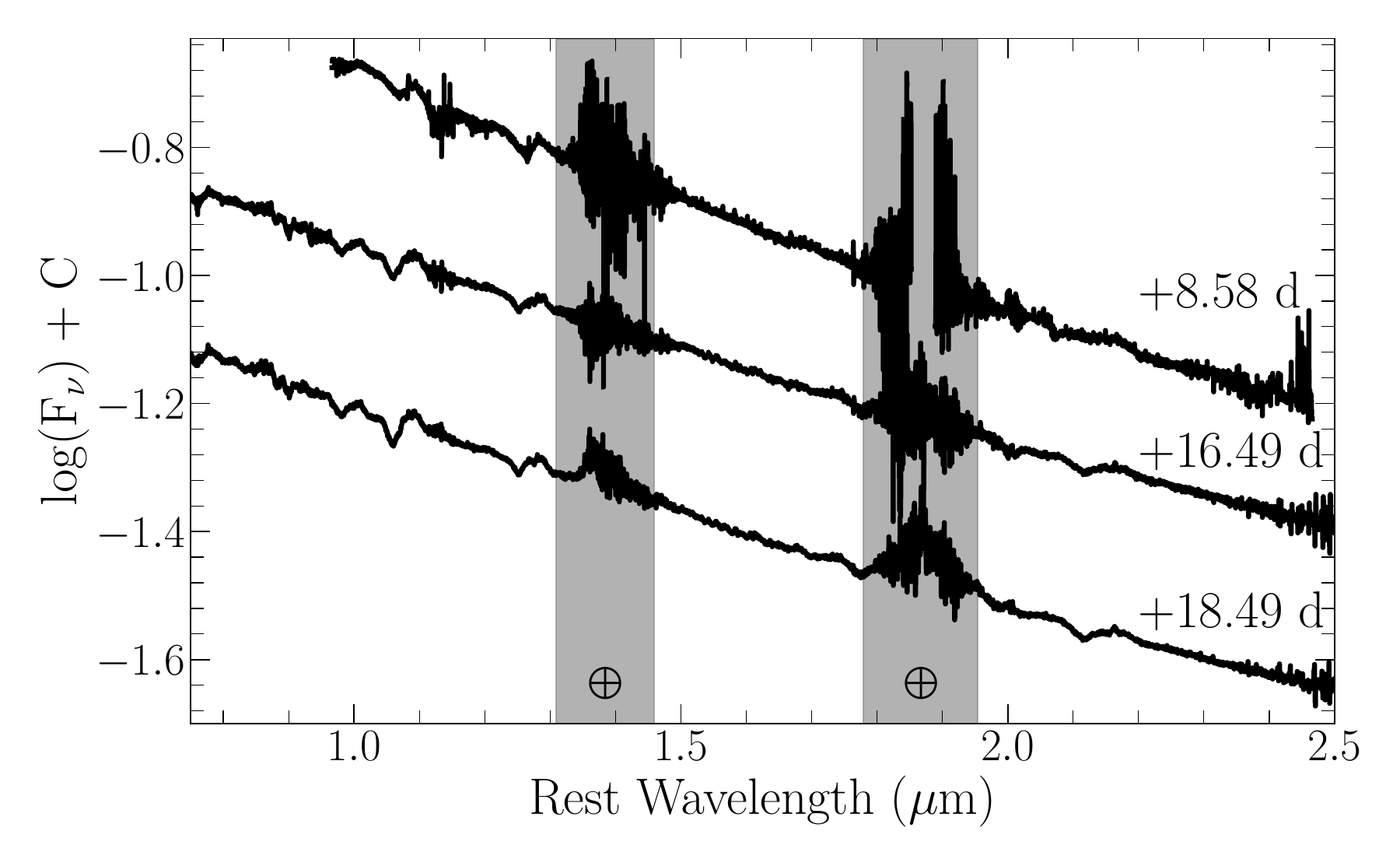}
    \caption{Ground-based NIR spectra obtained by HISS \citep{Medler2025_hiss} 
    using Keck-II/NIRES and IRTF/SpeX. Channel gaps arising from atmospheric
    absorptions are marked in grey.}
    \label{fig:hiss}
\end{figure}

We first look for evidence of HV features (HVFs) in our 
time-series spectroscopic
data shown in \autoref{fig:scat_opt} and \autoref{fig:hiss}.
These features can form a multitude of profiles as 
demonstrated by \citet{Chugai2007}. In their model including 
circumstellar interaction, HVFs can arise from ionization 
of the outer ejecta by the reverse shock. Further structure 
originating from absorption by the CDS and outward mixing of 
the CDS produce additional narrow and broad HVFs, respectively. 
These results have been verified in other NLTE models of \sneii
\citep{Dessart2013,Dessart2022}. The ``Cachito'' is one 
manifestation of this complexity, where early, shallow 
\Halpha HVFs likely result from the reverse shock; while later, 
narrow HVFs are connected to the CDS and forward shock.

\begin{figure*}[t!]
    \centering
    \includegraphics[trim=3.5cm 0cm 3.5cm 2cm,clip=True,width=0.9\textwidth]{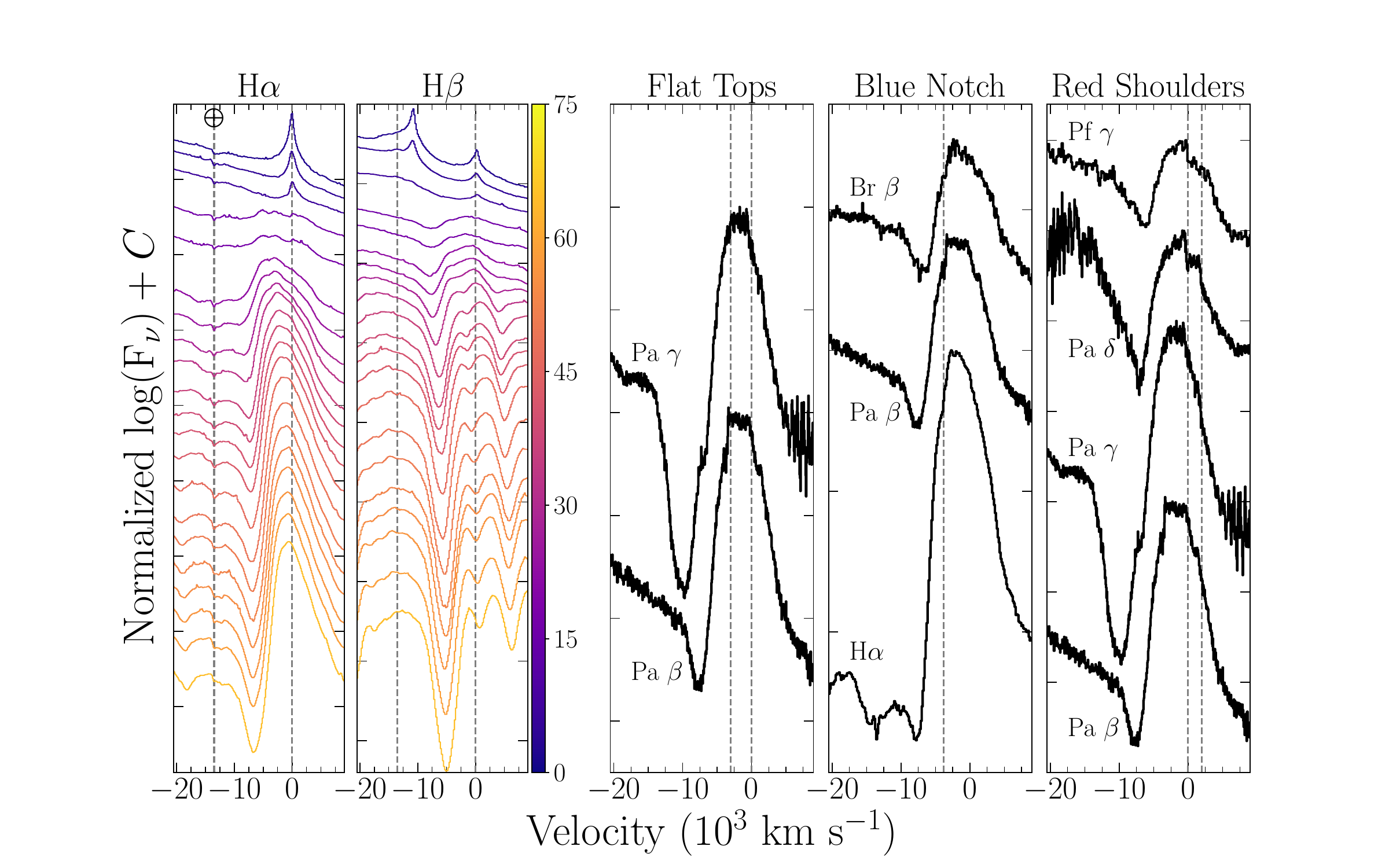}
    \caption{{\it Left panels}: SCAT time-series optical spectra of \Halpha 
             and \Hbeta, colored by days from the explosion. In contrast to 
             \citet{Singh2024}, we find no evidence of a HVF in \Hbeta 
             at $-13,500$~\kms.
             {\it Right panels}: Examples of line structure seen in IR 
             hydrogen lines at $\sim 33$~days. Vertical gray lines denote the velocity 
             extent of the structures. The ``flat-tops'' of lines extend 
             from $\sim -3,000 - 0$~\kms, the blue notch is centered at 
             $\approx -3,800$~\kms, and the red shoulders extend from 
             $\sim 0 - 2,000$~\kms, consistent across all lines 
             showing each structure within the resolution errors.}
    \label{fig:structure}
\end{figure*}

As seen in the left two panels of \autoref{fig:structure}, the 
absorption profile of \Halpha is complex in \ixf, showing both 
a shallow, broad HV component, in addition to the narrow, weak 
P~Cygni absorption. There is also strong contamination from the 
telluric O$_{2}$ $\gamma$-band at 6280~\AA\ ($-13,200$~\kms 
relative to \Halpha). A ``Cachito'' originating from \ion{Si}{2} 
$\lambda6355$ is ruled out, as the inferred velocity would be
less than the photospheric velocity measured from the \ion{Fe}{2} 
$\lambda5169$ line, consistent with other \ixf time-series analyses 
\citep{SinghTeja2023,Singh2024}. Time-series optical spectra of 
\ixf show that this HVF appears simultaneously with the P~Cygni 
component ($\sim+16.7$~days after explosion in our data); gradually 
weakening and slowing until around $+47$~days when its evolution 
becomes difficult to separate from the telluric contamination at 
$-13,200$~\kms. The weakening and slowing of the HVF is consistent 
with the behavior expected from the combination of cooling via 
geometrical dilution and the propagation of the reverse shock to 
deeper layers.

\citet{Chugai2007} suggest that such HVFs may also be visible in 
both \Hbeta and the \ion{He}{1} 1.083~\mic lines, and such HVFs have
been previously seen in SN~2017eaw \citep{Tinyanont2019}. We find no 
evidence for HVFs in either line in our data (but see \citealt{Singh2024} 
for an alternative view), nor evidence for HVFs in additional IR hydrogen 
lines. In particular, the \Hbeta  HVF is likely to be extremely weak 
due to the low Sobolev optical depth. For the majority of additional 
hydrogen lines covered by \jwst, the estimated Sobolev optical depths 
(see Eqns. (4) and (13) of \citealp{Jeffery1990}) will be weaker than 
that of \Hbeta, explaining the absence of HVFs in the IR H lines. 
Any weak HVFs which may be visible in strong NIR lines such as 
\hl{Pa}{\alpha} are obscured by blending.

\subsubsection{Clumping in \hl{Pa}{\alpha}?}

\begin{figure}[t!]
    \centering
    \includegraphics[width=\columnwidth]{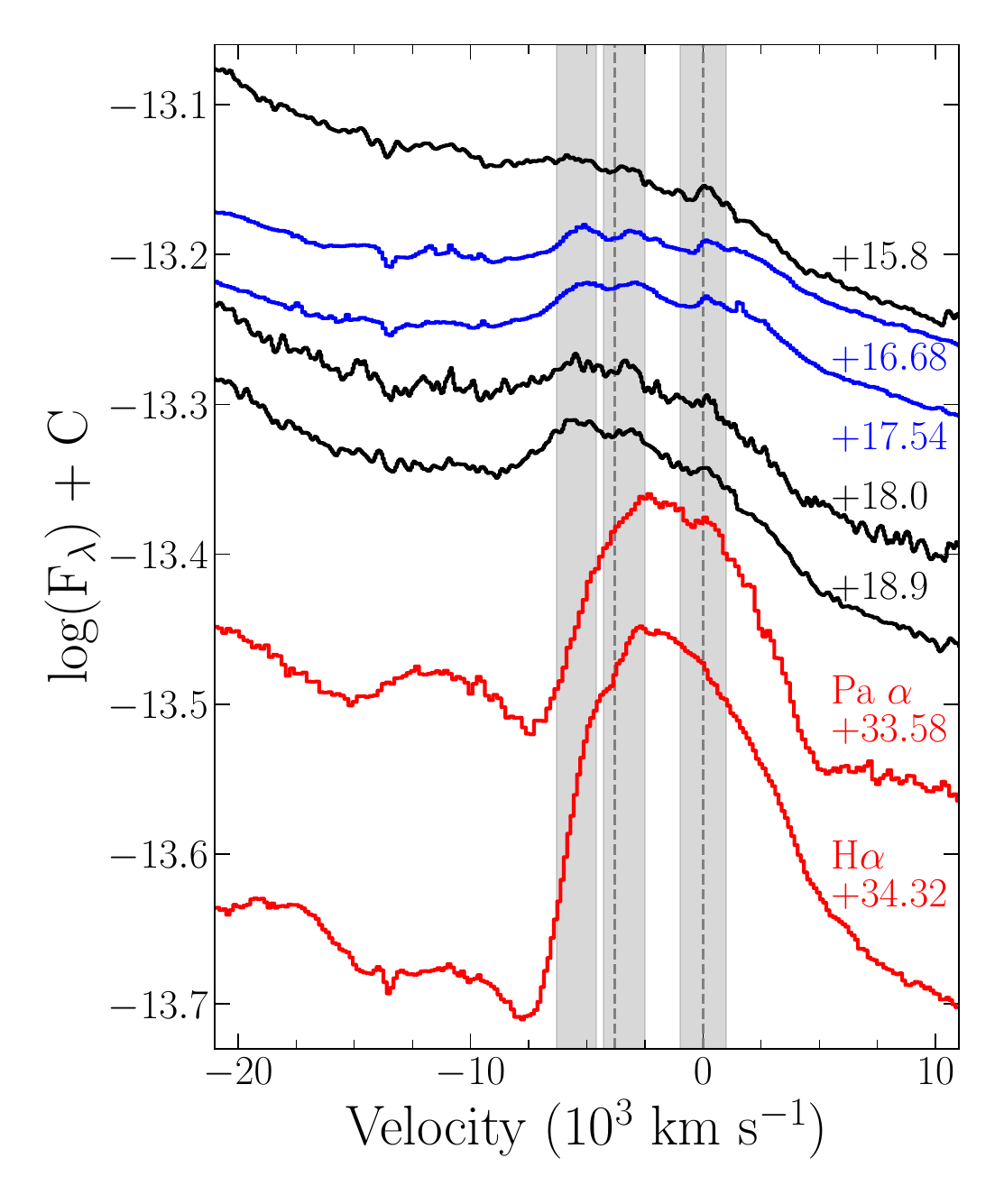}
    \caption{Clumpy structure in \Halpha from \citet{Singh2024} (black)
    and SCAT (blue) compared to the \Halpha and \hl{Pa}{\alpha} 
    profiles (red) from our near-contemporaneous SED of \ixf. Only the 
    structure arising from the clump at $v = 0$~\kms is seen atop 
    \hl{Pa}{\alpha}. All other clumps seen previously in \Halpha are no
    longer present.}
    \label{fig:clumps}
\end{figure}

\citet{Singh2024} note that between $+9.9$ and $+25.8$~days, 
additional intricate structures can be seen, appearing at similar 
velocities in both the \Halpha and \Hbeta emissions. They attribute 
these structures to multiple clumps in the interaction region. 
Consistent with their findings, by the time of our \jwst observations, 
these structures are absent from the \Halpha and \Hbeta lines. However, 
small-scale structure is seen atop the \hl{Pa}{\alpha} line from our 
\jwst/NIRSpec observations, with the $v \approx 0$~\kms clump appearing 
in \hl{Pa}{\alpha} with the same shape and velocity as first seen in 
\Halpha (see \autoref{fig:clumps}). No other structures matching those 
seen previously in \Halpha are identified in either \hl{Pa}{\alpha} or 
other strong hydrogen lines in our optical through MIR spectra. Nor can 
this structure be attributed to the nearby \hl{Br}{\delta} and 
\hl{Br}{\epsilon} lines, whose profiles only blend at the edges of the 
\hl{Pa}{\alpha} profile, as evidenced by the similarities between the 
\hl{Pa}{\alpha} profile and those of \Halpha and other Paschen lines. 
Furthermore, no other weaker hydrogen lines exist at these wavelengths 
originating from levels identified elsewhere in our spectra.

\subsubsection{Additional Structure in IR Hydrogen Lines}

As seen in \autoref{fig:structure}, there are multiple additional 
structures which appear in different subsets of the IR hydrogen lines.
These structures include: ``flat-topped'' emission profiles (such as 
those observed in the strong emissions of \hl{Pa}{\beta} and \hl{Pa}{\gamma}), 
a blue emission notch (observed in \Halpha, \hl{Pa}{\beta}, and \hl{Br}{\beta}), 
and a red emission shoulder (found in \hl{Pa}{\beta}, \hl{Pa}{\gamma}, 
\hl{Pa}{\delta}, and \hl{Pf}{\gamma}). Such structures may arise independently 
through a variety of effects, but can also emerge in combination through a 
single mechanism, in some instances (see, for example, the Bochum event). Among 
strong IR hydrogen lines deviating from the expected P~Cygni profile, only 
\hl{Pa}{\beta} shows all three structures. Below, we detail several possible 
origins of these additional structures and evaluate the consistency of these 
mechanisms with the known properties of \ixf.

``Flat-topped'' emission lines are common to many types of stellar explosions,
and have been observed in multiple \sneii \citep{Pastorello2009,Inserra2013,
Gutierrez2014,Gutierrez2020,Medler2023}. Also referred to as detached profiles 
\citep{Jeffery1990}, these flat-tops arise from shells of material above 
the photosphere. During the photospheric phases, lines still show an 
absorption trough; in contrast to their box-like appearance during
nebular phases. However, in \ixf, the edges of the flat-tops extend from 
$\sim -3000 - 0$~\kms, well below the inferred photospheric velocity of 6120~\kms.
Emission from the CDS could also produce a flat-topped profile, 
as suggested by \citet{Pastorello2009} in the case of SN~1999ga. However, 
as the CDS forms above the photosphere, the emission would appear at 
super-photospheric velocities, not the sub-photospheric velocities seen in
\ixf. This is confirmed by the models of \citet{Bostroem2024}, which show 
such a red shelf appearing at $\sim9200$~\kms in \Halpha around 50 days 
after explosion. \citet{Inserra2013} attribute the ``flat-topped'' lines 
they see during 
the photospheric phases of SNe~1995ad and 1996W to underlying \ion{H}{2}
regions. We can, however, rule out this possibility as extensive pre-explosion 
imaging reveals \ixf to be separated by $\sim$~1\arcsec\ from NGC~5461, the 
nearest observed \ion{H}{2} region in M~101 \citep{Pledger2023,VanDyk2023}.

Instead, we suggest that these flat-topped profiles 
likely arise from a combination of geometric and opacity 
effects. The inferred steep density profile from velocity 
measurements effectively creates a region with a small, 
non-zero constant opacity, resulting in the flat-topped 
profiles which are a special case of the detached profiles 
outlined by \citet{Jeffery1990}. Such flat-topped profiles 
are not seen in weaker lines originating from higher-level 
transitions, as they have effectively no opacity at the 
photosphere and therefore appear as the pure absorption 
lines described by \citet{Baron2025} as their 
detached lines case in SN~2024ggi. The stronger H lines 
(e.g \Halpha, \hl{Pa}{\alpha}, \hl{Br}{\alpha}) involve 
the only levels sufficiently populated across the steep 
density slope to have a varying opacity over the narrow 
velocity range and produce the typical P~Cygni emission 
peak.

Several IR lines show depressed flux in the red half of 
their emission peaks starting at zero velocity, with a 
corresponding shoulder at $\sim2000$~\kms. This is most 
prominently seen as an abrupt drop in flux in the Paschen 
series lines, but also as steeply-sloped flux decreases 
in the strong Brackett and Pfund lines (at lower S/N). 
At late times, one potential origin of such a depression 
is obscuration by newly formed dust in the SN ejecta 
\citep{Lucy1989,Bevan2019}. While our SED modeling in 
\autoref{sec:sed} cannot conclusively address the 
potential presence of dust; our combined analysis suggests 
the ejecta is too hot to form either molecules (see 
\autoref{sec:sed} and \autoref{sec:modeling}) or new dust, 
consistent with previous studies of other \sneii with early 
IR excesses \citep{Wooden1993,Sarangi2018_sn2010jl}. Therefore, 
we consider it highly unlikely that this red shoulder 
originates from newly formed dust in the ejecta.
Pre-existing heated dust, such as that found by
\citet{Medler2025_23ixf} is unable to produce this obscuration
as it is located exterior to the ejecta.

An additional notch is also seen on the blue side of the \Halpha 
emission peak at $\sim -3800$~\kms. This notch is visible in the
time-series of spectra of \citet{Singh2024} until $\sim32$~days after
explosion (see their Figure 6), consistent with the optical time series 
presented in \autoref{fig:scat_opt}. When combined with the red shoulder
observed in several other hydrogen lines, the combined structure is reminiscent 
of the Bochum event observed in SN~1987A \citep{Phillips1989,Hanuschik1990}. 
During the Bochum event, NIR profiles of the hydrogen lines showed 
a significant asymmetric, double-peaked profile 
(\citealp[see for example, Fig 1. of][]{Larson1987}), with the two
peaks of the \hl{Pa}{\alpha} line separated by $\sim 4000$~\kms. No
such double-peaked profiles are seen in \ixf,  seemingly ruling
out the presence of a Ni bullet.

The velocity of the blue notch does however, correspond to the minima
between the two higher velocity clumps identified in \citet{Singh2024}.
In their scenario, the clumps in the CSM are fully overrun by the SN ejecta
between 30-40 days, and the smooth nature of the P~Cygni profile emerges
in later epochs. However, the lack of strong, uncontaminated lines in 
the ground-based IR data and the sparsity of temporal coverage relative to 
optical wavelengths provides no strong evidence for the existence
of the clumps seen in the IR (see \citealt{Park2025} for discussions of 
small scale line structure in ground-based NIR spectra of \ixf).

As highlighted by \citet{Jeffery1990}, density peaks are far more likely to 
occur than detached atmospheres within the supernova ejecta. Such density peaks
can create profiles which are qualitatively similar to those seen in \ixf 
(see their Fig. 7), albeit with a secondary blue emission maximum due to the 
opacity jump from the density peak. However, when the opacity difference is 
small (as is expected if turbulent mixing occurs within the ejecta), these 
variations in the line profile may not be easily identifiable due to the
density peaks being smoothed out \citep{Jeffery1990}. However, the models 
they present place the density peak above the photosphere,
while these structures in \ixf lie below the inferred photospheric velocity. 
Such density peaks may be present in \ixf, possibly due to the existence of a 
dusty torus as inferred from polarimetric observations \citep{Vasylyev2023,Singh2024}.
But it is important to note that one-dimensional, spherically symmetric 
codes such as \texttt{SYNOW} can not fully capture the inherently 
multi-dimensional structure of the ejecta. Future detailed multi-dimensional 
radiation-hydrodynamic modeling of the asymmetric structures, any associated 
shocks, and viewing angle dependencies present in \ixf may distinguish between 
these possible scenarios.

\section{Limits on Carbon Monoxide} \label{sec:modeling}

\begin{figure*}[th!]
    \centering
    \includegraphics[width=0.9\textwidth]{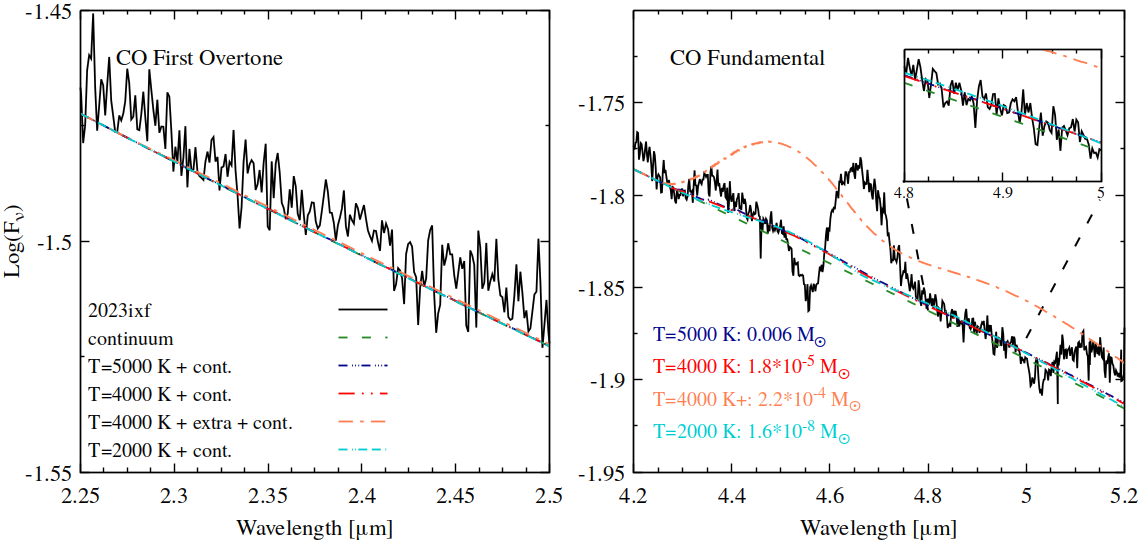}
    \caption{Temperature-dependent limits on CO emission in \ixf. Note
    the P Cygni profile at $\sim 4.2-4.3$, $4.5-4.8$, and $5-5.2$~\mic 
    are from \hl{Hu}{\zeta}, \hl{Pf}{\beta}$+$\hl{Hu}{\epsilon}, and 
    \hl{Hu}{\delta}, respectively.
    The representative examples of the emission from various temperature
    and mass combinations in the optically thin case are shown for 
    the first overtone band (left) and the fundamental band (right). The 
    maximum CO mass is determined based on the fundamental band, and 
    varies between $1.6 \times 10^{-8} - 0.006$~\Msol with temperature. 
    The orange dash-dotted 4000~K signal demonstrates a realistic 
    detection of the CO fundamental, including an extra continuum 
    (above that of a blackbody) to mimic effects of free-free emission 
    and lines. The inset shows the non-detections compared to 
    the continuum in a region of the fundamental free of H lines.}
    \label{fig:CO}
\end{figure*}

The spectrum of \ixf shows no evidence for the formation of 
molecules (e.g., CO or SiO). As seen in \autoref{sec:sed} the 
excess continuum is well fit by just a $\sim 1000$~K Planck 
function at wavelengths $>4$~\mic and by a $\sim 6000$~K 
Planck function at shorter wavelengths. Therefore we obtain upper 
limits on the amount of pre-existing molecules and dust, which 
sets a baseline for observations at later times. The in-depth 
procedure by which we place these upper-limits for the plateau 
phase of SNe~IIP is detailed in \citet{Shahbandeh2024_22acko}. 
We briefly summarize it here for clarity, and refer interested 
readers to that paper and references therein for further details.

To obtain upper limits on CO in \ixf during the plateau phase, we 
assumed an isothermal ejecta with a density structure typical for a 
\sniip. The density gradient was obtained as in \citet{Shahbandeh2024_22acko} 
and gives a power-law index of $n \approx -2.5$. \autoref{fig:CO} 
shows the CO emission for both the fundamental (right) and first 
overtone (left) of CO, and their flux for given temperatures in 
comparison with observation. The opacity peaks at approximately 
2500–3000~K and decreases by roughly six orders of magnitude at 
the recombination temperature of H. Because the opacity and the 
specific emissivity relative to the continuum flux in the first 
CO overtone is smaller than the fundamental band by a factor on 
the order of $\approx 100$, the fundamental band provides the upper 
limits, illustrating the importance of \jwst data. If CO exists 
in the CSM it would add opacity due to cold CO in the surroundings 
of \ixf. Since no CO is observed, any pre-existing CO is optically 
thin. The emission will be $\propto T$ in the fundamental band. If 
CO is observed in \ixf at later epochs, it will be due to molecules 
synthesized in the ejecta and not due to existing molecules in the 
CSM, which will not evolve in time. Our simple exercise does not 
constrain anything about the CO, because there is, in fact, no 
evidence in the data for the presence of CO at this epoch. 
\citet{Park2025} find about $2\times 10^{-4}$~\Msol of CO with 
$T\approx 3000$~K in \ixf at $+199$~days from 
ground-based spectra.

\section{Conclusion} \label{sec:conclusion}

Here, we present observations of \ixf with \jwst taken 33.6~days 
after explosion, which roughly corresponds to the mid-point
of the plateau phase. 
The NIR+MIR spectra are dominated by H lines, while contemporaneous
ground-based observations in the optical and NIR reveal spectra 
consistent with other \sneii at the observed epoch. The \jwst 
NIR spectra comparisons to \acko show that the correlations between 
H line widths and velocities and $V$-band peak magnitude continue
into the IR. MIR spectral comparisons to SNe~1987A and 2022acko at 
similar evolutionary phases reveal only the strongest lines such
as the \hl{Pf}{\alpha}+\hl{Hu}{\beta} blend and \hl{Hu}{\alpha}
lines are prominent at low spectral resolutions. 

Fits to the panchromatic (0.35--14~\mic) SED 
reveals the spectral continuum can be fit  to a blackbody with 
$T_{BB} = 6150 \pm 60$~K, with a small excess at $\lambda \gtrsim 4$~\mic. 
This excess can be fit by the inclusion of free-free
emission or  by adding a second blackbody component. We reject the 
explanation that the excess emission is caused by warm dust,
based on the physical processes occurring in the SN ejecta and 
the time evolution of the MIR flux. The nature and need 
of additional emission components varies across the small number
of \sneii with early MIR observations. Caution should be given to 
inferring dust excess during the plateau from single epoch 
spectroscopy. 

Furthermore, no observational signals of molecules (e.g., CO) 
are detected; and we place limits on the mass of pre-existing CO 
in the ejecta. Such measures are necessary to ensure the amount 
of newly synthesized molecules and dust can be accurately traced 
over time. This is of critical importance for \sneii 
assumed to arise from dusty RSG progenitors like \ixf, where 
such signals may also arise from heated CSM, as suggested by
later observations \citep{Medler2025_23ixf}.

These observations will form the basis of what is expected to 
be decades of follow-up of \ixf using \jwst. Paper II of this 
series \citep{Medler2025_23ixf} explores the nebular phase panchromatic 
spectral evolution with \jwst observations 
\citep{Ashall2023_cycle2_23ixf,Ashall2024_cycle3_ixf}, 
and future papers will focus on the evolution of molecules and 
dust in \ixf. These and other future observations will provide 
the opportunity to trace potential molecule formation and dust 
growth in what is likely to be one of the closest \sneii in the 
\jwst era, and provide valuable insight into the formation and 
origin of dust in the early universe.

\begin{acknowledgments}
We thank Alison Vick, Stephan Birkmann, George Chapman, Amanda Marrione,
Brian McLean, Ed Nelan, Alberto Noriega-Crespo, Beverly Owens, Scott Stallcup, 
and the entire \jwst operations and scheduling teams for their hard work in
scheduling and executing these time-critical observations.

J.D., C.A., K.M., P.H., E.B., T.M., and M.S. are supported in part by NASA 
grants JWST-GO-02114, JWST-GO-02122, JWST-GO-03726, JWST-DD-04436,
JWST-DD-04522, JWST-GO-4217, 
JWST-DD-04575, JWST-GO-5057, JWST-GO-5290, JWST-GO-6023, and
JWST-GO-6677.
J.D., C.A., and E.B. were supported in part by HST-AR-17555, 
Support for program Nos. 2114, 2122, 3726, 4436, 4522,
4575, 5057, 5290, 6023, 6677, and 17555 were provided by NASA through 
grants from the Space Telescope Science Institute, which is operated 
by the Association of Universities for Research in Astronomy, Inc., under 
NASA contract NAS 5-03127. 
PAH is supported in parts by the NSF grant AST-230639.
L.G. acknowledges financial support from AGAUR, CSIC, MCIN and 
AEI 10.13039/501100011033 under projects PID2023-151307NB-I00, 
PIE 20215AT016, CEX2020-001058-M, ILINK23001, COOPB2304, and 
2021-SGR-01270.
M.D.S is funded by the Independent Research Fund Denmark 
(IRFD, grant number  10.46540/2032-00022B) and by an Aarhus 
University Research Foundation Nova project (AUFF-E-2023-9-28).
J.T.H. was supported by NASA grant 80NSSC23K1431.
S.M. is funded by Leverhulme Trust grant RPG-2023-240.

This work is based on observations made with the NASA/ESA/CSA James 
Webb Space Telescope. The data were obtained from the Mikulski Archive 
for Space Telescopes at the Space Telescope Science Institute, which 
is operated by the Association of Universities for Research in Astronomy, Inc., 
under NASA contract NAS 5-03127 for \jwst. These observations are associated 
with program No. 4522. The specific observations analyzed in this work can be 
accessed via \dataset[doi: 10.17909/ekjp-5b33]{\doi{10.17909/ekjp-5b33}}.

This work was supported by a NASA Keck PI Data Award, administered 
by the NASA Exoplanet Science Institute. Data presented herein were 
obtained at the W. M. Keck Observatory from telescope time allocated 
to the National Aeronautics and Space Administration through the 
agency's scientific partnership with the California Institute of
Technology and the University of California. The Observatory was 
made possible by the generous financial support of the W. M. Keck 
Foundation. The authors wish to recognize and acknowledge the very 
significant cultural role and reverence that the summit of Maunakea 
has always had within the indigenous Hawaiian community. We are most 
fortunate to have the opportunity to conduct observations from this 
mountain.

Based on observations made with the Nordic Optical Telescope, 
owned in collaboration by the University of Turku and Aarhus 
University, and operated jointly by Aarhus University, the 
University of Turku and the University of Oslo, representing 
Denmark, Finland and Norway, the University of Iceland and 
Stockholm University at the Observatorio del Roque de los 
Muchachos, La Palma, Spain, of the Instituto de Astrofisica 
de Canarias. The NOT data were obtained under program ID P66-506.
\end{acknowledgments}

%

\vspace{5mm}
\facilities{\jwst (NIRSpec and MIRI), NOT (ALFOSC), Keck:II (NIRES), 
            IRTF (Spex), UH88 (SNIFS).}


\software{\jwst Science Calibration Pipeline (\citealp[version 1.18.0;][]{Bushouse2025}),
          \texttt{spextractor} \citep{Burrow2020},
          Astropy \citep{astropy:2013, astropy:2018, astropy:2022},
          NumPy \citep{numpy2020}, SciPy \citep{SciPy2020}, 
          Matplotlib \citep{matplotlib},
          \texttt{dust-extinction} \citep{Gordon2023b,Gordon2023a}.}

\bibliographystyle{aasjournal}
\bibliography{ms}

\begin{thebibliography}{}
\expandafter\ifx\csname natexlab\endcsname\relax\def\natexlab#1{#1}\fi
\providecommand{\url}[1]{\href{#1}{#1}}
\providecommand{\dodoi}[1]{doi:~\href{http://doi.org/#1}{\nolinkurl{#1}}}
\providecommand{\doeprint}[1]{\href{http://ascl.net/#1}{\nolinkurl{http://ascl.net/#1}}}
\providecommand{\doarXiv}[1]{\href{https://arxiv.org/abs/#1}{\nolinkurl{https://arxiv.org/abs/#1}}}

\bibitem[{{Abac} {et~al.}(2025){Abac}, {Abbott}, {Abouelfettouh}, {Acernese},
  {Ackley}, {Adhicary}, {Adhikari}, {Adhikari}, {Adkins}, {Agarwal}, {Agathos},
  {Aghaei Abchouyeh}, {Aguiar}, {Aguilar}, {Aiello}, {Ain}, {Akutsu},
  {Albanesi}, {Alfaidi}, {Al-Jodah}, {All{\'e}n{\'e}}, {Allocca},
  {Al-Shammari}, {Altin}, {Alvarez-Lopez}, {Amato}, {Amez-Droz}, {Amorosi},
  {Amra}, {Ananyeva}, {Anderson}, {Anderson}, {Andia}, {Ando}, {Andrade},
  {Andres}, {Andr{\'e}s-Carcasona}, {Andri{\'c}}, {Anglin}, {Ansoldi},
  {Antelis}, {Antier}, {Aoumi}, {Appavuravther}, {Appert}, {Apple}, {Arai},
  {Araya}, {Araya}, {Areeda}, {Argianas}, {Aritomi}, {Armato}, {Arnaud},
  {Arogeti}, {Aronson}, {Ashton}, {Aso}, {Assiduo}, {Assis de Souza Melo},
  {Aston}, {Astone}, {Attadio}, {Aubin}, {Aultoneal}, {Avallone}, {Babak},
  {Badaracco}, {Badger}, {Bae}, {Bagnasco}, {Bagui}, {Baier}, {Baiotti},
  {Bajpai}, {Baka}, {Ball}, {Ballardin}, {Ballmer}, {Banagiri}, {Banerjee},
  {Bankar}, {Baral}, {Barayoga}, {Barish}, {Barker}, {Barneo}, {Barone},
  {Barr}, {Barsotti}, {Barsuglia}, {Barta}, {Bartoletti}, {Barton}, {Bartos},
  {Basak}, {Basalaev}, {Bassiri}, {Basti}, {Bates}, {Bawaj}, {Baxi}, {Bayley},
  {Baylor}, {Baynard}, {Bazzan}, {Bedakihale}, {Beirnaert}, {Bejger},
  {Belardinelli}, {Bell}, {Benedetto}, {Benoit}, {Bentley}, {Ben Yaala},
  {Bera}, {Berbel}, {Bergamin}, {Berger}, {Bernuzzi}, {Beroiz}, {Bersanetti},
  {Bertolini}, {Betzwieser}, {Beveridge}, {Bevins}, {Bhandare}, {Bhardwaj},
  {Bhatt}, {Bhattacharjee}, {Bhaumik}, {Bhowmick}, {Bianchi}, {Bilenko},
  {Billingsley}, {Binetti}, {Bini}, {Birnholtz}, {Biscoveanu}, {Bisht},
  {Bitossi}, {Bizouard}, {Blackburn}, {Blagg}, {Blair}, {Blair}, {Bobba},
  {Bode}, {Boileau}, {Boldrini}, {Bolingbroke}, {Bolliand}, {Bonavena},
  {Bondarescu}, {Bondu}, {Bonilla}, {Bonilla}, {Bonino}, {Bonnand}, {Booker},
  {Borchers}, {Boschi}, {Bose}, {Bossilkov}, {Boudart}, {Boudon}, {Bozzi},
  {Bradaschia}, {Brady}, {Braglia}, {Branch}, {Branchesi}, {Brandt}, {Braun},
  {Breschi}, {Briant}, {Brillet}, {Brinkmann}, {Brockill}, {Brockmueller},
  {Brooks}, {Brown}, {Brown}, {Brozzetti}, {Brunett}, {Bruno}, {Bruntz},
  {Bryant}, {Bucci}, {Buchanan}, {Bulashenko}, {Bulik}, {Bulten}, {Buonanno},
  {Burtnyk}, {Buscicchio}, {Buskulic}, {Buy}, {Byer}, \& {Cabourn
  Davies}}]{LIGO2024}
{Abac}, A.~G., {Abbott}, R., {Abouelfettouh}, I., {et~al.} 2025, \apj, 985,
  183, \dodoi{10.3847/1538-4357/adc681}

\bibitem[{{Aitken} {et~al.}(1988{\natexlab{a}}){Aitken}, {Smith}, {James},
  {Roche}, {Hyland}, \& {McGregor}}]{Aitken1988b}
{Aitken}, D.~K., {Smith}, C.~H., {James}, S.~D., {et~al.} 1988{\natexlab{a}},
  \mnras, 231, 7P, \dodoi{10.1093/mnras/231.1.7P}

\bibitem[{{Aitken} {et~al.}(1988{\natexlab{b}}){Aitken}, {Smith}, {James},
  {Roche}, {Hyland}, \& {McGregor}}]{Aitken1988a}
---. 1988{\natexlab{b}}, \mnras, 235, 19P, \dodoi{10.1093/mnras/235.1.19P}

\bibitem[{{Anderson} {et~al.}(2014){Anderson}, {Dessart}, {Gutierrez}, {Hamuy},
  {Morrell}, {Phillips}, {Folatelli}, {Stritzinger}, {Freedman},
  {Gonz{\'a}lez-Gait{\'a}n}, {McCarthy}, {Suntzeff}, \&
  {Thomas-Osip}}]{Anderson2014}
{Anderson}, J.~P., {Dessart}, L., {Gutierrez}, C.~P., {et~al.} 2014, \mnras,
  441, 671, \dodoi{10.1093/mnras/stu610}

\bibitem[{{Ashall} {et~al.}(2023{\natexlab{a}}){Ashall}, {Baron}, {DerKacy},
  {Hoeflich}, {Shahbandeh}, {Baade}, {Brown}, {Burns}, {Engesser}, {Fox},
  {Galbany}, {Guolo}, {Hsiao}, {Kumar}, {Lu}, {Mazzali}, {Medler}, {Mera
  Evans}, {Morrell}, {Phillips}, {Rest}, {Stritzinger}, {Strolger}, {Suntzeff},
  {Temim}, {Tinyanont}, {Tucker}, {Wang}, \& {de
  Jaeger}}]{Ashall2023_cycle1_23ixf}
{Ashall}, C., {Baron}, E., {DerKacy}, J.~M., {et~al.} 2023{\natexlab{a}}, {Dust
  Our Luck? Measuring Molecule and Dust Formation in M101's Hydrogen-rich SN
  2023ixf}, JWST Proposal. Cycle 1, ID. \#4522

\bibitem[{{Ashall} {et~al.}(2023{\natexlab{b}}){Ashall}, {Baron}, {DerKacy},
  {Hoeflich}, {Shahbandeh}, {Baade}, {Brown}, {Burns}, {Engesser}, {Fox},
  {Galbany}, {Guolo}, {Hsiao}, {Kumar}, {Lu}, {Mazzali}, {Medler}, {Mera
  Evans}, {Morrell}, {Phillips}, {Rest}, {Stritzinger}, {Strolger}, {Suntzeff},
  {Temim}, {Tinyanont}, {Tucker}, {Wang}, \& {de
  Jaeger}}]{Ashall2023_cycle2_23ixf}
---. 2023{\natexlab{b}}, {Dust Our Luck? Measuring Molecule and Dust Formation
  in M101's Hydrogen-rich SN 2023ixf}, JWST Proposal. Cycle 2, ID. \#4575

\bibitem[{{Ashall} {et~al.}(2024){Ashall}, {Hoeflich}, {Shahbandeh}, {Baade},
  {Baron}, {Brown}, {Burns}, {DerKacy}, {Engesser}, {Fox}, {Galbany}, {Hsiao},
  {Johansson}, {Krisciunas}, {Kumar}, {Lu}, {Matsuura}, {Mazzali}, {Medler},
  {Mera Evans}, {Phillips}, {Rest}, {Sarangi}, {Stritzinger}, {Strolger},
  {Suntzeff}, {Szalai}, {Temim}, {Tinyanont}, {Tucker}, {Van Dyk}, {Wang},
  {Wesson}, {Zsiros}, \& {de Jaeger}}]{Ashall2024_cycle3_ixf}
{Ashall}, C., {Hoeflich}, P.~A., {Shahbandeh}, M., {et~al.} 2024, {Building the
  Legacy of Supernova 2023ixf: How Does Molecule Formation Lead to Dust?}, JWST
  Proposal. Cycle 3, ID. \#5290

\bibitem[{{Astropy Collaboration} {et~al.}(2013){Astropy Collaboration},
  {Robitaille}, {Tollerud}, {Greenfield}, {Droettboom}, {Bray}, {Aldcroft},
  {Davis}, {Ginsburg}, {Price-Whelan}, {Kerzendorf}, {Conley}, {Crighton},
  {Barbary}, {Muna}, {Ferguson}, {Grollier}, {Parikh}, {Nair}, {Unther},
  {Deil}, {Woillez}, {Conseil}, {Kramer}, {Turner}, {Singer}, {Fox}, {Weaver},
  {Zabalza}, {Edwards}, {Azalee Bostroem}, {Burke}, {Casey}, {Crawford},
  {Dencheva}, {Ely}, {Jenness}, {Labrie}, {Lim}, {Pierfederici}, {Pontzen},
  {Ptak}, {Refsdal}, {Servillat}, \& {Streicher}}]{astropy:2013}
{Astropy Collaboration}, {Robitaille}, T.~P., {Tollerud}, E.~J., {et~al.} 2013,
  \aap, 558, A33, \dodoi{10.1051/0004-6361/201322068}

\bibitem[{{Astropy Collaboration} {et~al.}(2018){Astropy Collaboration},
  {Price-Whelan}, {Sip{\H{o}}cz}, {G{\"u}nther}, {Lim}, {Crawford}, {Conseil},
  {Shupe}, {Craig}, {Dencheva}, {Ginsburg}, {Vand erPlas}, {Bradley},
  {P{\'e}rez-Su{\'a}rez}, {de Val-Borro}, {Aldcroft}, {Cruz}, {Robitaille},
  {Tollerud}, {Ardelean}, {Babej}, {Bach}, {Bachetti}, {Bakanov}, {Bamford},
  {Barentsen}, {Barmby}, {Baumbach}, {Berry}, {Biscani}, {Boquien}, {Bostroem},
  {Bouma}, {Brammer}, {Bray}, {Breytenbach}, {Buddelmeijer}, {Burke},
  {Calderone}, {Cano Rodr{\'\i}guez}, {Cara}, {Cardoso}, {Cheedella}, {Copin},
  {Corrales}, {Crichton}, {D'Avella}, {Deil}, {Depagne}, {Dietrich}, {Donath},
  {Droettboom}, {Earl}, {Erben}, {Fabbro}, {Ferreira}, {Finethy}, {Fox},
  {Garrison}, {Gibbons}, {Goldstein}, {Gommers}, {Greco}, {Greenfield},
  {Groener}, {Grollier}, {Hagen}, {Hirst}, {Homeier}, {Horton}, {Hosseinzadeh},
  {Hu}, {Hunkeler}, {Ivezi{\'c}}, {Jain}, {Jenness}, {Kanarek}, {Kendrew},
  {Kern}, {Kerzendorf}, {Khvalko}, {King}, {Kirkby}, {Kulkarni}, {Kumar},
  {Lee}, {Lenz}, {Littlefair}, {Ma}, {Macleod}, {Mastropietro}, {McCully},
  {Montagnac}, {Morris}, {Mueller}, {Mumford}, {Muna}, {Murphy}, {Nelson},
  {Nguyen}, {Ninan}, {N{\"o}the}, {Ogaz}, {Oh}, {Parejko}, {Parley}, {Pascual},
  {Patil}, {Patil}, {Plunkett}, {Prochaska}, {Rastogi}, {Reddy Janga},
  {Sabater}, {Sakurikar}, {Seifert}, {Sherbert}, {Sherwood-Taylor}, {Shih},
  {Sick}, {Silbiger}, {Singanamalla}, {Singer}, {Sladen}, {Sooley},
  {Sornarajah}, {Streicher}, {Teuben}, {Thomas}, {Tremblay}, {Turner},
  {Terr{\'o}n}, {van Kerkwijk}, {de la Vega}, {Watkins}, {Weaver}, {Whitmore},
  {Woillez}, {Zabalza}, \& {Astropy Contributors}}]{astropy:2018}
{Astropy Collaboration}, {Price-Whelan}, A.~M., {Sip{\H{o}}cz}, B.~M., {et~al.}
  2018, \aj, 156, 123, \dodoi{10.3847/1538-3881/aabc4f}

\bibitem[{{Astropy Collaboration} {et~al.}(2022){Astropy Collaboration},
  {Price-Whelan}, {Lim}, {Earl}, {Starkman}, {Bradley}, {Shupe}, {Patil},
  {Corrales}, {Brasseur}, {N{"o}the}, {Donath}, {Tollerud}, {Morris},
  {Ginsburg}, {Vaher}, {Weaver}, {Tocknell}, {Jamieson}, {van Kerkwijk},
  {Robitaille}, {Merry}, {Bachetti}, {G{"u}nther}, {Aldcroft},
  {Alvarado-Montes}, {Archibald}, {B{'o}di}, {Bapat}, {Barentsen}, {Baz{'a}n},
  {Biswas}, {Boquien}, {Burke}, {Cara}, {Cara}, {Conroy}, {Conseil}, {Craig},
  {Cross}, {Cruz}, {D'Eugenio}, {Dencheva}, {Devillepoix}, {Dietrich},
  {Eigenbrot}, {Erben}, {Ferreira}, {Foreman-Mackey}, {Fox}, {Freij}, {Garg},
  {Geda}, {Glattly}, {Gondhalekar}, {Gordon}, {Grant}, {Greenfield}, {Groener},
  {Guest}, {Gurovich}, {Handberg}, {Hart}, {Hatfield-Dodds}, {Homeier},
  {Hosseinzadeh}, {Jenness}, {Jones}, {Joseph}, {Kalmbach}, {Karamehmetoglu},
  {Ka{l}uszy{'n}ski}, {Kelley}, {Kern}, {Kerzendorf}, {Koch}, {Kulumani},
  {Lee}, {Ly}, {Ma}, {MacBride}, {Maljaars}, {Muna}, {Murphy}, {Norman},
  {O'Steen}, {Oman}, {Pacifici}, {Pascual}, {Pascual-Granado}, {Patil},
  {Perren}, {Pickering}, {Rastogi}, {Roulston}, {Ryan}, {Rykoff}, {Sabater},
  {Sakurikar}, {Salgado}, {Sanghi}, {Saunders}, {Savchenko}, {Schwardt},
  {Seifert-Eckert}, {Shih}, {Jain}, {Shukla}, {Sick}, {Simpson},
  {Singanamalla}, {Singer}, {Singhal}, {Sinha}, {Sip{H{o}}cz}, {Spitler},
  {Stansby}, {Streicher}, {{ {S}}umak}, {Swinbank}, {Taranu}, {Tewary},
  {Tremblay}, {Val-Borro}, {Van Kooten}, {Vasovi{'c}}, {Verma}, {de Miranda
  Cardoso}, {Williams}, {Wilson}, {Winkel}, {Wood-Vasey}, {Xue}, {Yoachim},
  {Zhang}, {Zonca}, \& {Astropy Project Contributors}}]{astropy:2022}
{Astropy Collaboration}, {Price-Whelan}, A.~M., {Lim}, P.~L., {et~al.} 2022,
  apj, 935, 167, \dodoi{10.3847/1538-4357/ac7c74}

\bibitem[{{Barlow}(1978{\natexlab{a}})}]{Barlow1978a}
{Barlow}, M.~J. 1978{\natexlab{a}}, \mnras, 183, 367,
  \dodoi{10.1093/mnras/183.3.367}

\bibitem[{{Barlow}(1978{\natexlab{b}})}]{Barlow1978b}
---. 1978{\natexlab{b}}, \mnras, 183, 397, \dodoi{10.1093/mnras/183.3.397}

\bibitem[{{Barlow}(1978{\natexlab{c}})}]{Barlow1978c}
---. 1978{\natexlab{c}}, \mnras, 183, 417, \dodoi{10.1093/mnras/183.3.417}

\bibitem[{{Baron} {et~al.}(2003){Baron}, {Nugent}, {Branch}, {Hauschildt},
  {Turatto}, \& {Cappellaro}}]{Baron2003}
{Baron}, E., {Nugent}, P.~E., {Branch}, D., {et~al.} 2003, \apj, 586, 1199,
  \dodoi{10.1086/367888}

\bibitem[{{Baron} {et~al.}(2025){Baron}, {Ashall}, {DerKacy}, {Hoeflich},
  {Medler}, {Shahbandeh}, {Fereidouni}, {Pfeffer}, {Mera}, {Hoogendam},
  {Shiber}, {Auchettl}, {Brown}, {Burns}, {Burrow}, {Coulter}, {Engesser},
  {Folatelli}, {Fox}, {Galbany}, {Guolo}, {Hinkle}, {Huber}, {Hsiao}, {de
  Jaeger}, {Jones}, {Kumar}, {Lu}, {Mazzali}, {Morrell}, {Phillips}, {Rest},
  {Suntzeff}, {Shappee}, {Shi}, {Stritzinger}, {Strolger}, {Temim},
  {Tinyanont}, {Tucker}, {Wang}, {Wang}, \& {Yang}}]{Baron2025}
{Baron}, E., {Ashall}, C., {DerKacy}, J.~M., {et~al.} 2025, arXiv e-prints,
  arXiv:2507.18753.
\newblock \doarXiv{2507.18753}

\bibitem[{{Berger} {et~al.}(2023){Berger}, {Keating}, {Margutti}, {Maeda},
  {Alexander}, {Cendes}, {Eftekhari}, {Gurwell}, {Hiramatsu}, {Ho}, {Laskar},
  {Rao}, \& {Williams}}]{Berger2023}
{Berger}, E., {Keating}, G.~K., {Margutti}, R., {et~al.} 2023, \apjl, 951, L31,
  \dodoi{10.3847/2041-8213/ace0c4}

\bibitem[{{Bersten} {et~al.}(2024){Bersten}, {Orellana}, {Folatelli},
  {Martinez}, {Piccirilli}, {Regna}, {Rom{\'a}n Aguilar}, \&
  {Ertini}}]{Bersten2023}
{Bersten}, M.~C., {Orellana}, M., {Folatelli}, G., {et~al.} 2024, \aap, 681,
  L18, \dodoi{10.1051/0004-6361/202348183}

\bibitem[{{Bertoldi} {et~al.}(2003){Bertoldi}, {Cox}, {Neri}, {Carilli},
  {Walter}, {Omont}, {Beelen}, {Henkel}, {Fan}, {Strauss}, \&
  {Menten}}]{Bertoldi2003}
{Bertoldi}, F., {Cox}, P., {Neri}, R., {et~al.} 2003, \aap, 409, L47,
  \dodoi{10.1051/0004-6361:20031345}

\bibitem[{{Bevan} {et~al.}(2019){Bevan}, {Wesson}, {Barlow}, {De Looze},
  {Andrews}, {Clayton}, {Krafton}, {Matsuura}, \& {Milisavljevic}}]{Bevan2019}
{Bevan}, A., {Wesson}, R., {Barlow}, M.~J., {et~al.} 2019, \mnras, 485, 5192,
  \dodoi{10.1093/mnras/stz679}

\bibitem[{{B{\"o}ker} {et~al.}(2023){B{\"o}ker}, {Beck}, {Birkmann},
  {Giardino}, {Keyes}, {Kumari}, {Muzerolle}, {Rawle}, {Zeidler}, {Abul-Huda},
  {Alves de Oliveira}, {Arribas}, {Bechtold}, {Bhatawdekar}, {Bonaventura},
  {Bunker}, {Cameron}, {Carniani}, {Charlot}, {Curti}, {Espinoza}, {Ferruit},
  {Franx}, {Jakobsen}, {Karakla}, {L{\'o}pez-Caniego}, {L{\"u}tzgendorf},
  {Maiolino}, {Manjavacas}, {Marston}, {Moseley}, {Ogle}, {Perna},
  {Pe{\~n}a-Guerrero}, {Pirzkal}, {Plesha}, {Proffitt}, {Rauscher}, {Rix},
  {Rodr{\'\i}guez del Pino}, {Rustamkulov}, {Sabbi}, {Sing}, {Sirianni}, {te
  Plate}, {{\'U}beda}, {Wahlgren}, {Wislowski}, {Wu}, \& {Willott}}]{Boker2023}
{B{\"o}ker}, T., {Beck}, T.~L., {Birkmann}, S.~M., {et~al.} 2023, \pasp, 135,
  038001, \dodoi{10.1088/1538-3873/acb846}

\bibitem[{{Bostroem} {et~al.}(2023){Bostroem}, {Pearson}, {Shrestha}, {Sand},
  {Valenti}, {Jha}, {Andrews}, {Smith}, {Terreran}, {Green}, {Dong},
  {Lundquist}, {Haislip}, {Hoang}, {Hosseinzadeh}, {Janzen}, {Jencson},
  {Kouprianov}, {Paraskeva}, {Meza Retamal}, {Reichart}, {Arcavi}, {Bonanos},
  {Coughlin}, {Dobson}, {Farah}, {Galbany}, {Guti{\'e}rrez}, {Hawley}, {Hebb},
  {Hiramatsu}, {Howell}, {Iijima}, {Ilyin}, {Jhass}, {McCully}, {Moran},
  {Morris}, {Mura}, {M{\"u}ller-Bravo}, {Munday}, {Newsome}, {Pabst}, {Ochner},
  {Gonzalez}, {Pastorello}, {Pellegrino}, {Piscarreta}, {Ravi}, {Reguitti},
  {Salo}, {Vink{\'o}}, {de Vos}, {Wheeler}, {Williams}, \&
  {Wyatt}}]{Bostroem2023}
{Bostroem}, K.~A., {Pearson}, J., {Shrestha}, M., {et~al.} 2023, \apjl, 956,
  L5, \dodoi{10.3847/2041-8213/acf9a4}

\bibitem[{{Bostroem} {et~al.}(2024){Bostroem}, {Sand}, {Dessart}, {Smith},
  {Jha}, {Valenti}, {Andrews}, {Dong}, {Filippenko}, {Gomez}, {Hiramatsu},
  {Hoang}, {Hosseinzadeh}, {Howell}, {Jencson}, {Lundquist}, {McCully},
  {Mehta}, {Meza-Retamal}, {Pearson}, {Ravi}, {Shrestha}, \&
  {Wyatt}}]{Bostroem2024}
{Bostroem}, K.~A., {Sand}, D.~J., {Dessart}, L., {et~al.} 2024, \apjl, 973,
  L47, \dodoi{10.3847/2041-8213/ad7855}

\bibitem[{{Brooker} {et~al.}(2022){Brooker}, {Stangl}, {Mauney}, \&
  {Fryer}}]{Brooker2022}
{Brooker}, E.~S., {Stangl}, S.~M., {Mauney}, C.~M., \& {Fryer}, C.~L. 2022,
  \apj, 931, 85, \dodoi{10.3847/1538-4357/ac57c3}

\bibitem[{{Burrow} {et~al.}(2020){Burrow}, {Baron}, {Ashall}, {Burns},
  {Morrell}, {Stritzinger}, {Brown}, {Folatelli}, {Freedman}, {Galbany},
  {Hoeflich}, {Hsiao}, {Krisciunas}, {Phillips}, {Piro}, {Suntzeff}, \&
  {Uddin}}]{Burrow2020}
{Burrow}, A., {Baron}, E., {Ashall}, C., {et~al.} 2020, \apj, 901, 154,
  \dodoi{10.3847/1538-4357/abafa2}

\bibitem[{{Bushouse} {et~al.}(2025){Bushouse}, {Eisenhamer}, {Dencheva},
  {Davies}, {Greenfield}, {Morrison}, {Hodge}, {Simon}, {Grumm}, {Droettboom},
  {Slavich}, {Sosey}, {Pauly}, {Miller}, {Jedrzejewski}, {Hack}, {Davis},
  {Crawford}, {Law}, {Gordon}, {Regan}, {Cara}, {MacDonald}, {Bradley},
  {Shanahan}, {Jamieson}, {Teodoro}, {Williams}, {Pena-Guerrero}, {Graham},
  {Molter}, {Brandt}, {Hayes}, {Cooper}, {Clarke}, \&
  {Filippazzo}}]{Bushouse2025}
{Bushouse}, H., {Eisenhamer}, J., {Dencheva}, N., {et~al.} 2025, {JWST
  Calibration Pipeline}, 1.18.0,  Zenodo, \dodoi{10.5281/zenodo.15178003}

\bibitem[{{Catchpole} {et~al.}(1988){Catchpole}, {Whitelock}, {Feast},
  {Menzies}, {Glass}, {Marang}, {Laing}, {Spencer Jones}, {Roberts}, {Balona},
  {Carter}, {Laney}, {Evans}, {Sekiguchi}, {Hutchinson}, {Maddison},
  {Albinson}, {Evans}, {Allen}, {Winkler}, {Fairall}, {Corbally}, {Davies}, \&
  {Parker}}]{Catchpole1988}
{Catchpole}, R.~M., {Whitelock}, P.~A., {Feast}, M.~W., {et~al.} 1988, \mnras,
  231, 75P, \dodoi{10.1093/mnras/231.1.75P}

\bibitem[{{Cernuschi} {et~al.}(1967){Cernuschi}, {Marsicano}, \&
  {Codina}}]{Cernuschi1967}
{Cernuschi}, F., {Marsicano}, F., \& {Codina}, S. 1967, Annales
  d'Astrophysique, 30, 1039

\bibitem[{{Chandra} {et~al.}(2024){Chandra}, {Chevalier}, {Maeda}, {Ray}, \&
  {Nayana}}]{Chandra2023}
{Chandra}, P., {Chevalier}, R.~A., {Maeda}, K., {Ray}, A.~K., \& {Nayana},
  A.~J. 2024, \apjl, 963, L4, \dodoi{10.3847/2041-8213/ad275d}

\bibitem[{{Chugai} {et~al.}(2007){Chugai}, {Chevalier}, \&
  {Utrobin}}]{Chugai2007}
{Chugai}, N.~N., {Chevalier}, R.~A., \& {Utrobin}, V.~P. 2007, \apj, 662, 1136,
  \dodoi{10.1086/518160}

\bibitem[{{Davis} {et~al.}(2019){Davis}, {Hsiao}, {Ashall}, {Hoeflich},
  {Phillips}, {Marion}, {Kirshner}, {Morrell}, {Sand}, {Burns}, {Contreras},
  {Stritzinger}, {Anderson}, {Baron}, {Diamond}, {Guti{\'e}rrez}, {Hamuy},
  {Holmbo}, {Kasliwal}, {Krisciunas}, {Kumar}, {Lu}, {Pessi}, {Piro}, {Prieto},
  {Shahbandeh}, \& {Suntzeff}}]{Davis2019}
{Davis}, S., {Hsiao}, E.~Y., {Ashall}, C., {et~al.} 2019, \apj, 887, 4,
  \dodoi{10.3847/1538-4357/ab4c40}

\bibitem[{{de Jaeger} {et~al.}(2020){de Jaeger}, {Galbany},
  {Gonz{\'a}lez-Gait{\'a}n}, {Kessler}, {Filippenko}, {F{\"o}rster}, {Hamuy},
  {Brown}, {Davis}, {Guti{\'e}rrez}, {Inserra}, {Lewis}, {M{\"o}ller},
  {Scolnic}, {Smith}, {Brout}, {Carollo}, {Foley}, {Glazebrook}, {Hinton},
  {Macaulay}, {Nichol}, {Sako}, {Sommer}, {Tucker}, {Abbott}, {Aguena},
  {Allam}, {Annis}, {Avila}, {Bertin}, {Bhargava}, {Brooks}, {Burke}, {Carnero
  Rosell}, {Carrasco Kind}, {Carretero}, {Costanzi}, {Crocce}, {da Costa}, {De
  Vicente}, {Desai}, {Diehl}, {Doel}, {Drlica-Wagner}, {Eifler}, {Estrada},
  {Everett}, {Flaugher}, {Fosalba}, {Frieman}, {Garc{\'\i}a-Bellido},
  {Gaztanaga}, {Gruen}, {Gruendl}, {Gschwend}, {Gutierrez}, {Hartley},
  {Hollowood}, {Honscheid}, {James}, {Kuehn}, {Kuropatkin}, {Li}, {Lima},
  {Maia}, {Menanteau}, {Miquel}, {Palmese}, {Paz-Chinch{\'o}n}, {Plazas},
  {Romer}, {Roodman}, {Sanchez}, {Scarpine}, {Schubnell}, {Serrano},
  {Sevilla-Noarbe}, {Soares-Santos}, {Suchyta}, {Swanson}, {Tarle}, {Thomas},
  {Tucker}, {Varga}, {Walker}, {Weller}, {Wilkinson}, \& {DES
  Collaboration}}]{deJaeger2020}
{de Jaeger}, T., {Galbany}, L., {Gonz{\'a}lez-Gait{\'a}n}, S., {et~al.} 2020,
  \mnras, 495, 4860, \dodoi{10.1093/mnras/staa1402}

\bibitem[{{de Vaucouleurs} {et~al.}(1991){de Vaucouleurs}, {de Vaucouleurs},
  {Corwin}, {Buta}, {Paturel}, \& {Fouque}}]{deVaucouleurs1991}
{de Vaucouleurs}, G., {de Vaucouleurs}, A., {Corwin}, Herold~G., J., {et~al.}
  1991, {Third Reference Catalogue of Bright Galaxies} (Berlin: Springer)

\bibitem[{{Dell'Agli} {et~al.}(2015){Dell'Agli}, {Ventura}, {Schneider}, {Di
  Criscienzo}, {Garc{\'\i}a-Hern{\'a}ndez}, {Rossi}, \&
  {Brocato}}]{DellAgli2015}
{Dell'Agli}, F., {Ventura}, P., {Schneider}, R., {et~al.} 2015, \mnras, 447,
  2992, \dodoi{10.1093/mnras/stu2559}

\bibitem[{{Dessart}(2025)}]{Dessart2025}
{Dessart}, L. 2025, arXiv e-prints, arXiv:2505.19818,
  \dodoi{10.48550/arXiv.2505.19818}

\bibitem[{{Dessart} \& {Hillier}(2005)}]{Dessart2005}
{Dessart}, L., \& {Hillier}, D.~J. 2005, \aap, 437, 667,
  \dodoi{10.1051/0004-6361:20042525}

\bibitem[{{Dessart} \& {Hillier}(2022)}]{Dessart2022}
---. 2022, \aap, 660, L9, \dodoi{10.1051/0004-6361/202243372}

\bibitem[{{Dessart} {et~al.}(2013){Dessart}, {Hillier}, {Waldman}, \&
  {Livne}}]{Dessart2013}
{Dessart}, L., {Hillier}, D.~J., {Waldman}, R., \& {Livne}, E. 2013, \mnras,
  433, 1745, \dodoi{10.1093/mnras/stt861}

\bibitem[{{Di Criscienzo} {et~al.}(2013){Di Criscienzo}, {Dell'Agli},
  {Ventura}, {Schneider}, {Valiante}, {La Franca}, {Rossi}, {Gallerani}, \&
  {Maiolino}}]{DiCriscienzo2013}
{Di Criscienzo}, M., {Dell'Agli}, F., {Ventura}, P., {et~al.} 2013, \mnras,
  433, 313, \dodoi{10.1093/mnras/stt732}

\bibitem[{{Dong} {et~al.}(2023){Dong}, {Sand}, {Valenti}, {Bostroem},
  {Andrews}, {Hosseinzadeh}, {Hoang}, {Janzen}, {Jencson}, {Lundquist}, {Meza
  Retamal}, {Pearson}, {Shrestha}, {Haislip}, {Kouprianov}, \&
  {Reichart}}]{Dong2023}
{Dong}, Y., {Sand}, D.~J., {Valenti}, S., {et~al.} 2023, \apj, 957, 28,
  \dodoi{10.3847/1538-4357/acef18}

\bibitem[{{Duschinger} {et~al.}(1995){Duschinger}, {Puls}, {Branch},
  {Hoeflich}, \& {Gabler}}]{Duschinger1995}
{Duschinger}, M., {Puls}, J., {Branch}, D., {Hoeflich}, P., \& {Gabler}, A.
  1995, \aap, 297, 802

\bibitem[{{Dwek}(1998)}]{Dwek1998}
{Dwek}, E. 1998, \apj, 501, 643, \dodoi{10.1086/305829}

\bibitem[{{Dwek} {et~al.}(2007){Dwek}, {Galliano}, \& {Jones}}]{Dwek2007}
{Dwek}, E., {Galliano}, F., \& {Jones}, A.~P. 2007, \apj, 662, 927,
  \dodoi{10.1086/518430}

\bibitem[{{Dwek} {et~al.}(2019){Dwek}, {Sarangi}, \& {Arendt}}]{Dwek2019}
{Dwek}, E., {Sarangi}, A., \& {Arendt}, R.~G. 2019, \apjl, 871, L33,
  \dodoi{10.3847/2041-8213/aaf9a8}

\bibitem[{{Ferrarotti} \& {Gail}(2006)}]{Ferrarotti2006}
{Ferrarotti}, A.~S., \& {Gail}, H.~P. 2006, \aap, 447, 553,
  \dodoi{10.1051/0004-6361:20041198}

\bibitem[{{Flinner} {et~al.}(2023){Flinner}, {Tucker}, {Beacom}, \&
  {Shappee}}]{Flinner2023}
{Flinner}, N., {Tucker}, M.~A., {Beacom}, J.~F., \& {Shappee}, B.~J. 2023,
  Research Notes of the American Astronomical Society, 7, 174,
  \dodoi{10.3847/2515-5172/acefc4}

\bibitem[{{Fox} {et~al.}(2010){Fox}, {Chevalier}, {Dwek}, {Skrutskie},
  {Sugerman}, \& {Leisenring}}]{Fox2010}
{Fox}, O.~D., {Chevalier}, R.~A., {Dwek}, E., {et~al.} 2010, \apj, 725, 1768,
  \dodoi{10.1088/0004-637X/725/2/1768}

\bibitem[{{Fox} {et~al.}(2011){Fox}, {Chevalier}, {Skrutskie}, {Soderberg},
  {Filippenko}, {Ganeshalingam}, {Silverman}, {Smith}, \& {Steele}}]{Fox2011}
{Fox}, O.~D., {Chevalier}, R.~A., {Skrutskie}, M.~F., {et~al.} 2011, \apj, 741,
  7, \dodoi{10.1088/0004-637X/741/1/7}

\bibitem[{{Gall} {et~al.}(2011){Gall}, {Hjorth}, \& {Andersen}}]{Gall2011}
{Gall}, C., {Hjorth}, J., \& {Andersen}, A.~C. 2011, \aapr, 19, 43,
  \dodoi{10.1007/s00159-011-0043-7}

\bibitem[{{Gall} {et~al.}(2014){Gall}, {Hjorth}, {Watson}, {Dwek}, {Maund},
  {Fox}, {Leloudas}, {Malesani}, \& {Day-Jones}}]{Gall2014}
{Gall}, C., {Hjorth}, J., {Watson}, D., {et~al.} 2014, \nat, 511, 326,
  \dodoi{10.1038/nature13558}

\bibitem[{{Gerardy} {et~al.}(2000){Gerardy}, {Fesen}, {H{\"o}flich}, \&
  {Wheeler}}]{Gerardy2000}
{Gerardy}, C.~L., {Fesen}, R.~A., {H{\"o}flich}, P., \& {Wheeler}, J.~C. 2000,
  \aj, 119, 2968, \dodoi{10.1086/301390}

\bibitem[{{Gordon}(2023)}]{Gordon2023b}
{Gordon}, K. 2023, {karllark/dust\_extinction: OneRelationForAllWaves}, v1.2,
  Zenodo,  Zenodo, \dodoi{10.5281/zenodo.7799360}

\bibitem[{{Gordon} {et~al.}(2023){Gordon}, {Clayton}, {Decleir}, {Fitzpatrick},
  {Massa}, {Misselt}, \& {Tollerud}}]{Gordon2023a}
{Gordon}, K.~D., {Clayton}, G.~C., {Decleir}, M., {et~al.} 2023, \apj, 950, 86,
  \dodoi{10.3847/1538-4357/accb59}

\bibitem[{{Grefenstette} {et~al.}(2023){Grefenstette}, {Brightman}, {Earnshaw},
  {Harrison}, \& {Margutti}}]{Grefenstette2023}
{Grefenstette}, B.~W., {Brightman}, M., {Earnshaw}, H.~P., {Harrison}, F.~A.,
  \& {Margutti}, R. 2023, \apjl, 952, L3, \dodoi{10.3847/2041-8213/acdf4e}

\bibitem[{{Guetta} {et~al.}(2023){Guetta}, {Langella}, {Gagliardini}, \&
  {Valle}}]{Guetta2023}
{Guetta}, D., {Langella}, A., {Gagliardini}, S., \& {Valle}, M.~D. 2023, \apjl,
  955, L9, \dodoi{10.3847/2041-8213/acf573}

\bibitem[{{Guti{\'e}rrez} {et~al.}(2014){Guti{\'e}rrez}, {Anderson}, {Hamuy},
  {Gonz{\'a}lez-Gait{\'a}n}, {Folatelli}, {Morrell}, {Stritzinger}, {Phillips},
  {McCarthy}, {Suntzeff}, \& {Thomas-Osip}}]{Gutierrez2014}
{Guti{\'e}rrez}, C.~P., {Anderson}, J.~P., {Hamuy}, M., {et~al.} 2014, \apjl,
  786, L15, \dodoi{10.1088/2041-8205/786/2/L15}

\bibitem[{{Guti{\'e}rrez} {et~al.}(2017){Guti{\'e}rrez}, {Anderson}, {Hamuy},
  {Morrell}, {Gonz{\'a}lez-Gaitan}, {Stritzinger}, {Phillips}, {Galbany},
  {Folatelli}, {Dessart}, {Contreras}, {Della Valle}, {Freedman}, {Hsiao},
  {Krisciunas}, {Madore}, {Maza}, {Suntzeff}, {Prieto}, {Gonz{\'a}lez},
  {Cappellaro}, {Navarrete}, {Pizzella}, {Ruiz}, {Smith}, \&
  {Turatto}}]{Gutierrez2017}
---. 2017, \apj, 850, 89, \dodoi{10.3847/1538-4357/aa8f52}

\bibitem[{{Guti{\'e}rrez} {et~al.}(2020){Guti{\'e}rrez}, {Pastorello},
  {Jerkstrand}, {Galbany}, {Sullivan}, {Anderson}, {Taubenberger},
  {Kuncarayakti}, {Gonz{\'a}lez-Gait{\'a}n}, {Wiseman}, {Inserra}, {Fraser},
  {Maguire}, {Smartt}, {M{\"u}ller-Bravo}, {Arcavi}, {Benetti}, {Bersier},
  {Bose}, {Bostroem}, {Burke}, {Chen}, {Chen}, {Della Valle}, {Dong},
  {Gal-Yam}, {Gromadzki}, {Hiramatsu}, {Holoien}, {Hosseinzadeh}, {Howell},
  {Kankare}, {Kochanek}, {McCully}, {Nicholl}, {Pignata}, {Prieto}, {Shappee},
  {Taggart}, {Tomasella}, {Valenti}, \& {Young}}]{Gutierrez2020}
{Guti{\'e}rrez}, C.~P., {Pastorello}, A., {Jerkstrand}, A., {et~al.} 2020,
  \mnras, 499, 974, \dodoi{10.1093/mnras/staa2763}

\bibitem[{{Hamuy} \& {Pinto}(2002)}]{Hamuy2002}
{Hamuy}, M., \& {Pinto}, P.~A. 2002, \apjl, 566, L63, \dodoi{10.1086/339676}

\bibitem[{{Hanuschik} \& {Thimm}(1990)}]{Hanuschik1990}
{Hanuschik}, R.~W., \& {Thimm}, G.~J. 1990, \aap, 231, 77

\bibitem[{{Harkness} {et~al.}(1987){Harkness}, {Wheeler}, {Margon}, {Downes},
  {Kirshner}, {Uomoto}, {Barker}, {Cochran}, {Dinerstein}, {Garnett}, \&
  {Levreault}}]{Harkness1987}
{Harkness}, R.~P., {Wheeler}, J.~C., {Margon}, B., {et~al.} 1987, \apj, 317,
  355, \dodoi{10.1086/165283}

\bibitem[{{Harris} {et~al.}(2020){Harris}, {Millman}, {van der Walt},
  {Gommers}, {Virtanen}, {Cournapeau}, {Wieser}, {Taylor}, {Berg}, {Smith},
  {Kern}, {Picus}, {Hoyer}, {van Kerkwijk}, {Brett}, {Haldane}, {del R{\'\i}o},
  {Wiebe}, {Peterson}, {G{\'e}rard-Marchant}, {Sheppard}, {Reddy}, {Weckesser},
  {Abbasi}, {Gohlke}, \& {Oliphant}}]{numpy2020}
{Harris}, C.~R., {Millman}, K.~J., {van der Walt}, S.~J., {et~al.} 2020, \nat,
  585, 357, \dodoi{10.1038/s41586-020-2649-2}

\bibitem[{{Hauschildt} \& {Baron}(1995)}]{Hauschildt1995}
{Hauschildt}, P.~H., \& {Baron}, E. 1995, \jqsrt, 54, 987,
  \dodoi{10.1016/0022-4073(95)00118-5}

\bibitem[{{Hiramatsu} {et~al.}(2023){Hiramatsu}, {Tsuna}, {Berger}, {Itagaki},
  {Goldberg}, {Gomez}, {Kishalay}, {Hosseinzadeh}, {Bostroem}, {Brown},
  {Arcavi}, {Bieryla}, {Blanchard}, {Esquerdo}, {Farah}, {Howell}, {Matsumoto},
  {McCully}, {Newsome}, {Gonzalez}, {Pellegrino}, {Rhee}, {Terreran},
  {Vink{\'o}}, \& {Wheeler}}]{Hiramatsu2023}
{Hiramatsu}, D., {Tsuna}, D., {Berger}, E., {et~al.} 2023, \apjl, 955, L8,
  \dodoi{10.3847/2041-8213/acf299}

\bibitem[{{Hosseinzadeh} {et~al.}(2023){Hosseinzadeh}, {Farah}, {Shrestha},
  {Sand}, {Dong}, {Brown}, {Bostroem}, {Valenti}, {Jha}, {Andrews}, {Arcavi},
  {Haislip}, {Hiramatsu}, {Hoang}, {Howell}, {Janzen}, {Jencson}, {Kouprianov},
  {Lundquist}, {McCully}, {Meza Retamal}, {Modjaz}, {Newsome}, {Padilla
  Gonzalez}, {Pearson}, {Pellegrino}, {Ravi}, {Reichart}, {Smith}, {Terreran},
  \& {Vink{\'o}}}]{Hosseinzadeh2023}
{Hosseinzadeh}, G., {Farah}, J., {Shrestha}, M., {et~al.} 2023, \apjl, 953,
  L16, \dodoi{10.3847/2041-8213/ace4c4}

\bibitem[{{Hoyle} \& {Wickramasinghe}(1970)}]{Hoyle1970}
{Hoyle}, F., \& {Wickramasinghe}, N.~C. 1970, \nat, 226, 62,
  \dodoi{10.1038/226062a0}

\bibitem[{{Hu} {et~al.}(2025){Hu}, {Wang}, \& {Wang}}]{Hu2024}
{Hu}, M., {Wang}, L., \& {Wang}, X. 2025, \apj, 984, 44,
  \dodoi{10.3847/1538-4357/adc802}

\bibitem[{{Hunter}(2007)}]{matplotlib}
{Hunter}, J.~D. 2007, Computing in Science and Engineering, 9, 90,
  \dodoi{10.1109/MCSE.2007.55}

\bibitem[{{Inserra} {et~al.}(2013){Inserra}, {Pastorello}, {Turatto}, {Pumo},
  {Benetti}, {Cappellaro}, {Botticella}, {Bufano}, {Elias-Rosa}, {Harutyunyan},
  {Taubenberger}, {Valenti}, \& {Zampieri}}]{Inserra2013}
{Inserra}, C., {Pastorello}, A., {Turatto}, M., {et~al.} 2013, \aap, 555, A142,
  \dodoi{10.1051/0004-6361/201220496}

\bibitem[{{Itagaki}(2023)}]{Itagaki2023}
{Itagaki}, K. 2023, Transient Name Server Discovery Report, 2023-1158, 1

\bibitem[{{Iwata} {et~al.}(2025){Iwata}, {Akimoto}, {Matsuoka}, {Maeda},
  {Yonekura}, {Tominaga}, {Moriya}, {Fujisawa}, {Niinuma}, {Yoon}, {Lee},
  {Jung}, \& {Byun}}]{Iwata2024}
{Iwata}, Y., {Akimoto}, M., {Matsuoka}, T., {et~al.} 2025, \apj, 978, 138,
  \dodoi{10.3847/1538-4357/ad9a62}

\bibitem[{{Jacobson-Gal{\'a}n} {et~al.}(2023){Jacobson-Gal{\'a}n}, {Dessart},
  {Margutti}, {Chornock}, {Foley}, {Kilpatrick}, {Jones}, {Taggart}, {Angus},
  {Bhattacharjee}, {Braff}, {Brethauer}, {Burgasser}, {Cao}, {Carlile},
  {Chambers}, {Coulter}, {Dominguez-Ruiz}, {Dickinson}, {de Boer}, {Gagliano},
  {Gall}, {Gao}, {Gates}, {Gomez}, {Guolo}, {Halford}, {Hjorth}, {Huber},
  {Johnson}, {Karpoor}, {Laskar}, {LeBaron}, {Li}, {Lin}, {Loch}, {Lynam},
  {Magnier}, {Maloney}, {Matthews}, {McDonald}, {Miao}, {Milisavljevic}, {Pan},
  {Pradyumna}, {Ransome}, {Rees}, {Rest}, {Rojas-Bravo}, {Sandford},
  {Ascencio}, {Sanjaripour}, {Savino}, {Sears}, {Sharei}, {Smartt}, {Softich},
  {Theissen}, {Tinyanont}, {Tohfa}, {Villar}, {Wang}, {Wainscoat},
  {Westerling}, {Wiston}, {Wozniak}, {Yadavalli}, \&
  {Zenati}}]{Jacobson-Galan2023}
{Jacobson-Gal{\'a}n}, W.~V., {Dessart}, L., {Margutti}, R., {et~al.} 2023,
  \apjl, 954, L42, \dodoi{10.3847/2041-8213/acf2ec}

\bibitem[{{Jakobsen} {et~al.}(2022){Jakobsen}, {Ferruit}, {Alves de Oliveira},
  {Arribas}, {Bagnasco}, {Barho}, {Beck}, {Birkmann}, {B{\"o}ker}, {Bunker},
  {Charlot}, {de Jong}, {de Marchi}, {Ehrenwinkler}, {Falcolini}, {Fels},
  {Franx}, {Franz}, {Funke}, {Giardino}, {Gnata}, {Holota}, {Honnen}, {Jensen},
  {Jentsch}, {Johnson}, {Jollet}, {Karl}, {Kling}, {K{\"o}hler}, {Kolm},
  {Kumari}, {Lander}, {Lemke}, {L{\'o}pez-Caniego}, {L{\"u}tzgendorf},
  {Maiolino}, {Manjavacas}, {Marston}, {Maschmann}, {Maurer}, {Messerschmidt},
  {Moseley}, {Mosner}, {Mott}, {Muzerolle}, {Pirzkal}, {Pittet}, {Plitzke},
  {Posselt}, {Rapp}, {Rauscher}, {Rawle}, {Rix}, {R{\"o}del}, {Rumler},
  {Sabbi}, {Salvignol}, {Schmid}, {Sirianni}, {Smith}, {Strada}, {te Plate},
  {Valenti}, {Wettemann}, {Wiehe}, {Wiesmayer}, {Willott}, {Wright}, {Zeidler},
  \& {Zincke}}]{Jakobsen2022}
{Jakobsen}, P., {Ferruit}, P., {Alves de Oliveira}, C., {et~al.} 2022, \aap,
  661, A80, \dodoi{10.1051/0004-6361/202142663}

\bibitem[{{Jeffery} \& {Branch}(1990)}]{Jeffery1990}
{Jeffery}, D.~J., \& {Branch}, D. 1990, in Supernovae, Jerusalem Winter School
  for Theoretical Physics, ed. J.~C. {Wheeler}, T.~{Piran}, \& S.~{Weinberg},
  Vol.~6, 149

\bibitem[{{Jencson} {et~al.}(2023){Jencson}, {Pearson}, {Beasor}, {Lau},
  {Andrews}, {Bostroem}, {Dong}, {Engesser}, {Gomez}, {Guolo}, {Hoang},
  {Hosseinzadeh}, {Jha}, {Karambelkar}, {Kasliwal}, {Lundquist}, {Meza
  Retamal}, {Rest}, {Sand}, {Shahbandeh}, {Shrestha}, {Smith}, {Strader},
  {Valenti}, {Wang}, \& {Zenati}}]{Jencson2023}
{Jencson}, J.~E., {Pearson}, J., {Beasor}, E.~R., {et~al.} 2023, \apjl, 952,
  L30, \dodoi{10.3847/2041-8213/ace618}

\bibitem[{{Jones}(2004)}]{Jones2004}
{Jones}, A.~P. 2004, in Astronomical Society of the Pacific Conference Series,
  Vol. 309, Astrophysics of Dust, ed. A.~N. {Witt}, G.~C. {Clayton}, \& B.~T.
  {Draine}, 347

\bibitem[{{Jones} {et~al.}(1996){Jones}, {Tielens}, \&
  {Hollenbach}}]{Jones1996}
{Jones}, A.~P., {Tielens}, A.~G.~G.~M., \& {Hollenbach}, D.~J. 1996, \apj, 469,
  740, \dodoi{10.1086/177823}

\bibitem[{{Jones} {et~al.}(2023){Jones}, {Kavanagh}, {Barlow}, {Temim},
  {Fransson}, {Larsson}, {Blommaert}, {Meixner}, {Lau}, {Sargent}, {Bouchet},
  {Hjorth}, {Wright}, {Coulais}, {Fox}, {Gastaud}, {Glasse}, {Habel},
  {Hirschauer}, {Jaspers}, {Krause}, {Lenki{\'c}}, {Nayak}, {Rest}, {Tikkanen},
  {Wesson}, {Colina}, {van Dishoeck}, {G{\"u}del}, {Henning}, {Lagage},
  {{\"O}stlin}, {Ray}, \& {Vandenbussche}}]{Jones2023}
{Jones}, O.~C., {Kavanagh}, P.~J., {Barlow}, M.~J., {et~al.} 2023, \apj, 958,
  95, \dodoi{10.3847/1538-4357/ad0036}

\bibitem[{{Kendrew} {et~al.}(2015){Kendrew}, {Scheithauer}, {Bouchet},
  {Amiaux}, {Azzollini}, {Bouwman}, {Chen}, {Dubreuil}, {Fischer}, {Glasse},
  {Greene}, {Lagage}, {Lahuis}, {Ronayette}, {Wright}, \&
  {Wright}}]{Kendrew2015}
{Kendrew}, S., {Scheithauer}, S., {Bouchet}, P., {et~al.} 2015, \pasp, 127,
  623, \dodoi{10.1086/682255}

\bibitem[{{Kheirandish} \& {Murase}(2023)}]{Kheirandish2023}
{Kheirandish}, A., \& {Murase}, K. 2023, \apjl, 956, L8,
  \dodoi{10.3847/2041-8213/acf84f}

\bibitem[{{Kilpatrick} {et~al.}(2023){Kilpatrick}, {Foley},
  {Jacobson-Gal{\'a}n}, {Piro}, {Smartt}, {Drout}, {Gagliano}, {Gall},
  {Hjorth}, {Jones}, {Mandel}, {Margutti}, {Ramirez-Ruiz}, {Ransome}, {Villar},
  {Coulter}, {Gao}, {Matthews}, {Taggart}, \& {Zenati}}]{Kilpatrick2023}
{Kilpatrick}, C.~D., {Foley}, R.~J., {Jacobson-Gal{\'a}n}, W.~V., {et~al.}
  2023, \apjl, 952, L23, \dodoi{10.3847/2041-8213/ace4ca}

\bibitem[{{Kotak} {et~al.}(2005){Kotak}, {Meikle}, {van Dyk}, {H{\"o}flich}, \&
  {Mattila}}]{Kotak2005}
{Kotak}, R., {Meikle}, P., {van Dyk}, S.~D., {H{\"o}flich}, P.~A., \&
  {Mattila}, S. 2005, \apjl, 628, L123, \dodoi{10.1086/432719}

\bibitem[{{Kotak} {et~al.}(2006){Kotak}, {Meikle}, {Pozzo}, {van Dyk},
  {Farrah}, {Fesen}, {Filippenko}, {Foley}, {Fransson}, {Gerardy},
  {H{\"o}flich}, {Lundqvist}, {Mattila}, {Sollerman}, \& {Wheeler}}]{Kotak2006}
{Kotak}, R., {Meikle}, P., {Pozzo}, M., {et~al.} 2006, \apjl, 651, L117,
  \dodoi{10.1086/509655}

\bibitem[{{Kotak} {et~al.}(2009){Kotak}, {Meikle}, {Farrah}, {Gerardy},
  {Foley}, {Van Dyk}, {Fransson}, {Lundqvist}, {Sollerman}, {Fesen},
  {Filippenko}, {Mattila}, {Silverman}, {Andersen}, {H{\"o}flich}, {Pozzo}, \&
  {Wheeler}}]{Kotak2009_04et}
{Kotak}, R., {Meikle}, W.~P.~S., {Farrah}, D., {et~al.} 2009, \apj, 704, 306,
  \dodoi{10.1088/0004-637X/704/1/306}

\bibitem[{{Lantz} {et~al.}(2004){Lantz}, {Aldering}, {Antilogus}, {Bonnaud},
  {Capoani}, {Castera}, {Copin}, {Dubet}, {Gangler}, {Henault}, {Lemonnier},
  {Pain}, {Pecontal}, {Pecontal}, \& {Smadja}}]{Lantz2004}
{Lantz}, B., {Aldering}, G., {Antilogus}, P., {et~al.} 2004, in Society of
  Photo-Optical Instrumentation Engineers (SPIE) Conference Series, Vol. 5249,
  Optical Design and Engineering, ed. L.~{Mazuray}, P.~J. {Rogers}, \&
  R.~{Wartmann}, 146--155, \dodoi{10.1117/12.512493}

\bibitem[{{Larson} {et~al.}(1987){Larson}, {Drapatz}, {Mumma}, \&
  {Weaver}}]{Larson1987}
{Larson}, H.~P., {Drapatz}, S., {Mumma}, M.~J., \& {Weaver}, H.~A. 1987, in
  European Southern Observatory Conference and Workshop Proceedings, Vol.~26,
  European Southern Observatory Conference and Workshop Proceedings, 147

\bibitem[{{Larsson} {et~al.}(2023){Larsson}, {Fransson}, {Sargent}, {Jones},
  {Barlow}, {Bouchet}, {Meixner}, {Blommaert}, {Coulais}, {Fox}, {Gastaud},
  {Glasse}, {Habel}, {Hirschauer}, {Hjorth}, {Jaspers}, {Kavanagh}, {Krause},
  {Lau}, {Lenki{\'c}}, {Nayak}, {Rest}, {Temim}, {Tikkanen}, {Wesson}, \&
  {Wright}}]{Larsson2023}
{Larsson}, J., {Fransson}, C., {Sargent}, B., {et~al.} 2023, \apjl, 949, L27,
  \dodoi{10.3847/2041-8213/acd555}

\bibitem[{{Li} {et~al.}(2024){Li}, {Hu}, {Li}, {Yang}, {Wang}, {Yan}, {Hu},
  {Zhang}, {Mao}, {Riise}, {Gao}, {Sun}, {Liu}, {Xiong}, {Wang}, {Mo},
  {Iskandar}, {Xi}, {Xiang}, {Wang}, {Sun}, {Zhang}, {Chen}, {Lin}, {Guo},
  {Liu}, {Cai}, {Zhou}, {Zhao}, {Chen}, {Zheng}, {Li}, {Zhang}, {Xu}, {Lyu},
  {Castro-Tirado}, {Chufarin}, {Potapov}, {Ionov}, {Korotkiy}, {Nazarov},
  {Sokolovsky}, {Hamann}, \& {Herman}}]{Li2023}
{Li}, G., {Hu}, M., {Li}, W., {et~al.} 2024, \nat, 627, 754,
  \dodoi{10.1038/s41586-023-06843-6}

\bibitem[{{Li} {et~al.}(2020){Li}, {Wang}, {Fan}, {Wu}, {Jiang}, {Ba{\~n}ados},
  {Venemans}, {Shao}, {Li}, {Zhang}, {Zhang}, {Wagg}, {Decarli},
  {Mazzucchelli}, {Omont}, \& {Bertoldi}}]{Li2020}
{Li}, Q., {Wang}, R., {Fan}, X., {et~al.} 2020, \apj, 900, 12,
  \dodoi{10.3847/1538-4357/aba52d}

\bibitem[{{Liu} {et~al.}(2023){Liu}, {Chen}, {Er}, {Zeimann}, {Vink{\'o}},
  {Wheeler}, {Cooper}, {Davis}, {Farrow}, {Gebhardt}, {Guo}, {Hill}, {House},
  {Kollatschny}, {Kong}, {Kumar}, {Liu}, {Tuttle}, {Endl}, {Duke}, {Cochran},
  {Zhang}, \& {Liu}}]{Liu2023}
{Liu}, C., {Chen}, X., {Er}, X., {et~al.} 2023, \apjl, 958, L37,
  \dodoi{10.3847/2041-8213/ad0da8}

\bibitem[{{Lucy}(1991)}]{Lucy1991}
{Lucy}, L.~B. 1991, \apj, 383, 308, \dodoi{10.1086/170787}

\bibitem[{{Lucy} {et~al.}(1989){Lucy}, {Danziger}, {Gouiffes}, \&
  {Bouchet}}]{Lucy1989}
{Lucy}, L.~B., {Danziger}, I.~J., {Gouiffes}, C., \& {Bouchet}, P. 1989, in IAU
  Colloq. 120: Structure and Dynamics of the Interstellar Medium, ed.
  G.~{Tenorio-Tagle}, M.~{Moles}, \& J.~{Melnick}, Vol. 350 (Springer), 164,
  \dodoi{10.1007/BFb0114861}

\bibitem[{{Maiolino} {et~al.}(2004){Maiolino}, {Schneider}, {Oliva}, {Bianchi},
  {Ferrara}, {Mannucci}, {Pedani}, \& {Roca Sogorb}}]{Maiolino2004}
{Maiolino}, R., {Schneider}, R., {Oliva}, E., {et~al.} 2004, \nat, 431, 533,
  \dodoi{10.1038/nature02930}

\bibitem[{{Mao} {et~al.}(2023){Mao}, {Zhang}, {Cai}, {Chen}, {Chen}, {Gao},
  {Li}, {Lyu}, {Qin}, {Sun}, {Xu}, {Zhang}, {Zhang}, {Zhao}, {Zheng}, {Zhou},
  \& {Ye}}]{Mao2023}
{Mao}, Y., {Zhang}, M., {Cai}, G., {et~al.} 2023, Transient Name Server
  AstroNote, 130, 1

\bibitem[{{Martinez} {et~al.}(2024){Martinez}, {Bersten}, {Folatelli},
  {Orellana}, \& {Ertini}}]{Martinez2023}
{Martinez}, L., {Bersten}, M.~C., {Folatelli}, G., {Orellana}, M., \& {Ertini},
  K. 2024, \aap, 683, A154, \dodoi{10.1051/0004-6361/202348142}

\bibitem[{{Matsuura} {et~al.}(2019){Matsuura}, {De Buizer}, {Arendt}, {Dwek},
  {Barlow}, {Bevan}, {Cigan}, {Gomez}, {Rho}, {Wesson}, {Bouchet}, {Danziger},
  \& {Meixner}}]{Matsuura2019}
{Matsuura}, M., {De Buizer}, J.~M., {Arendt}, R.~G., {et~al.} 2019, \mnras,
  482, 1715, \dodoi{10.1093/mnras/sty2734}

\bibitem[{{Mazzali} {et~al.}(1992){Mazzali}, {Lucy}, \& {Butler}}]{Mazzali1992}
{Mazzali}, P.~A., {Lucy}, L.~B., \& {Butler}, K. 1992, \aap, 258, 399

\bibitem[{{Medler} {et~al.}(2023){Medler}, {Mazzali}, {Ashall}, {Teffs},
  {Shahbandeh}, \& {Shappee}}]{Medler2023}
{Medler}, K., {Mazzali}, P.~A., {Ashall}, C., {et~al.} 2023, \mnras, 518, L40,
  \dodoi{10.1093/mnrasl/slac127}

\bibitem[{{Medler} {et~al.}(2025{\natexlab{a}}){Medler}, {Ashall}, {Hoeflich},
  {Baron}, {DerKacy}, {Shahbandeh}, {Mera}, {Pfeffer}, {Hoogendam}, {Jones},
  {Shiber}, {Fereidouni}, {Fox}, {Jencson}, {Galbany}, {Hinkle}, {Tucker},
  {Shappee}, {Huber}, {Auchettl}, {Angus}, {Desai}, {Do}, {Payne}, {Shi},
  {Kong}, {Romagnoli}, {Syncatto}, {Clayton}, {Dulude}, {Engesser},
  {Filippenko}, {Gomez}, {Hsiao}, {de Jaeger}, {Johansson}, {Krisciunas},
  {Kumar}, {Lu}, {Matsuura}, {Mazzali}, {Milisavljevic}, {Morrell}, {O'Steen},
  {Park}, {Phillips}, {Ravi}, {Rest}, {Rho}, {Suntzeff}, {Sarangi}, {Smith},
  {Stritzinger}, {Strolger}, {Szalai}, {Temim}, {Tinyanont}, {Van Dyk}, {Wang},
  {Wang}, {Wesson}, {Yang}, \& {Zsiros}}]{Medler2025_23ixf}
{Medler}, K., {Ashall}, C., {Hoeflich}, P., {et~al.} 2025{\natexlab{a}}, arXiv
  e-prints, arXiv:2507.19727.
\newblock \doarXiv{2507.19727}

\bibitem[{{Medler} {et~al.}(2025{\natexlab{b}}){Medler}, {Ashall},
  {Shahbandeh}, {DerKacy}, {Hoogendam}, {Jones}, {Shappee}, {Hinkle},
  {Pfeffer}, {Baron}, {Hoeflich}, \& {Hsiao}}]{Medler2025_hiss}
{Medler}, K., {Ashall}, C., {Shahbandeh}, M., {et~al.} 2025{\natexlab{b}},
  arXiv e-prints, arXiv:2505.18507, \dodoi{10.48550/arXiv.2505.18507}

\bibitem[{{Meikle} {et~al.}(1989){Meikle}, {Allen}, {Spyromilio}, \&
  {Varani}}]{Meikle1989}
{Meikle}, W.~P.~S., {Allen}, D.~A., {Spyromilio}, J., \& {Varani}, G.~F. 1989,
  \mnras, 238, 193, \dodoi{10.1093/mnras/238.1.193}

\bibitem[{{Meikle} {et~al.}(2011){Meikle}, {Kotak}, {Farrah}, {Mattila}, {Van
  Dyk}, {Andersen}, {Fesen}, {Filippenko}, {Foley}, {Fransson}, {Gerardy},
  {H{\"o}flich}, {Lundqvist}, {Pozzo}, {Sollerman}, \& {Wheeler}}]{Meikle2011}
{Meikle}, W.~P.~S., {Kotak}, R., {Farrah}, D., {et~al.} 2011, \apj, 732, 109,
  \dodoi{10.1088/0004-637X/732/2/109}

\bibitem[{{Michel} {et~al.}(2025){Michel}, {Mazzali}, {Perley}, {Hinds}, \&
  {Wise}}]{Michel2025}
{Michel}, P.~D., {Mazzali}, P.~A., {Perley}, D.~A., {Hinds}, K.~R., \& {Wise},
  J.~L. 2025, \mnras, 539, 633, \dodoi{10.1093/mnras/staf443}

\bibitem[{{Miller} {et~al.}(2010){Miller}, {Smith}, {Li}, {Bloom}, {Chornock},
  {Filippenko}, \& {Prochaska}}]{Miller2010}
{Miller}, A.~A., {Smith}, N., {Li}, W., {et~al.} 2010, \aj, 139, 2218,
  \dodoi{10.1088/0004-6256/139/6/2218}

\bibitem[{{Moriya} \& {Singh}(2024)}]{Moriya2024}
{Moriya}, T.~J., \& {Singh}, A. 2024, \pasj, 76, 1050,
  \dodoi{10.1093/pasj/psae070}

\bibitem[{{M{\"u}ller} {et~al.}(2016){M{\"u}ller}, {Heger}, {Liptai}, \&
  {Cameron}}]{Muller2016}
{M{\"u}ller}, B., {Heger}, A., {Liptai}, D., \& {Cameron}, J.~B. 2016, \mnras,
  460, 742, \dodoi{10.1093/mnras/stw1083}

\bibitem[{{Nayana} {et~al.}(2025){Nayana}, {Margutti}, {Wiston}, {Chornock},
  {Campana}, {Laskar}, {Murase}, {Krips}, {Migliori}, {Tsuna}, {Alexander},
  {Chandra}, {Bietenholz}, {Berger}, {Chevalier}, {De Colle}, {Dessart},
  {Diesing}, {Grefenstette}, {Jacobson-Gal{\'a}n}, {Maeda}, {Marcote},
  {Matthews}, {Milisavljevic}, {Ray}, {Reguitti}, \& {Polzin}}]{Nayana2024}
{Nayana}, A.~J., {Margutti}, R., {Wiston}, E., {et~al.} 2025, \apj, 985, 51,
  \dodoi{10.3847/1538-4357/adc2fb}

\bibitem[{{Neustadt} {et~al.}(2024){Neustadt}, {Kochanek}, \&
  {Smith}}]{Neustadt2023}
{Neustadt}, J.~M.~M., {Kochanek}, C.~S., \& {Smith}, M.~R. 2024, \mnras, 527,
  5366, \dodoi{10.1093/mnras/stad3073}

\bibitem[{{Niu} {et~al.}(2023){Niu}, {Sun}, {Maund}, {Zhang}, {Zhao}, \&
  {Liu}}]{Niu2023}
{Niu}, Z., {Sun}, N.-C., {Maund}, J.~R., {et~al.} 2023, \apjl, 955, L15,
  \dodoi{10.3847/2041-8213/acf4e3}

\bibitem[{{Panjkov} {et~al.}(2024){Panjkov}, {Auchettl}, {Shappee}, {Do},
  {Lopez}, \& {Beacom}}]{Panjkov2023}
{Panjkov}, S., {Auchettl}, K., {Shappee}, B.~J., {et~al.} 2024, \pasa, 41,
  e059, \dodoi{10.1017/pasa.2024.66}

\bibitem[{{Park} {et~al.}(2025){Park}, {Rho}, {Yoon}, {Pearson}, {Shrestha},
  {Tinyanont}, {Geballe}, {Foley}, {Ravi}, {Andrews}, {Sand}, {Bostroem},
  {Ashall}, {Hoeflich}, {Valenti}, {Dong}, {Meza Retamal}, {Hoang}, {Mehta},
  {Howell}, {Farah}, {Terreran}, {Padilla Gonzalez}, {Andrews}, {Newsome},
  {Shahbandeh}, {Smith}, {Kang}, {Suntzeff}, {Baron}, {Medler}, {Mera Evans},
  {DerKacy}, {Larison}, {Galbany}, \& {Jacobson-Galan}}]{Park2025}
{Park}, S.~H., {Rho}, J., {Yoon}, S.-C., {et~al.} 2025, arXiv e-prints,
  arXiv:2507.11877.
\newblock \doarXiv{2507.11877}

\bibitem[{{Pastorello} {et~al.}(2009){Pastorello}, {Crockett}, {Martin},
  {Smartt}, {Altavilla}, {Benetti}, {Botticella}, {Cappellaro}, {Mattila},
  {Maund}, {Ryder}, {Salvo}, {Taubenberger}, \& {Turatto}}]{Pastorello2009}
{Pastorello}, A., {Crockett}, R.~M., {Martin}, R., {et~al.} 2009, \aap, 500,
  1013, \dodoi{10.1051/0004-6361/200911993}

\bibitem[{{Perley} \& {Gal-Yam}(2023)}]{Perley2023}
{Perley}, D., \& {Gal-Yam}, A. 2023, Transient Name Server Classification
  Report, 2023-1164, 1

\bibitem[{{Phillips} \& {Heathcote}(1989)}]{Phillips1989}
{Phillips}, M.~M., \& {Heathcote}, S.~R. 1989, \pasp, 101, 137,
  \dodoi{10.1086/132414}

\bibitem[{{Pledger} \& {Shara}(2023)}]{Pledger2023}
{Pledger}, J.~L., \& {Shara}, M.~M. 2023, \apjl, 953, L14,
  \dodoi{10.3847/2041-8213/ace88b}

\bibitem[{{Pozzo} {et~al.}(2004){Pozzo}, {Meikle}, {Fassia}, {Geballe},
  {Lundqvist}, {Chugai}, \& {Sollerman}}]{Pozzo2004}
{Pozzo}, M., {Meikle}, W.~P.~S., {Fassia}, A., {et~al.} 2004, \mnras, 352, 457,
  \dodoi{10.1111/j.1365-2966.2004.07951.x}

\bibitem[{{Qin} {et~al.}(2024){Qin}, {Zhang}, {Bloom}, {Sollerman},
  {Zimmerman}, {Irani}, {Schulze}, {Gal-Yam}, {Kasliwal}, {Coughlin}, {Perley},
  {Fremling}, \& {Kulkarni}}]{Qin2023}
{Qin}, Y.-J., {Zhang}, K., {Bloom}, J., {et~al.} 2024, \mnras, 534, 271,
  \dodoi{10.1093/mnras/stae2012}

\bibitem[{{Rank} {et~al.}(1988){Rank}, {Bregman}, {Witteborn}, {Cohen},
  {Lynch}, \& {Russell}}]{Rank1988}
{Rank}, D.~M., {Bregman}, J., {Witteborn}, F.~C., {et~al.} 1988, \apjl, 325,
  L1, \dodoi{10.1086/185096}

\bibitem[{{Ransome} {et~al.}(2024){Ransome}, {Villar}, {Tartaglia}, {Gonzalez},
  {Jacobson-Gal{\'a}n}, {Kilpatrick}, {Margutti}, {Foley}, {Grayling}, {Ni},
  {Yarza}, {Ye}, {Auchettl}, {de Boer}, {Chambers}, {Coulter}, {Drout},
  {Farias}, {Gall}, {Gao}, {Huber}, {Ibik}, {Jones}, {Khetan}, {Lin},
  {Politsch}, {Raimundo}, {Rest}, {Wainscoat}, {Yadavalli}, \&
  {Zenati}}]{Ransome2023}
{Ransome}, C.~L., {Villar}, V.~A., {Tartaglia}, A., {et~al.} 2024, \apj, 965,
  93, \dodoi{10.3847/1538-4357/ad2df7}

\bibitem[{{Ravensburg} {et~al.}(2024){Ravensburg}, {Carenza}, {Eckner}, \&
  {Goobar}}]{Muller2023}
{Ravensburg}, E., {Carenza}, P., {Eckner}, C., \& {Goobar}, A. 2024, \prd, 109,
  023018, \dodoi{10.1103/PhysRevD.109.023018}

\bibitem[{{Rho} {et~al.}(2018){Rho}, {Geballe}, {Banerjee}, {Dessart}, {Evans},
  \& {Joshi}}]{Rho2018}
{Rho}, J., {Geballe}, T.~R., {Banerjee}, D.~P.~K., {et~al.} 2018, \apjl, 864,
  L20, \dodoi{10.3847/2041-8213/aad77f}

\bibitem[{{Riess} {et~al.}(2022){Riess}, {Yuan}, {Macri}, {Scolnic}, {Brout},
  {Casertano}, {Jones}, {Murakami}, {Anand}, {Breuval}, {Brink}, {Filippenko},
  {Hoffmann}, {Jha}, {D'arcy Kenworthy}, {Mackenty}, {Stahl}, \&
  {Zheng}}]{Riess2022}
{Riess}, A.~G., {Yuan}, W., {Macri}, L.~M., {et~al.} 2022, \apjl, 934, L7,
  \dodoi{10.3847/2041-8213/ac5c5b}

\bibitem[{{Roche} {et~al.}(1993){Roche}, {Aitken}, \& {Smith}}]{Roche1993}
{Roche}, P.~F., {Aitken}, D.~K., \& {Smith}, C.~H. 1993, \mnras, 261, 522,
  \dodoi{10.1093/mnras/261.3.522}

\bibitem[{{Sarangi} \& {Cherchneff}(2013)}]{Sarangi2013}
{Sarangi}, A., \& {Cherchneff}, I. 2013, \apj, 776, 107,
  \dodoi{10.1088/0004-637X/776/2/107}

\bibitem[{{Sarangi} {et~al.}(2018){Sarangi}, {Dwek}, \&
  {Arendt}}]{Sarangi2018_sn2010jl}
{Sarangi}, A., {Dwek}, E., \& {Arendt}, R.~G. 2018, \apj, 859, 66,
  \dodoi{10.3847/1538-4357/aabfc3}

\bibitem[{{Sarmah}(2024)}]{Sarmah2023}
{Sarmah}, P. 2024, \jcap, 2024, 083, \dodoi{10.1088/1475-7516/2024/04/083}

\bibitem[{{Schlafly} \& {Finkbeiner}(2011)}]{Schlafly2011}
{Schlafly}, E.~F., \& {Finkbeiner}, D.~P. 2011, \apj, 737, 103,
  \dodoi{10.1088/0004-637X/737/2/103}

\bibitem[{{Serrano-Hern{\'a}ndez} {et~al.}(2025){Serrano-Hern{\'a}ndez},
  {Mart{\'\i}nez-Gonz{\'a}lez}, {Jim{\'e}nez}, {Silich}, \&
  {W{\"u}nsch}}]{Serrano-Hernandez2025}
{Serrano-Hern{\'a}ndez}, D.~B., {Mart{\'\i}nez-Gonz{\'a}lez}, S.,
  {Jim{\'e}nez}, S., {Silich}, S., \& {W{\"u}nsch}, R. 2025, \aap, 695, A271,
  \dodoi{10.1051/0004-6361/202449717}

\bibitem[{{Shahbandeh} {et~al.}(2022){Shahbandeh}, {Hsiao}, {Ashall}, {Teffs},
  {Hoeflich}, {Morrell}, {Phillips}, {Anderson}, {Baron}, {Burns}, {Contreras},
  {Davis}, {Diamond}, {Folatelli}, {Galbany}, {Gall}, {Hachinger}, {Holmbo},
  {Karamehmetoglu}, {Kasliwal}, {Kirshner}, {Krisciunas}, {Kumar}, {Lu},
  {Marion}, {Mazzali}, {Piro}, {Sand}, {Stritzinger}, {Suntzeff}, {Taddia}, \&
  {Uddin}}]{Shahbandeh2022}
{Shahbandeh}, M., {Hsiao}, E.~Y., {Ashall}, C., {et~al.} 2022, \apj, 925, 175,
  \dodoi{10.3847/1538-4357/ac4030}

\bibitem[{{Shahbandeh} {et~al.}(2023){Shahbandeh}, {Sarangi}, {Temim},
  {Szalai}, {Fox}, {Tinyanont}, {Dwek}, {Dessart}, {Filippenko}, {Brink},
  {Foley}, {Jencson}, {Pierel}, {Zs{\'\i}ros}, {Rest}, {Zheng}, {Andrews},
  {Clayton}, {De}, {Engesser}, {Gezari}, {Gomez}, {Gonzaga}, {Johansson},
  {Kasliwal}, {Lau}, {De Looze}, {Marston}, {Milisavljevic}, {O'Steen},
  {Siebert}, {Skrutskie}, {Smith}, {Strolger}, {Van Dyk}, {Wang}, {Williams},
  {Williams}, {Xiao}, \& {Yang}}]{Shahbandeh2023}
{Shahbandeh}, M., {Sarangi}, A., {Temim}, T., {et~al.} 2023, \mnras, 523, 6048,
  \dodoi{10.1093/mnras/stad1681}

\bibitem[{{Shahbandeh} {et~al.}(2024){Shahbandeh}, {Ashall}, {Hoeflich},
  {Baron}, {Fox}, {Mera}, {DerKacy}, {Stritzinger}, {Shappee}, {Law},
  {Morrison}, {Pauly}, {Pierel}, {Medler}, {Andrews}, {Baade}, {Bostroem},
  {Brown}, {Burns}, {Burrow}, {Cikota}, {Cross}, {Davis}, {de Jaeger}, {Do},
  {Dong}, {Hsiao}, {Dominguez}, {Galbany}, {Janzen}, {Jencson}, {Hoang},
  {Karamehmetoglu}, {Khaghani}, {Krisciunas}, {Kumar}, {Lu}, {Mazzali},
  {Morrell}, {Patat}, {Pearson}, {Pfeffer}, {Wang}, {Yang}, {Cai},
  {Camacho-Neves}, {Elias-Rosa}, {Lundquist}, {Maund}, {Phillips}, {Rest},
  {Retamal}, {Stangl}, {Shrestha}, {Stevens}, {Suntzeff}, {Telesco}, {Tucker},
  {Foley}, {Jha}, {Kwok}, {Larison}, {LeBaron}, {Moran}, {Rho}, {Salmaso},
  {Schmidt}, \& {Tinyanont}}]{Shahbandeh2024_22acko}
{Shahbandeh}, M., {Ashall}, C., {Hoeflich}, P., {et~al.} 2024, arXiv e-prints,
  arXiv:2401.14474, \dodoi{10.48550/arXiv.2401.14474}

\bibitem[{{Shahbandeh} {et~al.}(2025){Shahbandeh}, {Fox}, {Temim}, {Dwek},
  {Sarangi}, {Smith}, {Dessart}, {Nickson}, {Engesser}, {Filippenko}, {Brink},
  {Zheng}, {Szalai}, {Johansson}, {Rest}, {Van Dyk}, {Andrews}, {Ashall},
  {Clayton}, {De Looze}, {DerKacy}, {Dulude}, {Foley}, {Gezari}, {Gomez},
  {Gonzaga}, {Indukuri}, {Jencson}, {Kasliwal}, {Lane}, {Lau}, {Law},
  {Marston}, {Milisavljevic}, {O'Steen}, {Pierel}, {Siebert}, {Skrutskie},
  {Strolger}, {Tinyanont}, {Wang}, {Williams}, {Xiao}, {Yang}, \&
  {Zs{\'\i}ros}}]{Shahbandeh2025_05ip}
{Shahbandeh}, M., {Fox}, O.~D., {Temim}, T., {et~al.} 2025, \apj, 985, 262,
  \dodoi{10.3847/1538-4357/adce77}

\bibitem[{{Singh} {et~al.}(2024){Singh}, {Teja}, {Moriya}, {Maeda}, {Kawabata},
  {Tanaka}, {Imazawa}, {Nakaoka}, {Gangopadhyay}, {Yamanaka}, {Swain}, {Sahu},
  {Anupama}, {Kumar}, {Anche}, {Sano}, {Raj}, {Agnihotri}, {Bhalerao}, {Bisht},
  {Bisht}, {Belwal}, {Chakrabarti}, {Fujii}, {Nagayama}, {Matsumoto}, {Hamada},
  {Kawabata}, {Kumar}, {Kumar}, {Malkan}, {Smith}, {Sakagami}, {Taguchi},
  {Tominaga}, \& {Watanabe}}]{Singh2024}
{Singh}, A., {Teja}, R.~S., {Moriya}, T.~J., {et~al.} 2024, \apj, 975, 132,
  \dodoi{10.3847/1538-4357/ad7955}

\bibitem[{{Slavin} {et~al.}(2015){Slavin}, {Dwek}, \& {Jones}}]{Slavin2015}
{Slavin}, J.~D., {Dwek}, E., \& {Jones}, A.~P. 2015, \apj, 803, 7,
  \dodoi{10.1088/0004-637X/803/1/7}

\bibitem[{{Slavin} {et~al.}(2020){Slavin}, {Dwek}, {Mac Low}, \&
  {Hill}}]{Slavin2020}
{Slavin}, J.~D., {Dwek}, E., {Mac Low}, M.-M., \& {Hill}, A.~S. 2020, \apj,
  902, 135, \dodoi{10.3847/1538-4357/abb5a4}

\bibitem[{{Smartt}(2015)}]{Smartt2015}
{Smartt}, S.~J. 2015, \pasa, 32, e016, \dodoi{10.1017/pasa.2015.17}

\bibitem[{{Smith} {et~al.}(2023){Smith}, {Pearson}, {Sand}, {Ilyin},
  {Bostroem}, {Hosseinzadeh}, \& {Shrestha}}]{Smith2023}
{Smith}, N., {Pearson}, J., {Sand}, D.~J., {et~al.} 2023, \apj, 956, 46,
  \dodoi{10.3847/1538-4357/acf366}

\bibitem[{{Smith} {et~al.}(2008){Smith}, {Foley}, {Bloom}, {Li}, {Filippenko},
  {Gavazzi}, {Ghez}, {Konopacky}, {Malkan}, {Marshall}, {Pooley}, {Treu}, \&
  {Woo}}]{Smith2008}
{Smith}, N., {Foley}, R.~J., {Bloom}, J.~S., {et~al.} 2008, \apj, 686, 485,
  \dodoi{10.1086/590141}

\bibitem[{{Smith} {et~al.}(2009){Smith}, {Silverman}, {Chornock}, {Filippenko},
  {Wang}, {Li}, {Ganeshalingam}, {Foley}, {Rex}, \& {Steele}}]{Smith2009}
{Smith}, N., {Silverman}, J.~M., {Chornock}, R., {et~al.} 2009, \apj, 695,
  1334, \dodoi{10.1088/0004-637X/695/2/1334}

\bibitem[{{Soker}(2023)}]{Soker2023}
{Soker}, N. 2023, Research in Astronomy and Astrophysics, 23, 081002,
  \dodoi{10.1088/1674-4527/ace51f}

\bibitem[{{Soraisam} {et~al.}(2023){Soraisam}, {Szalai}, {Van Dyk}, {Andrews},
  {Srinivasan}, {Chun}, {Matheson}, {Scicluna}, \&
  {Vasquez-Torres}}]{Soraisam2023}
{Soraisam}, M.~D., {Szalai}, T., {Van Dyk}, S.~D., {et~al.} 2023, \apj, 957,
  64, \dodoi{10.3847/1538-4357/acef22}

\bibitem[{{Spyromilio} {et~al.}(1988){Spyromilio}, {Meikle}, {Learner}, \&
  {Allen}}]{Spyromilio1988}
{Spyromilio}, J., {Meikle}, W.~P.~S., {Learner}, R.~C.~M., \& {Allen}, D.~A.
  1988, \nat, 334, 327, \dodoi{10.1038/334327a0}

\bibitem[{{Stritzinger} {et~al.}(2023){Stritzinger}, {Valerin}, {Elias-Rosa},
  {Fraser}, {Galbany}, {Gutierrez}, {Kankare}, {Kotak}, {Moran}, {Lundqvist},
  {Matilainen}, {Reguitti}, {Reynolds}, {Salmaso}, \&
  {Shappee}}]{Stritzinger2023}
{Stritzinger}, M., {Valerin}, G., {Elias-Rosa}, N., {et~al.} 2023, Transient
  Name Server AstroNote, 145, 1

\bibitem[{{Szalai} \& {Vink{\'o}}(2013)}]{Szalai2013}
{Szalai}, T., \& {Vink{\'o}}, J. 2013, \aap, 549, A79,
  \dodoi{10.1051/0004-6361/201220015}

\bibitem[{{Szalai} {et~al.}(2019){Szalai}, {Vink{\'o}}, {K{\"o}nyves-T{\'o}th},
  {Nagy}, {Bostroem}, {S{\'a}rneczky}, {Brown}, {Pejcha}, {B{\'o}di}, {Cseh},
  {Cs{\"o}rnyei}, {Dencs}, {Hanyecz}, {Ign{\'a}cz}, {Kalup}, {Kriskovics},
  {Ordasi}, {P{\'a}l}, {Seli}, {S{\'o}dor}, {Szak{\'a}ts}, {Vida}, {Zsidi},
  {Konkoly Team}, {Arcavi}, {Ashall}, {Burke}, {Galbany}, {Hiramatsu},
  {Hosseinzadeh}, {Hsiao}, {Howell}, {McCully}, {Moran}, {Rho}, {Sand},
  {Shahbandeh}, {Valenti}, {Wang}, {Wheeler}, \& {Supernova
  Project}}]{Szalai2019}
{Szalai}, T., {Vink{\'o}}, J., {K{\"o}nyves-T{\'o}th}, R., {et~al.} 2019, \apj,
  876, 19, \dodoi{10.3847/1538-4357/ab12d0}

\bibitem[{{Teffs} {et~al.}(2020){Teffs}, {Ertl}, {Mazzali}, {Hachinger}, \&
  {Janka}}]{Teffs2020}
{Teffs}, J., {Ertl}, T., {Mazzali}, P., {Hachinger}, S., \& {Janka}, H.~T.
  2020, \mnras, 499, 730, \dodoi{10.1093/mnras/staa2549}

\bibitem[{{Teja} {et~al.}(2023){Teja}, {Singh}, {Basu}, {Anupama}, {Sahu},
  {Dutta}, {Swain}, {Nakaoka}, {Pathak}, {Bhalerao}, {Barway}, {Kumar},
  {A.~J.}, {Imazawa}, {Kumar}, \& {Kawabata}}]{SinghTeja2023}
{Teja}, R.~S., {Singh}, A., {Basu}, J., {et~al.} 2023, \apjl, 954, L12,
  \dodoi{10.3847/2041-8213/acef20}

\bibitem[{{Tinyanont} {et~al.}(2019){Tinyanont}, {Kasliwal}, {Krafton}, {Lau},
  {Rho}, {Leonard}, {De}, {Jencson}, {Mawet}, {Millar-Blanchaer}, {Nilsson},
  {Yan}, {Gehrz}, {Helou}, {Van Dyk}, {Serabyn}, {Fox}, \&
  {Clayton}}]{Tinyanont2019}
{Tinyanont}, S., {Kasliwal}, M.~M., {Krafton}, K., {et~al.} 2019, \apj, 873,
  127, \dodoi{10.3847/1538-4357/ab0897}

\bibitem[{{Tsuna} {et~al.}(2025){Tsuna}, {Fuller}, \& {Lu}}]{Tsuna2025}
{Tsuna}, D., {Fuller}, J., \& {Lu}, W. 2025, arXiv e-prints, arXiv:2508.21116,
  \dodoi{10.48550/arXiv.2508.21116}

\bibitem[{{Tucker} {et~al.}(2022){Tucker}, {Shappee}, {Huber}, {Payne}, {Do},
  {Hinkle}, {de Jaeger}, {Ashall}, {Desai}, {Hoogendam}, {Aldering},
  {Auchettl}, {Baranec}, {Bulger}, {Chambers}, {Chun}, {Hodapp}, {Lowe},
  {McKay}, {Rampy}, {Rubin}, \& {Tonry}}]{Tucker2022}
{Tucker}, M.~A., {Shappee}, B.~J., {Huber}, M.~E., {et~al.} 2022, \pasp, 134,
  124502, \dodoi{10.1088/1538-3873/aca719}

\bibitem[{{Utrobin} {et~al.}(1995){Utrobin}, {Chugai}, \&
  {Andronova}}]{Utrobin1995}
{Utrobin}, V.~P., {Chugai}, N.~N., \& {Andronova}, A.~A. 1995, \aap, 295, 129

\bibitem[{{Van Dyk} {et~al.}(2024{\natexlab{a}}){Van Dyk}, {Szalai}, {Cutri},
  {Kirkpatrick}, {Grillmair}, {Fajardo-Acosta}, {Masiero}, {Mainzer}, {Gelino},
  {Vink{\'o}}, {Jo{\'o}}, {P{\'a}l}, {K{\"o}nyves-T{\'o}th}, {Kriskovics},
  {Szak{\'a}ts}, {Vida}, {Zheng}, {Brink}, \& {Filippenko}}]{VanDyk2024}
{Van Dyk}, S.~D., {Szalai}, T., {Cutri}, R.~M., {et~al.} 2024{\natexlab{a}},
  \apj, 977, 98, \dodoi{10.3847/1538-4357/ad8cd8}

\bibitem[{{Van Dyk} {et~al.}(2024{\natexlab{b}}){Van Dyk}, {Srinivasan},
  {Andrews}, {Soraisam}, {Szalai}, {Howell}, {Isaacson}, {Matheson},
  {Petigura}, {Scicluna}, {Stephens}, {Van Zandt}, {Zheng}, {Chun}, \&
  {Fillippenko}}]{VanDyk2023}
{Van Dyk}, S.~D., {Srinivasan}, S., {Andrews}, J.~E., {et~al.}
  2024{\natexlab{b}}, \apj, 968, 27, \dodoi{10.3847/1538-4357/ad414b}

\bibitem[{{Vasylyev} {et~al.}(2023){Vasylyev}, {Yang}, {Filippenko}, {Patra},
  {Brink}, {Wang}, {Chornock}, {Margutti}, {Gates}, {Burgasser}, {Karpoor},
  {LeBaron}, {Softich}, {Theissen}, {Wiston}, \& {Zheng}}]{Vasylyev2023}
{Vasylyev}, S.~S., {Yang}, Y., {Filippenko}, A.~V., {et~al.} 2023, \apjl, 955,
  L37, \dodoi{10.3847/2041-8213/acf1a3}

\bibitem[{{Virtanen} {et~al.}(2020){Virtanen}, {Gommers}, {Oliphant},
  {Haberland}, {Reddy}, {Cournapeau}, {Burovski}, {Peterson}, {Weckesser},
  {Bright}, {van der Walt}, {Brett}, {Wilson}, {Millman}, {Mayorov}, {Nelson},
  {Jones}, {Kern}, {Larson}, {Carey}, {Polat}, {Feng}, {Moore}, {VanderPlas},
  {Laxalde}, {Perktold}, {Cimrman}, {Henriksen}, {Quintero}, {Harris},
  {Archibald}, {Ribeiro}, {Pedregosa}, {van Mulbregt}, \& {SciPy 1. 0
  Contributors}}]{SciPy2020}
{Virtanen}, P., {Gommers}, R., {Oliphant}, T.~E., {et~al.} 2020, Nature
  Methods, 17, 261, \dodoi{10.1038/s41592-019-0686-2}

\bibitem[{{Wooden} {et~al.}(1993){Wooden}, {Rank}, {Bregman}, {Witteborn},
  {Tielens}, {Cohen}, {Pinto}, \& {Axelrod}}]{Wooden1993}
{Wooden}, D.~H., {Rank}, D.~M., {Bregman}, J.~D., {et~al.} 1993, \apjs, 88,
  477, \dodoi{10.1086/191830}

\bibitem[{{Woosley} {et~al.}(2002){Woosley}, {Heger}, \&
  {Weaver}}]{Woosley2002}
{Woosley}, S.~E., {Heger}, A., \& {Weaver}, T.~A. 2002, Reviews of Modern
  Physics, 74, 1015, \dodoi{10.1103/RevModPhys.74.1015}

\bibitem[{{Yamanaka} {et~al.}(2023){Yamanaka}, {Fujii}, \&
  {Nagayama}}]{Yamanaka2023}
{Yamanaka}, M., {Fujii}, M., \& {Nagayama}, T. 2023, \pasj, 75, L27,
  \dodoi{10.1093/pasj/psad051}

\bibitem[{{Yaron} {et~al.}(2023){Yaron}, {Bruch}, {Chen}, {Irani}, {Zimmerman},
  {Gal-Yam}, \& {Qin}}]{Yaron2023}
{Yaron}, O., {Bruch}, R., {Chen}, P., {et~al.} 2023, Transient Name Server
  AstroNote, 133, 1

\bibitem[{{Zapartas} {et~al.}(2021){Zapartas}, {de Mink}, {Justham}, {Smith},
  {Renzo}, \& {de Koter}}]{Zapartas2021}
{Zapartas}, E., {de Mink}, S.~E., {Justham}, S., {et~al.} 2021, \aap, 645, A6,
  \dodoi{10.1051/0004-6361/202037744}

\bibitem[{{Zhang} {et~al.}(2023){Zhang}, {Lin}, {Wang}, {Zhao}, {Li}, {Liu},
  {Yan}, {Xiang}, {Wang}, \& {Bai}}]{Zhang2023}
{Zhang}, J., {Lin}, H., {Wang}, X., {et~al.} 2023, Science Bulletin, 68, 2548,
  \dodoi{10.1016/j.scib.2023.09.015}

\bibitem[{{Zimmerman} {et~al.}(2024){Zimmerman}, {Irani}, {Chen}, {Gal-Yam},
  {Schulze}, {Perley}, {Sollerman}, {Filippenko}, {Shenar}, {Yaron}, {Shahaf},
  {Bruch}, {Ofek}, {De Cia}, {Brink}, {Yang}, {Vasylyev}, {Ben Ami}, {Aubert},
  {Badash}, {Bloom}, {Brown}, {De}, {Dimitriadis}, {Fransson}, {Fremling},
  {Hinds}, {Horesh}, {Johansson}, {Kasliwal}, {Kulkarni}, {Kushnir}, {Martin},
  {Matuzewski}, {McGurk}, {Miller}, {Morag}, {Neil}, {Nugent}, {Post},
  {Prusinski}, {Qin}, {Raichoor}, {Riddle}, {Rowe}, {Rusholme}, {Sfaradi},
  {Sjoberg}, {Soumagnac}, {Stein}, {Strotjohann}, {Terwel}, {Wasserman},
  {Wise}, {Wold}, {Yan}, \& {Zhang}}]{Zimmerman2023}
{Zimmerman}, E.~A., {Irani}, I., {Chen}, P., {et~al.} 2024, \nat, 627, 759,
  \dodoi{10.1038/s41586-024-07116-6}

\end{thebibliography}



\end{document}